\documentclass[aps,prb,twocolumn,superscriptaddress,showpacs]{revtex4}
\usepackage{graphicx} %Include figure files
\usepackage{amsmath}
\usepackage{color}
\newcommand{\be}{\begin{equation}}
\newcommand{\ee}{\end{equation}}
\newcommand{\bea}{\begin{eqnarray}}
\newcommand{\eea}{\end{eqnarray}}
\newcommand{\bse}{\begin{subequations}}
\newcommand{\ese}{\end{subequations}}

\begin{document}

\title{Thermodynamics of the Noninteracting Bose Gas in a Two-Dimensional Box}

\author{Heqiu Li}
\author{Qiujiang Guo}
\author{Ji Jiang}
\affiliation{Department of Physics, Zhejiang University, Hangzhou 310027, China}
\author{D. C. Johnston}
\altaffiliation{johnston@ameslab.gov}
\affiliation{Department of Physics and Astronomy and Ames Laboratory, Iowa State University, Ames, Iowa 50011, USA}

\date{\today}

\begin{abstract}

Bose-Einstein condensation (BEC) of a noninteracting Bose gas of $N$ particles in a two-dimensional box with Dirichlet boundary conditions is studied.  Confirming previous work, we find that BEC occurs at finite $N$ at low temperatures~$T$ without the occurrence of a phase transition.  The conventionally-defined transition temperature $T_{\rm E}$ for an infinite 3D system is shown to correspond in a 2D system with finite $N$ to a crossover temperature between a slow and rapid increase in the fractional boson occupation $N_0/N$ of the ground state with decreasing~$T$\@.  We further show that $T_{\rm E}\sim 1/\log N$ at fixed area per boson, so in the thermodynamic limit there is no significant BEC in 2D at finite~$T$\@.  Thus, paradoxically, BEC only occurs in 2D at finite~$N$ with no phase transition associated with it.  Calculations of thermodynamic properties versus $T$ and area~$A$ are presented, including Helmholtz free energy, entropy~$S$, pressure~$p$, ratio of $p$ to the energy density $U/A$, heat capacity at constant volume (area) $C_{\rm V}$ and at constant pressure $C_{\rm p}$, isothermal compressibility $\kappa_{\rm T}$ and thermal expansion coefficient $\alpha_{\rm p}$, obtained using both the grand canonical ensemble (GCE) and canonical ensemble (CE) formalisms.  The GCE formalism gives acceptable predictions for $S$, $p$, $p/(U/A)$, $\kappa_{\rm T}$ and $\alpha_{\rm p}$ at large~$N$, $T$ and $A$, but fails for smaller values of these three parameters for which BEC becomes significant, whereas the CE formalism gives accurate results for all thermodynamic properties of finite systems even at low~$T$ and/or~$A$ where BEC occurs.

\end{abstract}

\pacs{05.30.Ch, 05.30.Jp, 03.74.Hh}

\maketitle

\section{Introduction}

Some of the thermodynamic properties of a noninteracting three-dimensional (3D) Bose gas without internal degrees of freedom and confined in a cubic box are well known, where macroscopic Bose-Einstein condensation (BEC) occurs in the thermodynamic limit below a phase transition temperature $T_{\rm E}$.\cite{London1938, London1938b, deGroot1950, Huang1963, Ziff1977, Kittel1980,Schroeder2000, Griffin1995} The experimental work on BEC greatly increased after the initial discoveries in 1995 of BEC in ultracold atomic gases confined to harmonic potential traps.\cite{Anderson1995, Bradley1995, Davis1995}  These discoveries led to much additional theoretical work on BEC in ultracold gases in harmonic traps that are in general anisotropic along the three Cartesian axes, especially including the effects of boson interactions.\cite{Griffin1995, Pitaevskii2003, Petrov2004, Leggett2006, Pethick2008} 

Theoretical studies of BEC have been carried out for 1D and 2D Bose gases,\cite{Petrov2004} which are relevant to the above experiments on ultracold trapped atomic gases.  Here we take the order parameter of a BEC phase transition to be the fraction $N_0/N$ of the boson occupation of the ground state $N_0$.  A BEC phase transition occurs if the dependences of $N_0/N$ and associated thermodynamic properties of the Bose gas on temperature~$T$ are nonanalytic at a temperature defined as the BEC transition temperature $T_{\rm E}$\@.  In 1967, Hohenberg showed that BEC cannot occur in 1D or 2D at finite temperature $T$ in the thermodynamic limit for a homogeneous Bose gas.\cite{Hohenberg1967}  However, this result does not rule out BEC in the thermodynamic limit in inhomogenous Bose gases.  Indeed, Bagnato and Kleppner showed in 1991 that a BEC phase transition occurs in 1D and 2D noninteracting Bose gases in power-law-potential traps.\cite{Bagnato1991}  Furthermore, BEC can occur in a finite 2D system containing a finite number~$N$ of bosons at finite $T$\@.  In such a system, BEC entails a smooth increase in the boson occupation of the ground state (and also excited states) with decreasing $T$ with no BEC phase transition occurring.  In particular, Ketterle and van Druten studied 3D and 1D boson systems in a harmonic potential for finite~$N$.\cite{Ketterle1996}  In addition to confirming that a BEC phase transition occurs in the thermodynamic limit in 1D, they found that a smooth increase of BEC occurs in 1D systems with decreasing~$T$ and finite~$N$, i.e., without a BEC phase transition occurring.  

Less well studied is BEC of noninteracting bosons confined to a 2D box with Dirichlet boundary conditions where the wave function of each boson is zero at the edges of the box.  Ziff et al.\ presented results for the pressure versus volume $p(V)$ and heat capacity at constant volume $C_{\rm V}(T)$ in the thermodynamic limit for dimensions 1 to~5.\cite{Ziff1977}  Ingold and Lambrecht\cite{Ingold1998} found that a BEC phase transition does not occur in 1D or 2D for a finite number $N$ of bosons with $N = 10^2-10^7$, even though BEC itself does occur at low~$T$\@.  Deng and Hui calculated $C_{\rm V}(T)$ for finite 2D systems with $1\leq N \leq 10^3$ and also found a smooth increase of BEC with decreasing~$T$, with no evidence for a temperature-induced phase transition.\cite{Deng1997}  With Dirichlet boundary conditions, one expects a nonzero pressure at $T=0$ in a finite 2D Bose gas,\cite{Grossman1995} whereas a zero pressure is obtained by setting the ground state energy to zero or by using the grand canonical ensemble (GCE) formalism instead of the canonical ensemble (CE) formalism.  Such studies of bosons in a 2D box are not just of pedagogical interest, since over the past few years cold-atom traps have been constructed with 2D box-like potentials.\cite{Gaunt2013, Gotlibovych2014, Chomaz2015}

Most theoretical studies of BEC in 2D have been on systems confined to harmonic traps.  In these systems pressure and volume are not relevant thermodynamic variables, and hence the associated thermodynamic properties isothermal compressibility $\kappa_{\rm T}$, coefficient of thermal expansion $\alpha_{\rm p}$ and heat capacity at constant pressure $C_{\rm p}$ are also not relevant. On the other hand, for bosons in a 2D box these thermodynamic variables and properties are appropriate.  Here we report a comprehensive study of the thermodynamics of the noninteracting Bose gas in a 2D box with finite~$N$ and Dirichlet boundary conditions using both the GCE and CE formalisms.  We present studies of the various thermodynamic properties versus $T$, $A$ and $N$, as well as of the populations of the ground and low-lying excited states.  For finite $N$ and~$A$, all properties versus $T$ and $A$ must be analytic and finite as discussed above.  Of special interest is how these properties behave for $T\to0$ and $T\to\infty$, where in the former limit the properties should be physically acceptable and in latter limit should correspond to those of the (classical) ideal gas.  As is well known, the GCE formalism gives unphysically large fluctuations in $N$ at $T\lesssim T_{\rm E}$.\cite{Grossmann1997, Weiss1997, Wilkins1997, Holthaus2001, Mullin2003, Zannetti2015}  Hence in addition to extensive calculations of the thermodynamics using the GCE formalism, we also present calculations performed using exact iterative expressions within the CE formalism.  The latter calculations are reported for properties for which the GCE gives incorrect results for the parameter regimes in which significant BEC occurs, and we compare the results of the two approaches.  The CE formalism gives numerically exact and some analytically exact results for all properties.

The calculation methods are discussed in Sec.~\ref{Sec:Methods}.  In Sec.~\ref{Sec:TE} we calculate a quantity $T_{\rm E}$ which is later shown to be the $N$-dependent crossover temperature between weak and strong increases in the boson populations of the ground and low-lying excited states with decreasing $T$ and/or $A$\@.  We find that $T_{\rm E}\to0$ in the thermodynamic limit $N\to\infty$ at fixed $A/N$\@.  Hence in this limit significant BEC condensation does not occur in the ground state or excited states at any finite $T$ or~$A$\@.  Our results obtained using the GCE formalism are presented in Sec.~\ref{Sec:GCEResults}.  The calculations of the fugacity and fractions of condensed bosons in the ground state and in a state in each of the first four energy levels are presented in Sec.~\ref{Sec:muN0}, of $C_{\rm V}$ in Sec.~\ref{Sec:CV}, of the Helmholtz free energy $F$ and entropy~$S$ in Sec.~\ref{Sec:FS}, of the pressure~$p$ in Sec.~\ref{Sec:pressure}, and of the isothermal compressibility $\kappa_{\rm T}$, thermal expansion coefficient~$\alpha_{\rm p}$ and heat capacity at constant pressure $C_{\rm p}$ in Sec.~\ref{Sec:kapalphCp}.

Our calculations of properties for $N=1,$ 10, 100 and~1000 within the CE formalism are described in Sec.~\ref{Sec:CE}, which begin with calculations of the quantum state boson population statistics in Sec.~\ref{Sec:popstats}.  Calculations of $F$ and~$S$ are presented in Sec.~\ref{Sec:FSCE}, where we show that the finite values of $S$ at $T\to0$ present in the GCE calculations for finite $N$ are incorrect because exact CE calculations show that the entropy at $T\to0$ is identically zero for all finite~$N$\@.  The pressure is calculated for various~$N$ in Sec.~\ref{Sec:pCE}, where we find nonzero values at $T=0$ as expected, in contrast to the null values obtained using the GCE formalism, and quantify $p(T=0)$ versus $N$ and $A$\@.  The ratio of $p$ to the energy density is found to be $p/(U/A)=1$ exactly, in contrast to the strong deviatiations that occur with the GCE formalism in the small $T$ and/or $A$ ranges in which BEC occurs.  The calculations of $\kappa_{\rm T}$, $\alpha_{\rm p}$ and $C_{\rm p}$ are presented in Sec.~\ref{Sec:kTapCpCE}, where we show that these properties are finite and positive  at low $T$ and/or $A$, in contrast to the divergent and/or negative values obtained from calculations of these properties using the GCE formalism.  A  brief summary of our results is given in Sec.~\ref{Summary}.

\section{\label{Sec:Methods} Methods}

\subsection{Single-Boson Wave Functions and Energies}

The wavefunction~$\psi$ of a particle of mass~$m$ in a 2D square of side-length $L$ and area $A=L^2$ with zero potential energy inside and infinite potential energy outside the square in the $xy$~plane is given by the Schr\"odinger equation as
\bse
\label{Eqs:PsiE}
\be
\psi(x,y)=C \sin(k_xx)\sin(k_yy),
\ee
where the edges of the square are at $x = 0,\ L$ and $y = 0,\ L$, and $C$ is the normalization constant.  Continuity of the wave function requires the wave function to be zero on all edges of the square (Dirichlet boundary conditions on the wave function), yielding the quantized wave vectors
\be
k_x = \frac{n_x\pi}{L},\qquad k_y = \frac{n_y\pi}{L},
\ee
\ese
where $n_x,\ n_y = 1,\ 2,\ \ldots$.  Thus the spatial distribution of the number density $\sim\psi^2$ of the bosons inside the square is inhomogeneous.  Periodic boundary conditions give incorrect energies for the low-energy quantum states which can modify their statistical and thermodynamic properties at low $T$ and/or $A$ for finite $N$\@.

The kinetic energy $E$ of a boson in the 2D box is quantized according to
\be
E = \frac{\hbar^2k^2}{2m} = \frac{\pi^2\hbar^2}{2mA}(n_x^2 + n_y^2).
\label{Eq:E1}
\ee
We put $N$ noninteracting bosons into the square box and write $A = N/(N/A)$.  Thus we use the parameters $N$ and $A/N$ (the area per boson) as independent variables instead of $N$ and $A$\@.  Then Eq.~(\ref{Eq:E1}) becomes
\be
E = \frac{\pi^2\hbar^2}{2mN (A/N)}(n_x^2 + n_y^2).
\label{Eq:E2}
\ee
The thermodynamic limit corresponds to $N\to\infty$ at fixed $A/N$\@.  We do not shift the energy scale so that the ground-state energy becomes zero, as usually done in calculations of the properties of the Bose gas, except when calculating the fugacity $z$ using the GCE formalism where the energy shift does not affect the calculated values of $z$ (see Sec.~\ref{Sec:z}) or in some cases where we use the continuum representation for the high-energy quantum states.

We introduce the dimensionless reduced parameter $\gamma$ defined by
\be
\gamma=\frac{A/N}{(A/N)_{\rm E}} = \frac{A}{A_{\rm E}},
\label{Eq:gammaDef}
\ee
where $(A/N)_{\rm E}$ is the value of $A/N$ of the Bose gas at the Bose-Einstein crossover temperature $T_{\rm E}$ to be defined in Sec.~\ref{Sec:TE}.  Thus $\gamma$ is the area per boson in units of the area per boson at $T = T_{\rm E}$, or equivalently the area normalized by the area~$A_{\rm E}$ at $T = T_{\rm E}$. The reduced energy $\bar{E}$ is defined using Eqs.~(\ref{Eq:E2}) and~(\ref{Eq:gammaDef}) by
\bse
\be
\bar{E} \equiv \frac{E}{k_{\rm B}T_{\rm E}} = \frac{a}{N\gamma}(n_x^2 + n_y^2),
\label{Eq:barEDef}
\ee
where $k_{\rm B}$ is Boltzmann's constant and the parameter $a$ is defined as
\be
a = \frac{\pi^2\hbar^2}{2mk_{\rm B}T_{\rm E}(A/N)_{\rm E}}.
\label{Eq:aDef}
\ee
\ese
In Sec.~\ref{Sec:TE} we show that $a=a(N)$ and at large~$N$ one obtains $a\sim \ln N$\@.  The reduced temperature $t$ is defined as
\be
t = \frac{T}{T_{\rm E}},
\label{Eq:tDef}
\ee
so from Eq.~(\ref{Eq:barEDef}) one obtains
\be
\frac{E}{k_{\rm B}T} = \frac{\bar{E}}{t} = \frac{a}{N\gamma t}(n_x^2 + n_y^2).
\label{Eq:EonT}
\ee
Thus $E/(k_{\rm B}T)$ is a function of the product $\gamma t$, so we define the additional reduced parameter $x$ as
\be
x = \gamma t,
\label{Eq:xDef}
\ee
and Eq.~(\ref{Eq:EonT}) becomes
\be
\frac{E}{k_{\rm B}T} = \frac{a(N)}{N x}(n_x^2 + n_y^2).
\label{Eq:EonT2}
\ee

Using Eqs.~(\ref{Eq:E1}) and~(\ref{Eq:EonT2}), we also write
\bse
\label{Eqs:EonkBTg}
\be
\frac{E}{k_{\rm B}T} = \frac{n_x^2 + n_y^2}{g},
\ee
where
\be
g = \frac{N x}{a(N)} = \frac{2mk_{\rm B}TA}{\pi^2\hbar^2}
\ee
\ese
Thus if we need to hold both $T$ and~$A$ constant in a calculation such as in Sec.~\ref{Sec:FSCE} to obtain Eq.~(\ref{Eq:St0}), one must hold $g$ constant.

\subsection{\label{Sec:GCE} Grand Canonical Ensemble}

\subsubsection{Distribution Function}

The Bose-Einstein distribution function for the average number of bosons with fugacity~$z$ in a quantum state with energy~$E$ at absolute temperature~$T$ is
\bse
\be
f_{\rm BE}(E,T) = \frac{1}{z^{-1}e^{E/k_{\rm B}T}-1},
\label{Eq:BEDist}
\ee
which in reduced parameters is
\be
f_{\rm BE}(\bar{E},t) = \frac{1}{z^{-1}e^{\bar{E}/t}-1}.
\label{Eq:fBERed}
\ee
\ese
The fugacity is related to the chemical potential~$\mu$ by
\be
z = e^{\mu/k_{\rm B}T} = e^{\bar{\mu}/t},
\label{Eq:zmu}
\ee
where the reduced chemical potential $\bar{\mu}$ is defined as
\be
\bar{\mu} = \frac{\mu}{k_{\rm B}T_{\rm E}}.
\ee
Thus Eq.~(\ref{Eq:fBERed}) can be written
\be
f_{\rm BE}(\bar{E},t) = \frac{1}{e^{(\bar{E}-\bar{\mu})/t}-1}.
\label{Eq:fBERed2}
\ee
An important consequence of Eq.~(\ref{Eq:fBERed2}) is that $f_{\rm BE}$ is invariant under a uniform shift of all energies, including both $E$ and $\mu$, by the same amount.  Hence for all calculations involving the  factor $z^{-1}e^{\bar{E}/t}= e^{(\bar{E}-\bar{\mu})/t}$, for convenience we set the ground state energy $E_0$ with $n_x=n_y=1$ to be zero, and then $z$ is calculated with reference to this energy.  Then Eq.~(\ref{Eq:EonT2}) becomes
\be
\frac{E}{k_{\rm B}T} = \frac{a(N)}{Nx}(n_x^2 + n_y^2-2),
\label{Eq:EonTzcalc2}
\ee 
and the Bose-Einstein distribution function~(\ref{Eq:BEDist}) becomes
\be
f_{\rm BE}(n_x,n_y,x,N) = \frac{1}{z^{-1}\exp[\frac{a}{N x}(n_x^2 + n_y^2-2)]-1}.
\label{Eq:fBERed3}
\ee

\subsubsection{Density of Orbital States}

In a continuum enumeration of the orbital quantum states, the number of these states $N_{\rm states}$ in a quadrant of a circle in $n$~space is $N_{\rm states} = \frac{\pi}{4}n^2$.  However, this includes states with $(n_x=0,\ n_y\neq0$), ($n_y=0,\ n_x\neq0)$ and $n_x=n_y=0$ for which the wave function in Eqs.~(\ref{Eqs:PsiE}) is zero.  The number of such states is $(2n+\frac{1}{2})/2$, where the states with $n_x=0,\ n_y>0$ and $n_y=0,\ n_x>0$ are shared by two quadrants and the state with $n_x=0,\ n_y = 0$ is shared by four quadrants.  Correcting for these terms gives
\be
N_{\rm states} = \frac{\pi}{4}n^2 - n - \frac{1}{4}.
\ee
The density of orbital states in $n$~space in the continuum representation is then
\be
{\cal D}(n) \equiv \frac{dN_{\rm states}}{dn} = \frac{\pi}{2}n - 1.
\label{Eq:Dofn}
\ee

\subsubsection{\label{Sec:z} Fugacity}

The fugacity~$z$ is determined from the requirement that the average number of bosons~$N$ in the system is equal to the sum of the average number of bosons in each quantum state, i.e.,
\be
N = \sum_{n_x,n_y=1}^\infty f_{\rm BE}(n_x,n_y,x,N).
\label{Eq:Findz2}
\ee
Using the energy expression in Eq.~(\ref{Eq:EonTzcalc2}), this becomes
\be
N = \sum_{n_x,n_y=1}^\infty \frac{1}{z^{-1}\exp\left[\frac{a(N)}{Nx}(n_x^2 + n_y^2-2)\right]-1}.
\label{Eq:Findz}
\ee
By specifying given values of~$N$ and of $a(N)$ derived later in Sec.~\ref{Sec:TE}, one can solve this equation for $z(x,N)$.  Since $x=\gamma t$, one also has $z=z(\gamma t,N)$.  On the other hand, when the fugacity appears by itself in an expression such as in Eq.~(\ref{Eq:Fcalc}) below where the fugacity is not multiplying the exponential of energy divided by $k_{\rm B}T$, one must use the fugacity $z_{\rm unshifted}$ calculated from the unshifted energy levels by solving
\be
N = \sum_{n_x,n_y=1}^\infty \frac{1}{z_{\rm unshifted}^{-1}\exp\left[\frac{a(N)}{Nx}(n_x^2 + n_y^2)\right]-1}.
\label{Eq:Findzunshifted}
\ee
Comparison of Eqs.~(\ref{Eq:Findz}) and~(\ref{Eq:Findzunshifted}) gives
\be
z_{\rm unshifted}=z\exp\left[\frac{2a(N)}{N x}\right],
\label{Eq:zunshifted}
\ee
so it is not necessary to do a separate calculation of $z_{\rm unshifted}(x,N)$ if $z(x,N)$ is already known.

The boson occupation number $N_0$ of the nondegenerate ground state with $n_x=n_y=1$ is given by Eq.~(\ref{Eq:fBERed3}) as
\bse
\label{Eqs:N0_N}
\be
N_0 = \frac{z}{1-z}.
\label{Eq:N0}
\ee
The requirement that $0<N_0 \leq N$ gives the allowed range
\be
0< z \leq \frac{N}{N+1}
\label{Eq:zLimits}
\ee
for $x=\infty$ and $x=0$, respectively.  The fractional occupation of the ground state by the $N$ bosons in the system is then
\be
\frac{N_0}{N} = \frac{z}{N(1-z)}.
\label{Eq:N0Nfromz}
\ee
\ese
In the continuum representation of the energy level distribution,  we use the density of states in Eq.~(\ref{Eq:Dofn}) and Eq.~(\ref{Eq:EonTzcalc2}) becomes
\be
\frac{E}{k_{\rm B}T} = \frac{a}{N x}(n^2-2),
\label{Eq:barEDef2}
\ee
where $n^2 = n_x^2 + n_y^2$. Thus the Bose-Einstein distribution function~(\ref{Eq:fBERed3}) becomes
\be
f_{\rm BE}(n,x,N) = \frac{1}{z^{-1}\exp\left[\frac{a (n^2-2)}{N x}\right]-1}.
\label{Eq:fBERed4}
\ee

In 2D with finite~$N$, Eq.~(\ref{Eq:Findz}) does not have an analytic solution for~$z$ and therefore must be solved numerically.  Furthermore, one cannot break up sums such as in Eq.~(\ref{Eq:Findz}) into the contribution of only the ground state plus an integral over the remainder such as is done for 3D Bose gases in the thermodynamic limit because as we will see for the 2D Bose gas with finite~$N$, in general significant BEC occurs in excited states in addition to the ground state.  Therefore, one must include a significant number of states above the ground state in the sum and then carry out an integral over the remainder, and Eq.~(\ref{Eq:Findz}) for solving for~$z$ becomes
\bea
N &=& \sum_{n_x=1}^{n_{\rm max}}\sum_{n_y=1}^{\sqrt{n_{\rm max}^2-n_x^2}}\frac{1}{z^{-1}\exp\left[\frac{a}{N x}(n_x^2 + n_y^2 - 2)\right]-1}\nonumber\\*
&& +\ \int_{n_{\rm max}}^\infty\frac{{\cal D}(n)}{z^{-1}\exp\left[\frac{a}{N x}(n^2-2)\right]-1}\,dn.
\label{Eq:NEqual}
\eea

After $z(x,N)$ is determined, the fractional populations $N_1/N,\ \ldots,\ N_4/N$ of a quantum state in each of the first four excited energy levels, respectively, versus $x$ and~$N$ are obtained using
\be
\frac{N_i}{N}(x,N) = \frac{1}{N} f_{\rm BE}(n_{x_i},n_{y_i},x,N),
\label{Eq:Ni}
\ee
where $f_{\rm BE}(n_x,n_y,x,N)$ is given in Eq.~(\ref{Eq:fBERed3}) and \mbox{$n_{x_i}^2+n_{y_i}^2=5$}, 8, 10 and 13 for the first four excited energy levels, respectively.  To calculate these populations  versus reduced temperature~$t$ at fixed reduced area per boson~$\gamma$ or vice versa one uses the definition of $x$ in Eq.~(\ref{Eq:xDef}) to replace $x$ in the results by $\gamma t$.

The grand partition function ${\cal Z}$ is given by\cite{Huang1963}
\be
\ln{\cal Z} = -\sum_i\ln\left(1 - ze^{-E_i/k_{\rm B}T}\right),
\ee
which for our system reads
\be
\ln{\cal Z} = -\sum_{n_x,n_y=1}^\infty\ln\left[1 - ze^{-\frac{a}{Nx}(n_x^2+n_y^2-2)}\right],
\label{Eq:lnZGCE}
\ee
where as discussed above we set the ground state energy to zero in multiplicative factors $z^{-1}e^{E_i/k_{\rm B}T}$ (or $ze^{-E_i/k_{\rm B}T}$) that appear in a calculation.  The sum is evaluated by first calculating $z(x,N)$ as described above and then replacing the sum to $\infty$ by a sum from 1 to $n_{\rm max}$ plus a numerical integral from $n_{\rm max}$ to $\infty$ similar to the procedure used to arrive at Eq.~(\ref{Eq:NEqual}).

\subsection{\label{Sec:CEMethods} Canonical Ensemble}

The partition function $Q(N)$ within the canonical ensemble formalism for a system containing $N$ noninteracting bosons is given by the recursion relation\cite{Mullin2003, Borrmann1993}
\bse
\label{Eqs:Q}
\be
Q(N) = \frac{1}{N}\sum_{k=1}^N Q_1(k)Q(N-k)\qquad [Q(0)=1],
\label{Eq:QNcalc}
\ee
where $Q_1(k)$ is the single-boson partition function for a modified temperature $T/k$ given by
\be
Q_1(k) = \sum_i \exp\left[-\frac{kE_i}{k_{\rm B}T}\right]
\ee
and the sum is over all quantum states~$i$.  Using Eq.~(\ref{Eq:EonT2}) for $E_i/k_{\rm B}T$ gives
\be
Q_1(k) = \sum_{n_x,\,n_y=1}^\infty\exp\left[-\frac{ka}{Nx}(n_x^2+n_y^2)\right].
\label{Eq:Q1Def}
\ee
This double sum can be expressed analytically as
\be
Q_1(k) = \frac{1}{4}\left[\theta_3(0,e^{-ka/Nx}) - 1\right]^2,
\ee
\ese
where $\theta_a(u,q)$ is a theta function that {\tt Mathematica} denotes as {\tt EllipticTheta}.

Another useful recursion relation within the canonical ensemble is for the average number $\bar{n}_i(N)$ of bosons occupying a quantum state with energy $E_i$, given by\cite{Mullin2003, Schmidt1989}
\bse
\be
\bar{n}_i(N) = \frac{1}{Q(N)}\sum_{k=1}^N \exp\left[-\frac{kE_i}{k_{\rm B}T}\right]Q(N-k).
\ee
For the 2D Bose gas under consideration one obtains
\be
\bar{n}_i(N) = \frac{1}{Q(N)}\sum_{k=1}^N \exp\left[-\frac{ka}{Nx}(n_{x_i}^2 + n_{y_i}^2)\right]Q(N-k).
\label{Eq:barni}
\ee
\ese

The computations in this paper were carried out using laptop or desktop computers and {\tt Mathematica} software.

\section{\label{Sec:TE} Crossover Temperature for Bose-Einstein Condensation in a 2D Box}

The usual prescription for calculating statistical and thermodynamic properties of a three-dimensional (3D) Bose gas is to utilize integral representations of all sums over quantum states except possibly for the ground state.  Thus the number of bosons in excited energy states above the nondegenerate ground state in $n$ space is
\be
N_{\rm exc} = \int_0^\infty {\cal D}(n) f_{\rm BE}(n,T,N) dn.
\ee
To obtain $T_{\rm E}$, one sets $T=T_{\rm E}$, $z=1$ and $N_{\rm exc}=N$.\cite{Ketterle1996, Kittel1980}  In 2D for $n\to0$, the integrand becomes $\frac{\pi N}{2a}\,\frac{1}{n}$, so evaluation of the integral at the lower limit $n\to0$ gives $\frac{\pi N}{2a}\,\ln n|_{n\to0} = -\infty$.  Thus the integral diverges logarithmically for $n\to0$.  However, this very slow divergence suggests that one should use a discrete sum over $n_x$ and $n_y$ for the lowest energy levels instead of an integral over all $n$ to determine $T_{\rm E}$.

We confirmed that a finite $T_{\rm E}$ can be obtained in 2D for finite~$N$ if the integral over energies of the Bose-Einstein distribution function is replaced by a sum over the lowest energy levels with small~$n_x,n_y$ and the integral formulation is used to sum over larger~$n$ as in Eq.~(\ref{Eq:NEqual}).  As discussed in the Introduction and  will be demonstrated in Sec.~\ref{Sec:muN0}, this $T_{\rm E}$ is a crossover temperature between weak and strong increases in $N_0/N$ with decreasing $T$ in a 2D boson gas with finite~$N$, and is not a BEC transition temperature.

\begin{figure}
\includegraphics[width=3.3in]{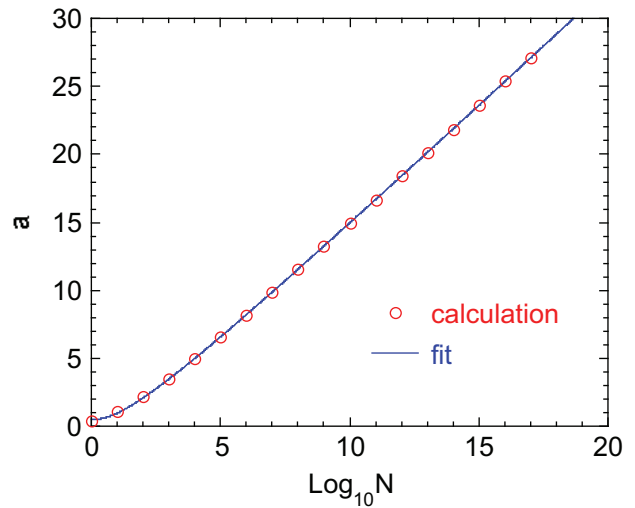}
\caption{(Color online) Parameter $a$ in Eq.~(\ref{Eq:aDef}) versus $\log_{10}$ of the number~$N$ of bosons in the 2D system (open red circles).  The empirical three-parameter fit of the data by Eqs.~(\ref{Eq:FitFcn}) is shown as the solid blue curve.}
\label{Fig:a_vs_Log10NFitPlot}
\end{figure}

\begin{table}
\caption{\label{Tab:aVSn} Parameter $a$ in Eq.~(\ref{Eq:aDef}) versus $\log_{10}$ of the number~$N$ of bosons in the system obtained using Eq.~(\ref{Eq:NEqual}) with $n_{\rm max}=500$ and $z=x=1$.  From calculations of $a$ versus $n_{\rm max}$ we infer that the quoted values are accurate to $\approx \pm1$ in the last decimal place.  The fitted values obtained from empirical Eqs.~(\ref{Eq:FitFcn}) are also shown, together with the percent deviations of the fit from the calculated $a$ values.}
\begin{ruledtabular}
\begin{tabular}{cccr}
$\log_{10}N$ 	& $a$  &  $a_{\rm fit}$  & $\frac{a - a_{\rm fit}}{a}$ (\%)\\
\hline
0  &  0.41539  &  0.54089  &  $-$30.21  \\
1  &  1.11125  &  0.96884  &  12.82  \\
2  &  2.17711  &  2.08950  &  4.02  \\
3  &  3.50392  &  3.47528  &  0.82  \\
4  &  4.98840  &  4.99046  &  $-$0.04  \\
5  &  6.56312  &  6.57845  &  $-$0.23  \\
6  &  8.19079  &  8.21147  &  $-$0.25  \\
7  &  9.85178  &  9.87427  &  $-$0.23  \\
8  &  11.5355  &  11.5578  &  $-$0.19  \\
9  &  13.2356  &  13.2563  &  $-$0.16  \\
10  &  14.9484  &  14.9660  &  $-$0.12  \\
11  &  16.6917  &  16.6843  &  0.04  \\
12  &  18.4020  &  18.4093  &  $-$0.04  \\
13  &  20.1444  &  20.1397  &  0.02  \\
14  &  21.8735  &  21.8745  &  $-$0.00  \\
15  &  23.6312  &  23.6128  &  0.08  \\
16  &  25.3885  &  25.3541  &  0.14  \\
17  &  27.0824  &  27.0980  &  $-$0.06  \\
\end{tabular}
\end{ruledtabular}
\end{table}

We set $z=1$ as in the 3D case and $x=1$ in Eq.~(\ref{Eq:fBERed3}) to obtain the Bose-Einstein distribution function for the calculation of $T_{\rm E}$ 
\be
f_{\rm BE}(n_x,n_y,T_{\rm E})=\frac{1}{e^{\frac{a}{N}(n_x^2 + n_y^2 - 2)}-1}
\label{Eq:BE2}
\ee
for the sum and 
\be
f_{\rm BE}(n,T_{\rm E})=\frac{1}{e^{\frac{a}{N}(n^2-2)}-1}
\label{Eq:BE22}
\ee
for the integral.  Then we numerically solved Eq.~(\ref{Eq:NEqual}) for the parameter $a$ as a function of $N$\@.  The $a$ values reached constant values with increasing $n_{\rm max}$ by $n_{\rm max}\sim500$.  A plot of $a$ versus $\log_{10}N$ for $n_{\rm max}=500$ and $\log_{10}N = 0,\ 1,\ \ldots,\ 17$ is shown in Fig.~\ref{Fig:a_vs_Log10NFitPlot} and the values are given in Table~\ref{Tab:aVSn}.  One sees that $a$ approaches linearity in $\log_{10}N$ at large~$N$\@.  Therefore we fitted the eighteen $\{\log_{10}N,a(N)\}$ data points by an emprical three-parameter Pad\'e approximant
\bse
\label{Eq:FitFcn}
\be
a_{\rm fit} = \frac{P_0 + P_2 \left(\log_{10}N\right)^2}{1 + D_1\log_{10}N},
\ee
and obtained the fitting parameters
\be
P_0 = 0.540886, \quad P_2 = 0.948778  ,\quad D_1 = 0.537570.
\ee
\ese
In the limit of large~$N$ the fit gives gives $a_{\rm fit} = -3.2832 + 1.7649 \log_{10}N$\@.  The fit is shown in Fig.~\ref{Fig:a_vs_Log10NFitPlot} and the fit values and deviations of the fit from the data are shown in Table~\ref{Tab:aVSn}.  The magnitude of the deviation is seen to be $\lesssim 0.2$\% for $10^4\leq N \leq 10^{17}$ with the deviation increasing to $30$\% for $N = 1$.

Equation~(\ref{Eq:aDef}) gives $T_{\rm E}$ for finite~$N$ to be
\be
T_{\rm E} = \frac{\pi^2\hbar^2}{2mk_{\rm B}(A/N)_{\rm E}a(N)},
\label{Eq:TE}
\ee
where the parameter~$a(N)$ diverges for $N\to\infty$ as shown above.  Therefore from Eq.~(\ref{Eq:TE}) and Fig.~\ref{Fig:a_vs_Log10NFitPlot}, $T_{\rm E}$ decreases monotonically with increasing~$N$, and $T_{\rm E}(N\to\infty) = 0$ at fixed $(A/N)_{\rm E}$, i.e., in the thermodynamic limit.

\section{\label{Sec:GCEResults}  Results: Grand Canonical Ensemble}

\subsection{\label{Sec:muN0} Fugacity and  Fraction of Condensed Bosons}

\begin{figure}
\includegraphics[width=3.3in]{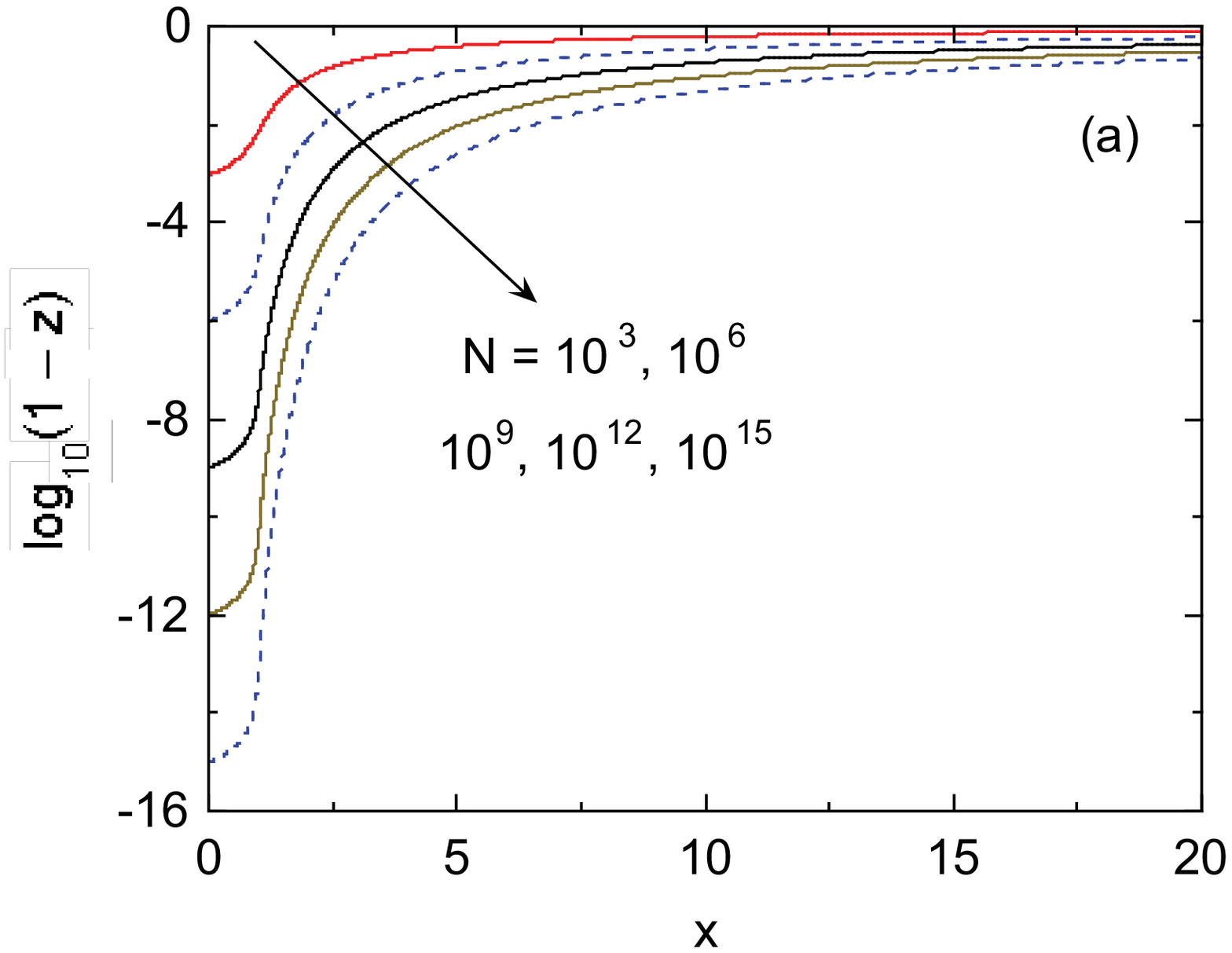}
\includegraphics[width=3.3in]{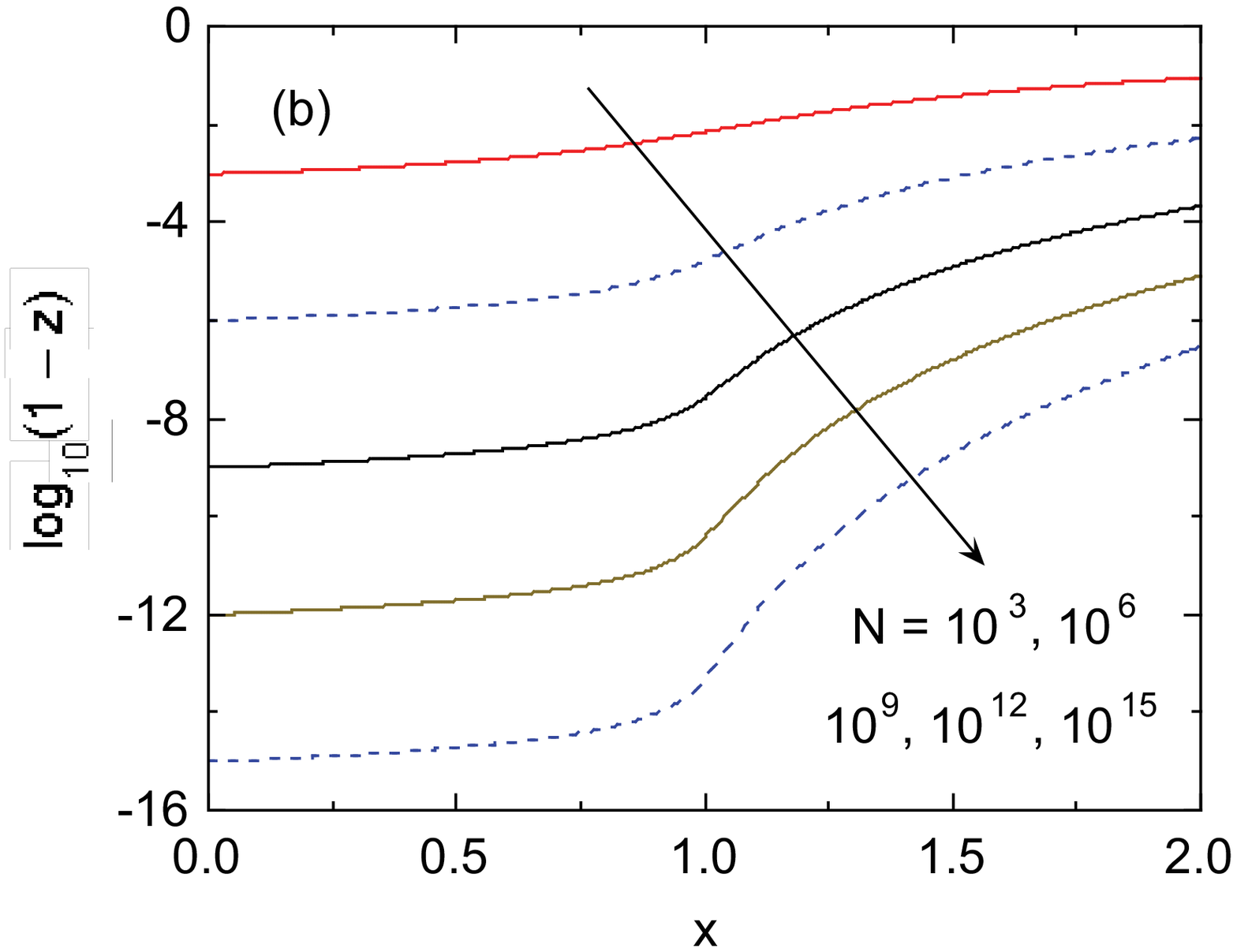}
\caption{(Color online) (a) Logarithm to the base 10 of $1-z$ versus $x$ for a variety of boson number $N$ values, where $z$ is the fugacity, $x=\gamma t$, $\gamma$ is the reduced area and $t$ is the reduced temperature of the Bose gas.  (b) Expanded plot of the data in (a) for $x = 0$ to~2.}
\label{Fig:AllDataLoT_N3}
\end{figure}

The fugacities~$z$ calculated versus the parameter~$x = \gamma t$ for finite systems with specific values of~$N$ obtained by solving Eq.~(\ref{Eq:NEqual}) for~$z$ using $n_{\rm max} = 200$--1000 and the respective $a(N)$ values in Table~\ref{Tab:aVSn} are presented as $\log_{10}(1-z)$ versus~$x$ in Fig.~\ref{Fig:AllDataLoT_N3}(a), with expanded plots for $x\leq 2$ in Fig.~\ref{Fig:AllDataLoT_N3}(b).  One sees for these finite systems that $\log_{10}(1-z)$ shows noticeable increases with increasing~$x$ near $x=1$ which become more pronounced as $N$ increases.  Since $x=\gamma t$, if $A = A_{\rm E}\ (\gamma=1)$, which is the value of $A$ at $T_{\rm E}$, then the $x=1$ crossover occurs at $T=T_{\rm E}\ (t=1)$.  From Eq.~(\ref{Eq:zLimits}), $\log_{10}[\lim_{x\to0}(1-z)] = -\log_{10}(N_0+1) = -\log_{10}(N+1)$, which is verified in Fig.~\ref{Fig:AllDataLoT_N3}.

\begin{figure}
\includegraphics[width=3.3in]{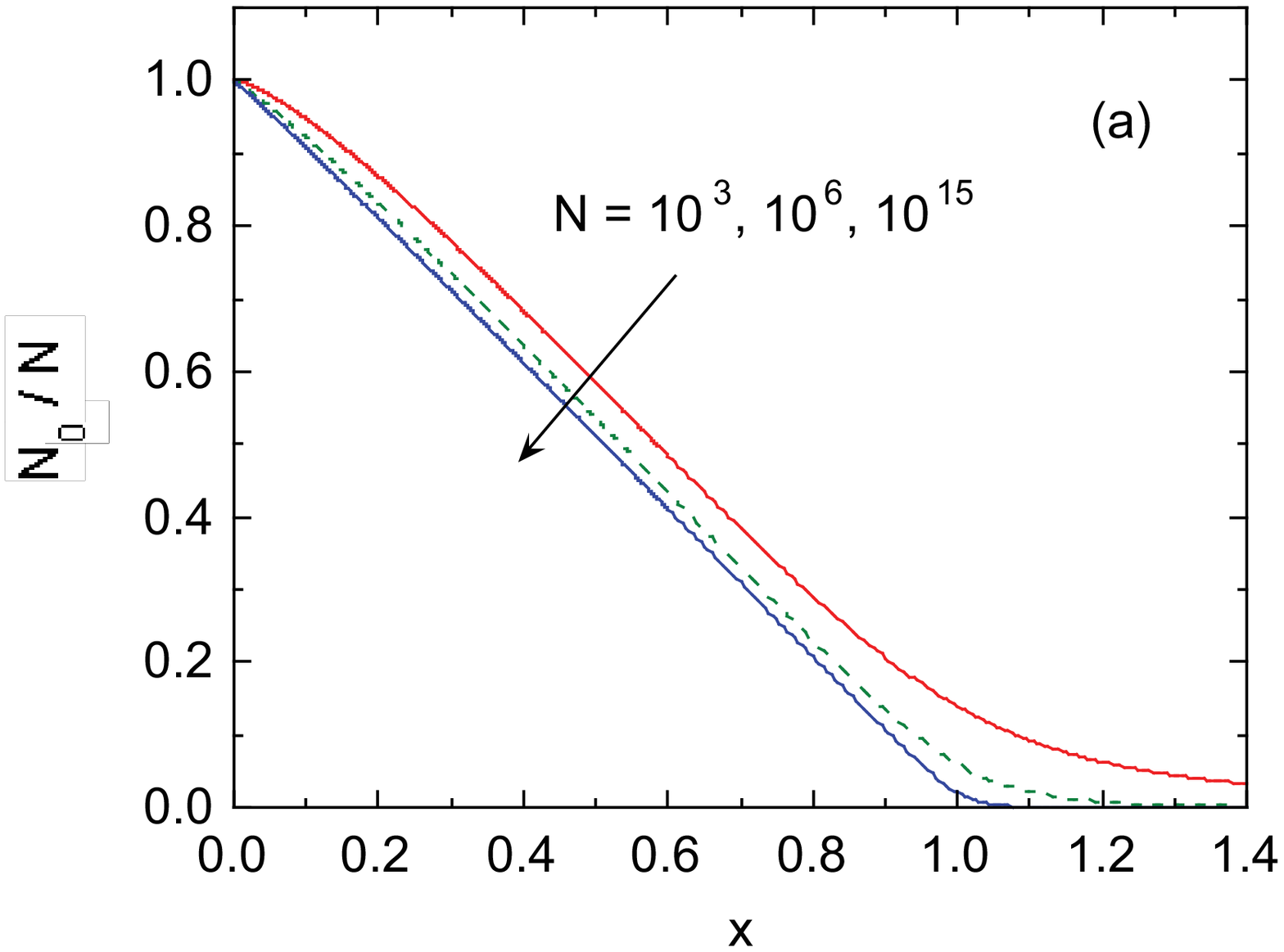}
\includegraphics[width=3.3in]{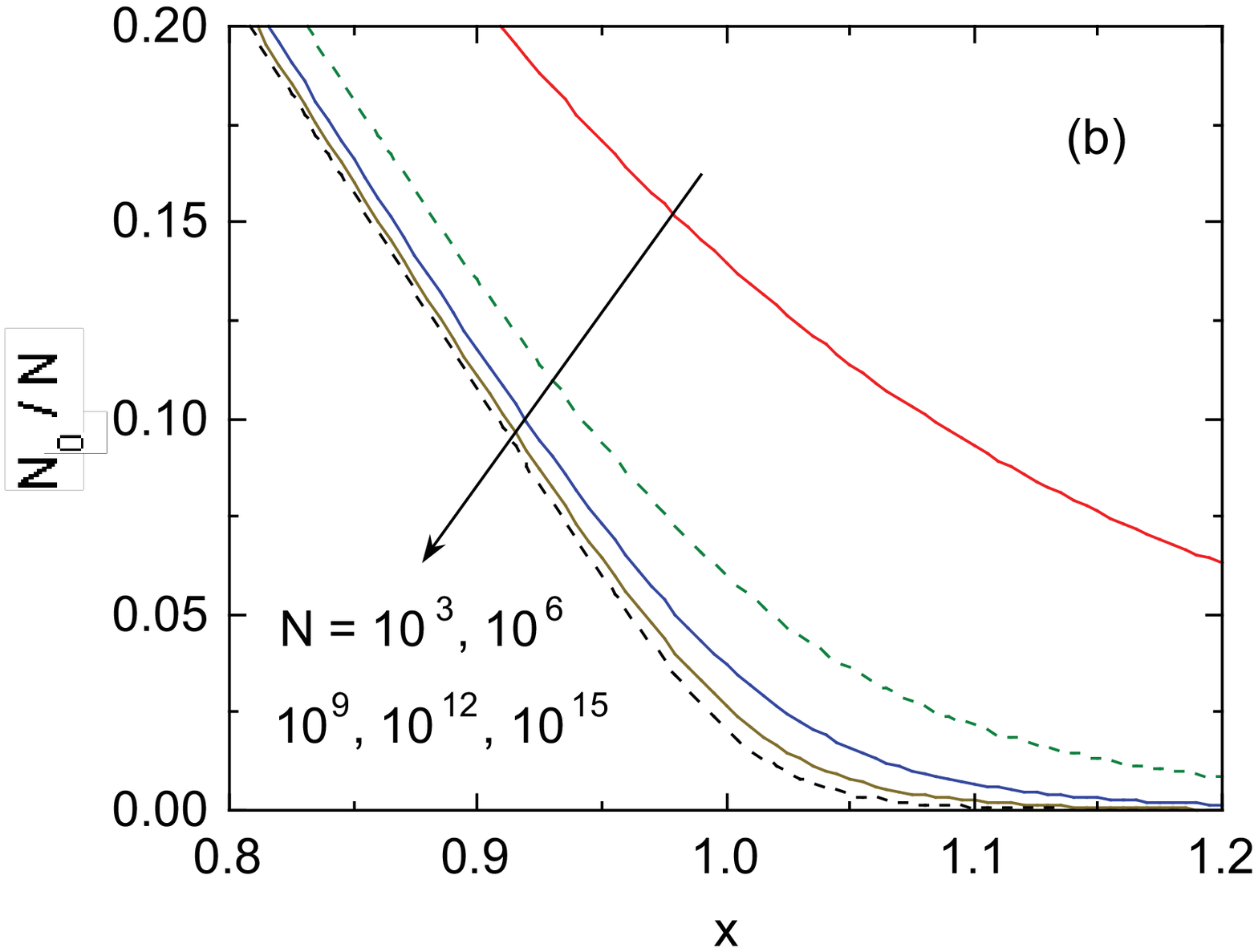}
\caption{(Color online) (a) Fractional occupation $N_0/N$ of the ground state versus $x = \gamma t$ for three $N$ values. (b) Expanded plots of $N_0/N$ versus~$x$ for $N = 10^3$ to $10^{15}$ and $x = 0.8$ to~1.2.}
\label{Fig:AllDataLoT_N0N}
\end{figure}

The fraction $N_0/N$ of condensed bosons in the ground state versus $x$ is now obtained from Eq.~(\ref{Eq:N0Nfromz}) as shown in Fig.~\ref{Fig:AllDataLoT_N0N}(a) for $N=10^3,\ 10^6$ and $10^{15}$.  With decreasing~$x$, the data approach a linear behavior in $x$ for $x \lesssim 1$ as $N$ increases, described by $\frac{N_0}{N} = 1 - x$.  Expanded plots of data near $x=1$ for $N=10^3$ to $10^{15}$ are shown in Fig.~\ref{Fig:AllDataLoT_N0N}(b).  The fraction of bosons in excited states for $x\lesssim 1$ is given by $\frac{N_{\rm exc}}{N} = 1 - \frac{N_0}{N} = x$ (not shown).  The approximately linear decrease of $N_0/N$ versus~$x$ for $x\lesssim1$ at large $N$ is different from the behavior of the 3D Bose gas in the thermodynamic limit which shows $N_0/N = 1 - (T/T_{\rm E})^{3/2}$.

The data in Fig.~\ref{Fig:AllDataLoT_N0N} show that for these finite systems, BEC of bosons into the ground state occurs and increases smoothly and continuously with decreasing~$x$\@. Hence there is no phase transition associated with BEC into the ground state (and low excited states, see below).  Furthermore, according to Eq.~(\ref{Eq:TE}) and the $a(N)$ behavior in Fig.~\ref{Fig:a_vs_Log10NFitPlot}, $T_{\rm E}\to0$ for $N\to\infty$ and hence BEC does not occur in the thermodynamic limit.  On the other hand, real systems do not contain an infinite number of bosons, and hence potentially observable BEC is expected to occur for finite $N$, but with no BEC phase transition associated with it.

\begin{figure}
\includegraphics[width=3.in]{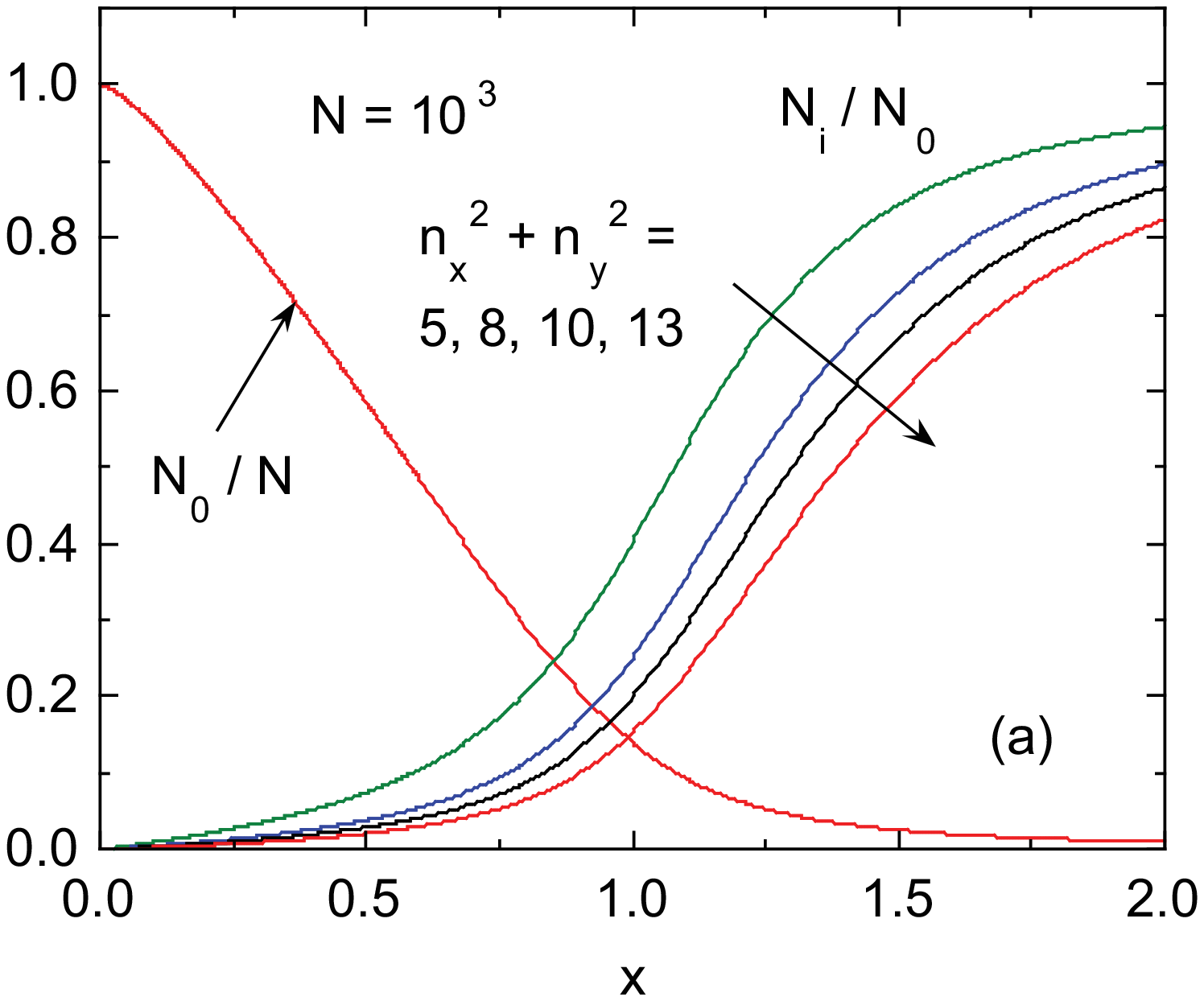}
\includegraphics[width=3.in]{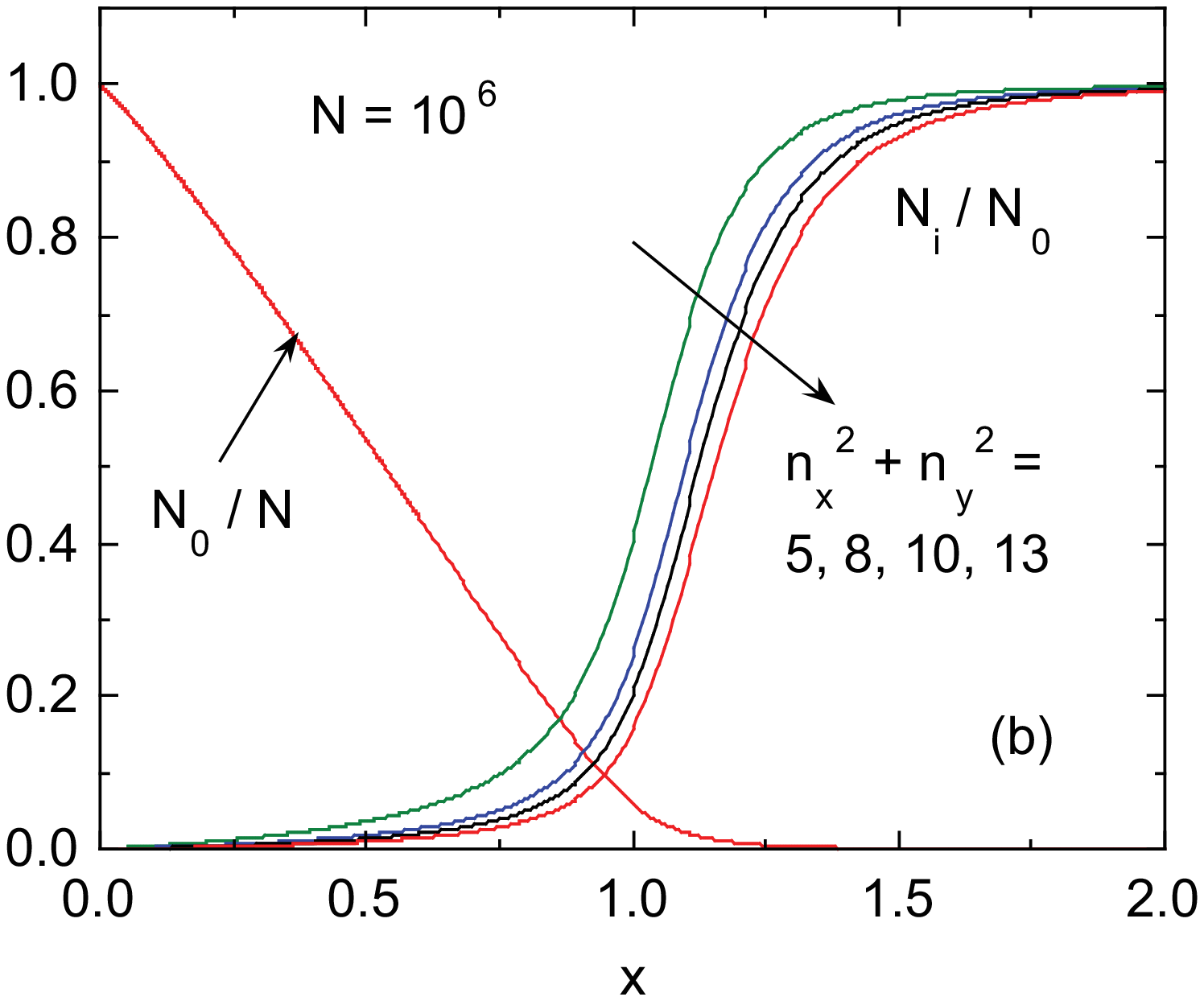}
\includegraphics[width=3.in]{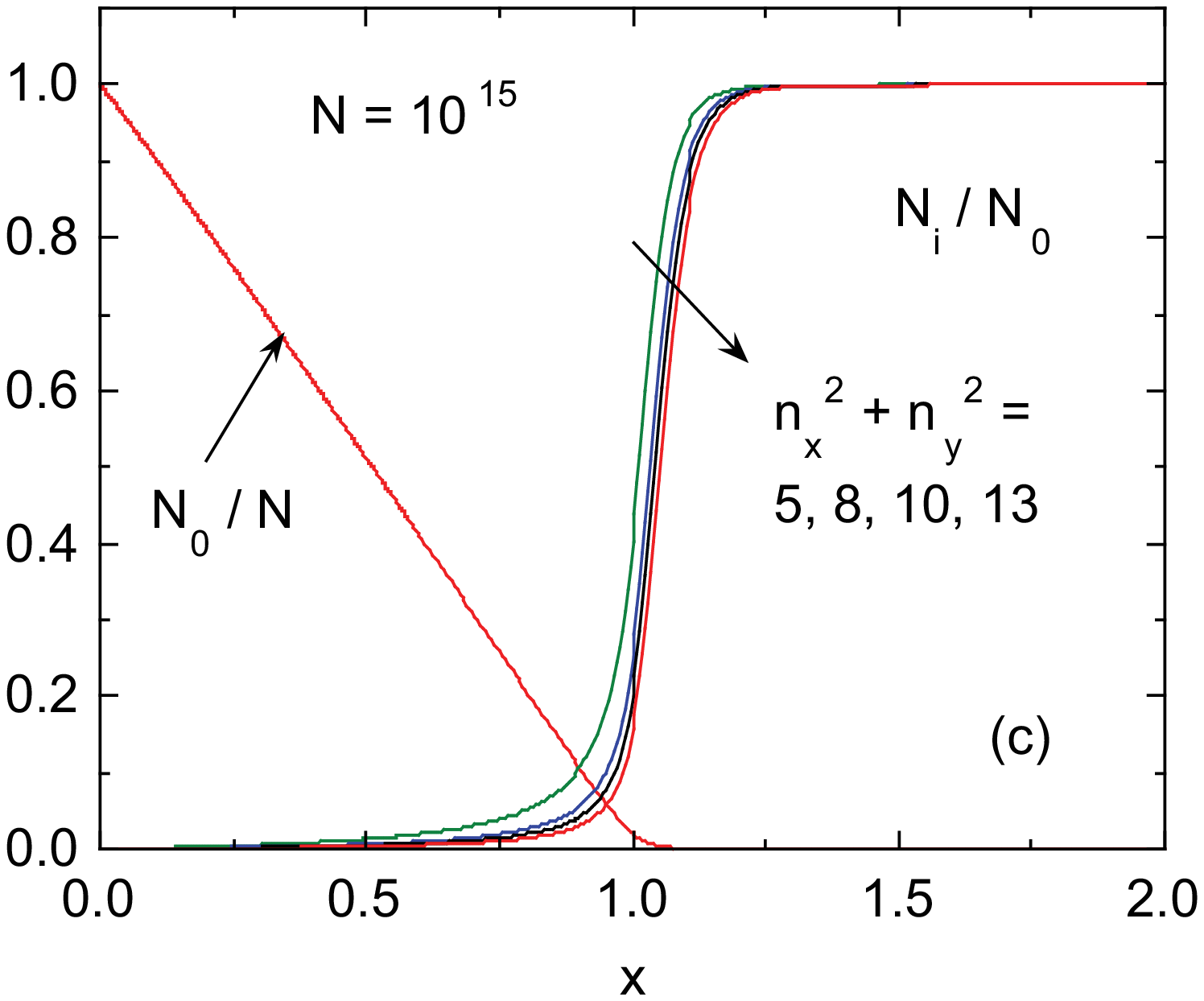}
\caption{(Color online) The ratio $N_0/N$ versus $x$ from Fig.~\ref{Fig:AllDataLoT_N0N} and the ratios  $N_i/N_0$ of the number of bosons $N_i$ occupying a state in each of the first four excited energy levels $E_i$ $(i=1$~to~4) with respective quantum numbers $n_{x_i}^2+n_{y_i}^2 = 5$, 8, 10 and~13 in Eq.~(\ref{Eq:EonT}) for (a) $N = 10^3$, (b) $N=10^6$ and (c) $N = 10^{15}$.  For the ground state $n_{x_0}^2+n_{y_0}^2 = 2$.  At fixed volume $\gamma=1$ one has $x=t=T/T_{\rm E}$ and hence the plots are then versus~$t$.}
\label{Fig:AllDataLoT_NiN0}
\end{figure}

The 3D Bose gas in the thermodynamic limit at $T=0$ has all $N$ bosons in the ground state and none in the excited states.  With increasing $T$, a macroscopic occupation of the ground state still occurs until the temperature (almost) reaches the BEC transition temperature, i.e., $N_0 = {\cal O}(N)$, but the occupation of any excited state $i$ is $N_i = {\cal O}(1)$.  For $T>T_{\rm E}$, the occupations of all low-lying states states are about the same and of ${\cal O}(1)$.  In reduced dimensions with a finite number of bosons and no BEC phase transition, one expects this behavior to change.  The ratio $N_i/N$ of the number of bosons in an excited state with energy $E_i$ is given by Eq.~(\ref{Eq:Ni}) as
\be
\frac{N_i}{N}(x,N) =\frac{1}{N}\ \frac{z}{\exp\left[\frac{a}{Nx}(n_{x_i}^2+n_{y_i}^2-2)\right]  - z},
\label{Eq:Ni2}
\ee
where $a(N)$ is given in Table~\ref{Tab:aVSn}.  For $i=0$ with $n_{x} = n_{y} = 1$ one obtains Eq.~(\ref{Eq:N0Nfromz}).  Then from Eqs.~(\ref{Eq:N0Nfromz}) and~(\ref{Eq:Ni2}) one obtains the additional ratios
\be
\frac{N_i}{N_0} = \frac{1-z}{\left\{\exp\left[\frac{a}{Nx}(n_{x_i}^2+n_{y_i}^2-2)\right] - 1\right\} + (1-z)},
\ee
where $1-z$ is plotted versus~$x$ in Fig.~\ref{Fig:AllDataLoT_N3}.  The $N_i/N_0$ ratios versus $x$ for $n_{x_i}^2+n_{y_i}^2 = 5,$ 8, 10 and 13 and for $N=10^3,\ 10^6$ and~$10^{15}$ are shown in Figs.~\ref{Fig:AllDataLoT_NiN0}(a), \ref{Fig:AllDataLoT_NiN0}(b) and \ref{Fig:AllDataLoT_NiN0}(c), respectively,  along with the respective $N_0/N$ versus~$x$ plots from Fig.~\ref{Fig:AllDataLoT_N0N}.  If the reduced volume per boson is fixed at $\gamma=1$, the parameter $x$ is simply $x=t=T/T_{\rm E}$.  In that case, one sees that for finite~$N$, the four excited states show $N_i = {\cal O}(N)$ even when $t\ll 1$.  Furthermore, with increasing $N$ the occupation of the excited states occurs more rapidly with increasing $t$ near $t=1$ and the $N_i/N_0$ values of the excited states at low temperatures decrease substantially.  For the 3D Bose gas in the thermodynamic limit with $t<1$ one has $N_i = {\cal O}(1)$ and hence $N_i/N_0 = {\cal O}(1/N)$, which qualitatively differs from the 2D case especially for the smaller values of $N$, whereas for $t>1$ one has $N_i/N_0 \approx 1$ as in the 2D case at sufficiently high $t$\@.

\subsection{\label{Sec:CV} Internal Energy and Heat Capacity at Constant Volume}

The reduced internal energy is defined as 
\be
\bar{U} = \frac{U}{k_{\rm B}T_{\rm E}}.
\ee
The internal energy per boson divided by $k_{\rm B}T$ is then
\bea
\frac{U}{Nk_{\rm B}T} &=& \frac{\bar{U}}{Nt} \label{Eq:UonNt}\\*
&=& \frac{1}{N}\sum_{n_x,n_y=1}^\infty \frac{E(n_x,n_y,N)}{k_{\rm B}T}f_{\rm BE}(z,x,n_x,n_y,N).\nonumber
\eea
Using Eqs.~(\ref{Eq:EonT2}) and~(\ref{Eq:fBERed3}) this becomes
\be
\frac{\bar{U}}{Nt} = \frac{1}{N} \frac{a(N)}{N x}\sum_{n_x,n_y=1}^\infty\frac{n_x^2 + n_y^2}{z^{-1}\exp\left[\frac{a}{N x}(n_x^2 + n_y^2-2)\right]-1}.
\label{Eq:Ucalc0}
\ee
Similar to Eq.~(\ref{Eq:NEqual}), we reformulate the sum as
\bea
\frac{\bar{U}}{Nt} &=& \label{Eq:Ucalc}\\*
&&\hspace{-0.5in} \frac{1}{N} \frac{a(N)}{N x}\Bigg\{\sum_{n_x=1}^{n_{\rm max}}\sum_{n_y=1}^{\sqrt{n_{\rm max}^2-n_x^2}}\frac{n_x^2 + n_y^2}{z^{-1}\exp\left[\frac{a}{N x}(n_x^2 + n_y^2-2)\right]-1} \nonumber\\*
&&\hspace{0.5in} +\ \int_{n_{\rm max}}^\infty \frac{{\cal D}(n)n^2}{z^{-1}\exp\left[\frac{a(n^2-2)}{N x}\right]-1}dn\Bigg\}, \nonumber
\eea
where $n_{\rm max}$ was in the range 200 to 1000.   For the large-$x$ region $x\geq2$, we replaced $n^2-2$ in the expression for the energy in the integral by $n^2$ and the density of states ${\cal D}(n)$ in $n$~space given in Eq.~(\ref{Eq:Dofn}) is replaced by ${\cal D}(n) = \frac{\pi n}{2}$ so that the integral could be evaluated analytically in terms of polylogarithm functions.  We utilized the same strategy for calculations of ${\cal Z}$ and other thermodynamic properties for $x\geq 2$ when integrals such as in Eq.~(\ref{Eq:Ucalc}) were to be evaluated.

\begin{figure}
\includegraphics[width=3.3in]{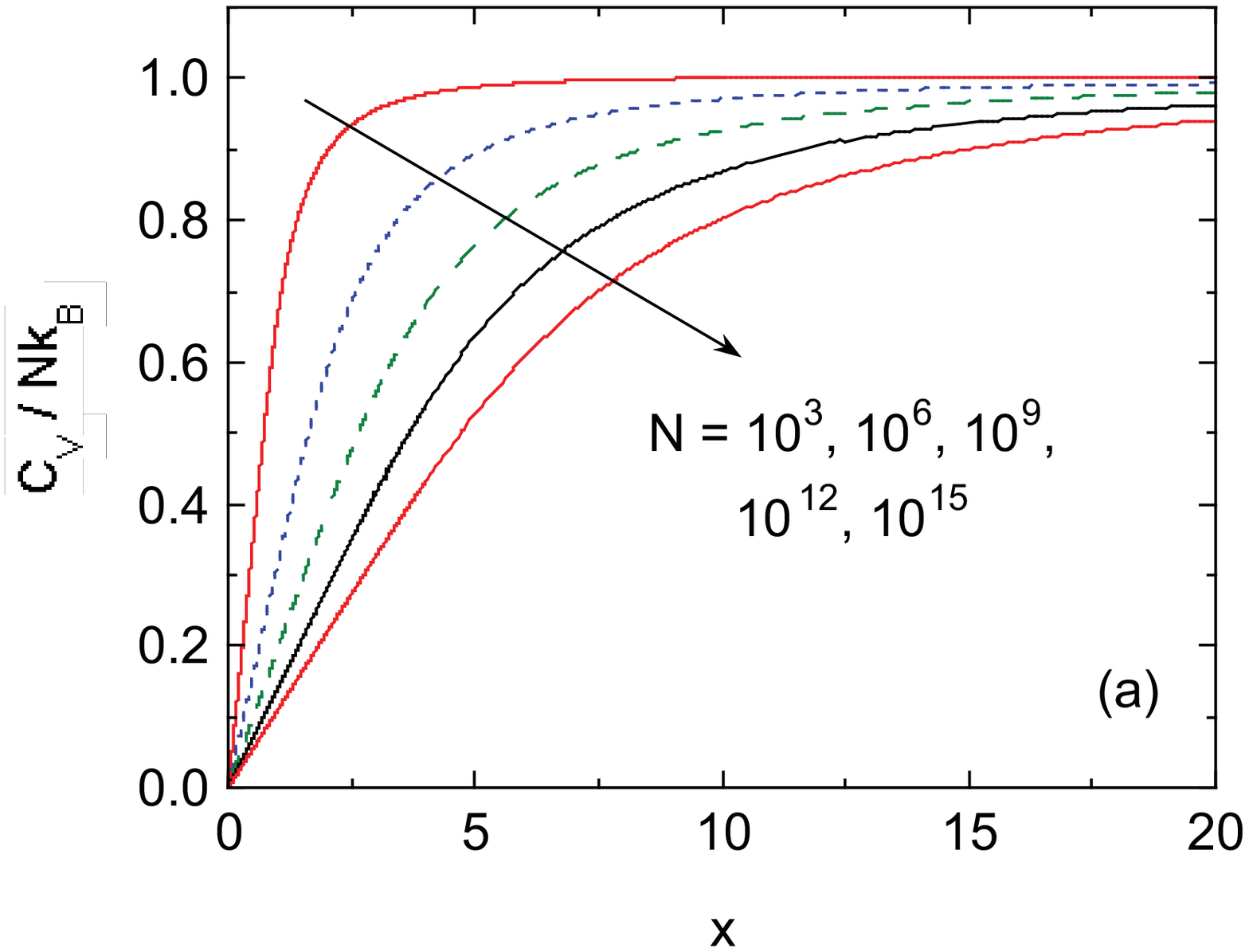}
\includegraphics[width=3.3in]{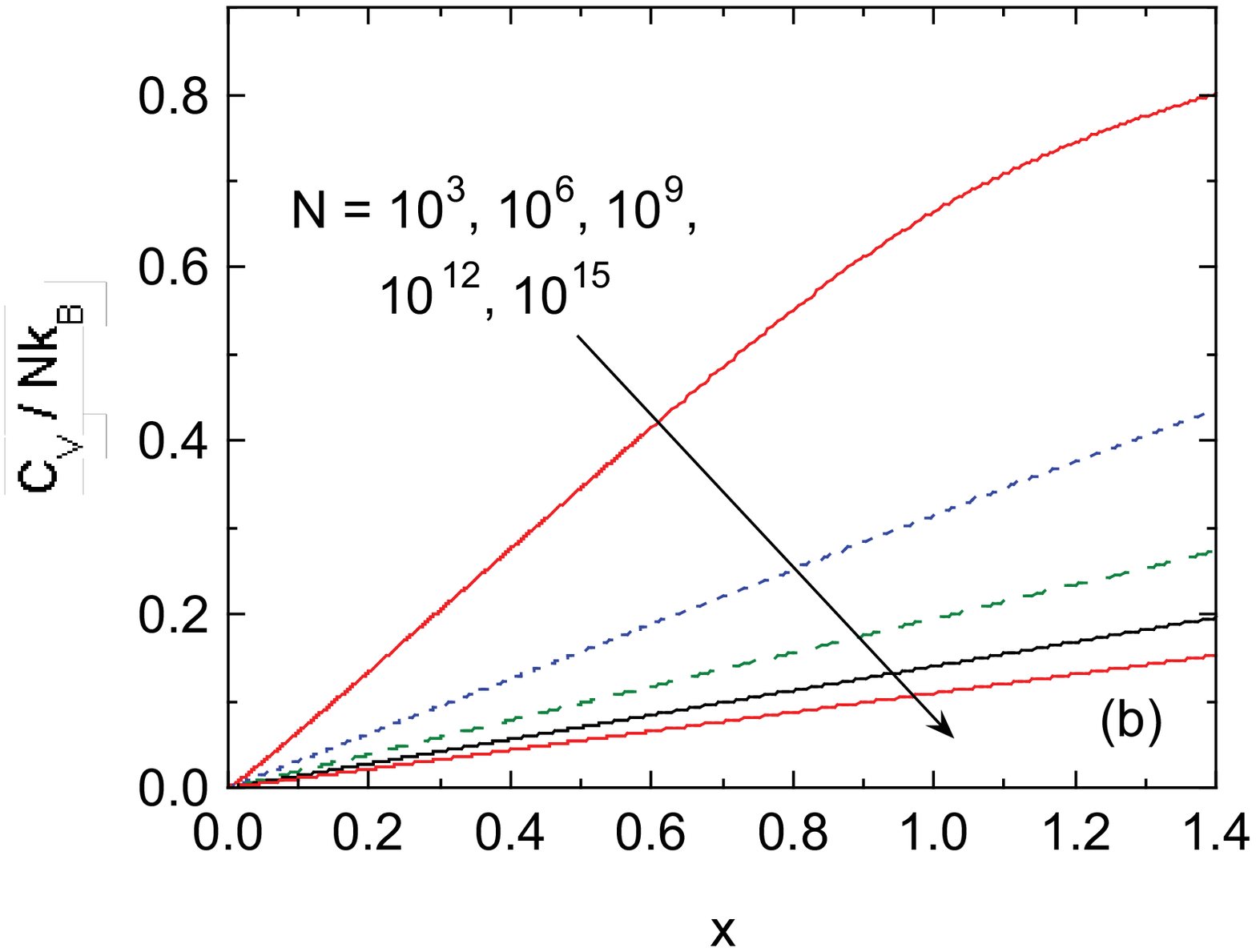}
\includegraphics[width=3.3in]{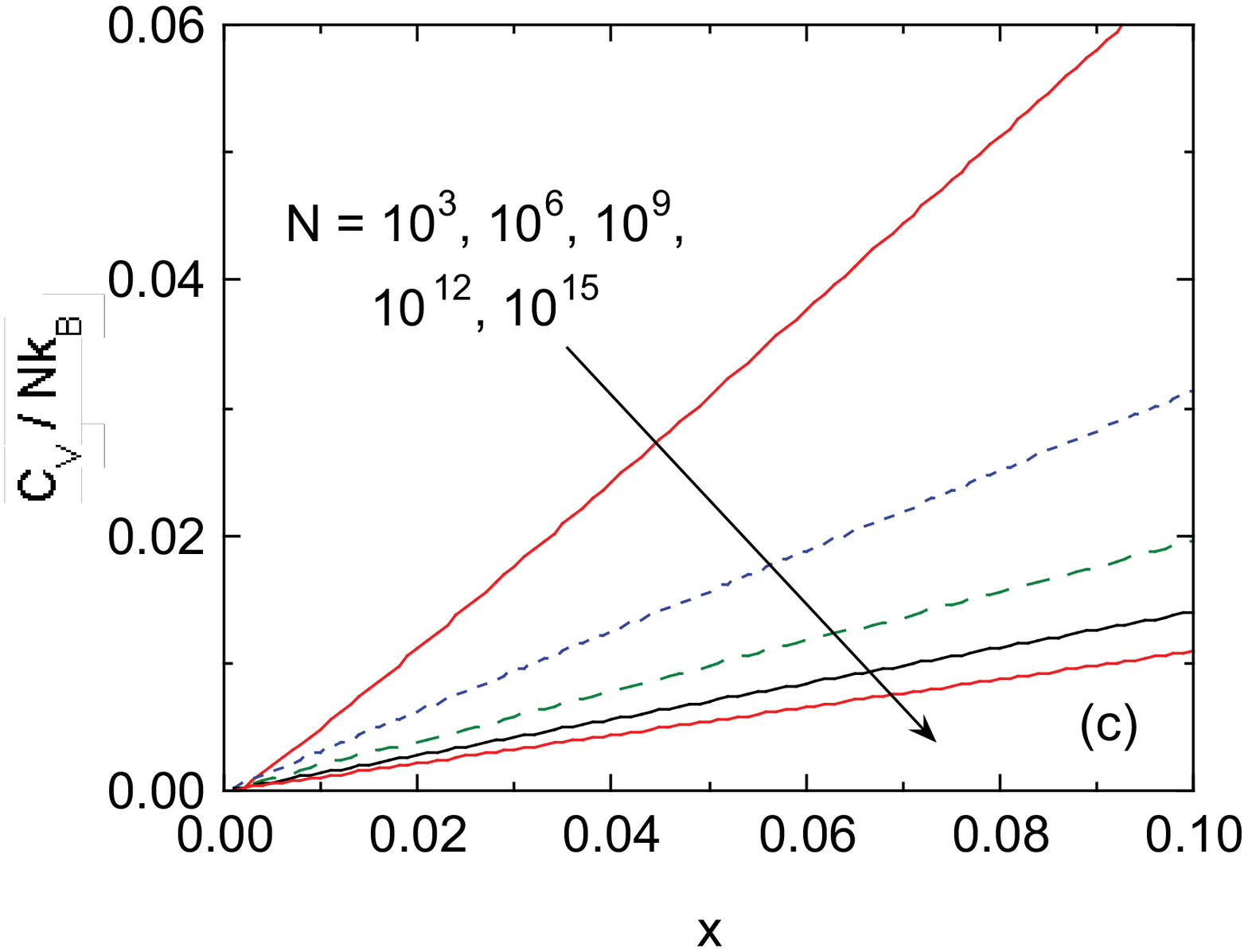}
\caption{(Color online) (a) Normalized heat capacity per boson at constant volume (area) $C_{\rm V}/N k_{\rm B}$ versus $x = \gamma t$ for several $N$ values and $0.001\leq x\leq 20$.  Expanded plots of the data in (a) are shown for (b)~$x\leq1.4$ and (c)~$x\leq0.1$.  }
\label{Fig:AllDataLoTHiT_CV}
\end{figure}

The heat capacity at constant volume (area) per boson is given in reduced units by
\be
\frac{C_{\rm V}}{Nk_{\rm B}}(x,N) = \frac{\partial \left[x{\frac{\bar{U}}{Nt}}(x,N)\right]}{\partial x},
\ee
where $\frac{\bar{U}}{Nt}(x,N)$ is obtained from Eq.~(\ref{Eq:Ucalc}) and the partial derivative is obtained as the $x$~derivative of a spline function of a list of closely-spaced $x\frac{\bar{U}}{Nt}(x,N)$ versus~$x$ data.  Shown in Fig.~\ref{Fig:AllDataLoTHiT_CV}(a) are plots of $C_{\rm V}/Nk_{\rm B}$ versus $x=\gamma t$ for $N=10^3$ to $10^{15}$ and $0.001\leq x\leq20$.  For each $N$, the large-$x$ (high temperature and/or large area) data approach unity as predicted by the classical equipartition theorem for the two translational degrees of freedom of a boson in 2D\@.  Successively expanded plots of the low-$x$ data for $x\leq 1.4$ and $x\leq 0.1$ are shown in Figs.~\ref{Fig:AllDataLoTHiT_CV}(b) and~\ref{Fig:AllDataLoTHiT_CV}(c), respectively.  No sharp features are visible at $x = 1$, which corresponds to $t=T/T_{\rm E} =1$ and $\gamma = A/A_{\rm E} = 1$, as expected for BEC in these finite systems where $T_{\rm E}$ is a crossover temperature rather than a phase transition temperature.  From Fig.~\ref{Fig:AllDataLoTHiT_CV}(c) one sees that $C_{\rm V}\sim x$ at small~$x$ for $N\geq10^6$, whereas the data for $N=10^3$ show positive curvature.  At sufficiently smaller~$x$ one expects an exponential dependence of $C_{\rm V}$ on $x$ due to the energy gap between the ground and first excited energy levels.

\subsection{\label{Sec:FS} Helmholtz Free Energy and Entropy}

\begin{figure}
\includegraphics[width=3.3in]{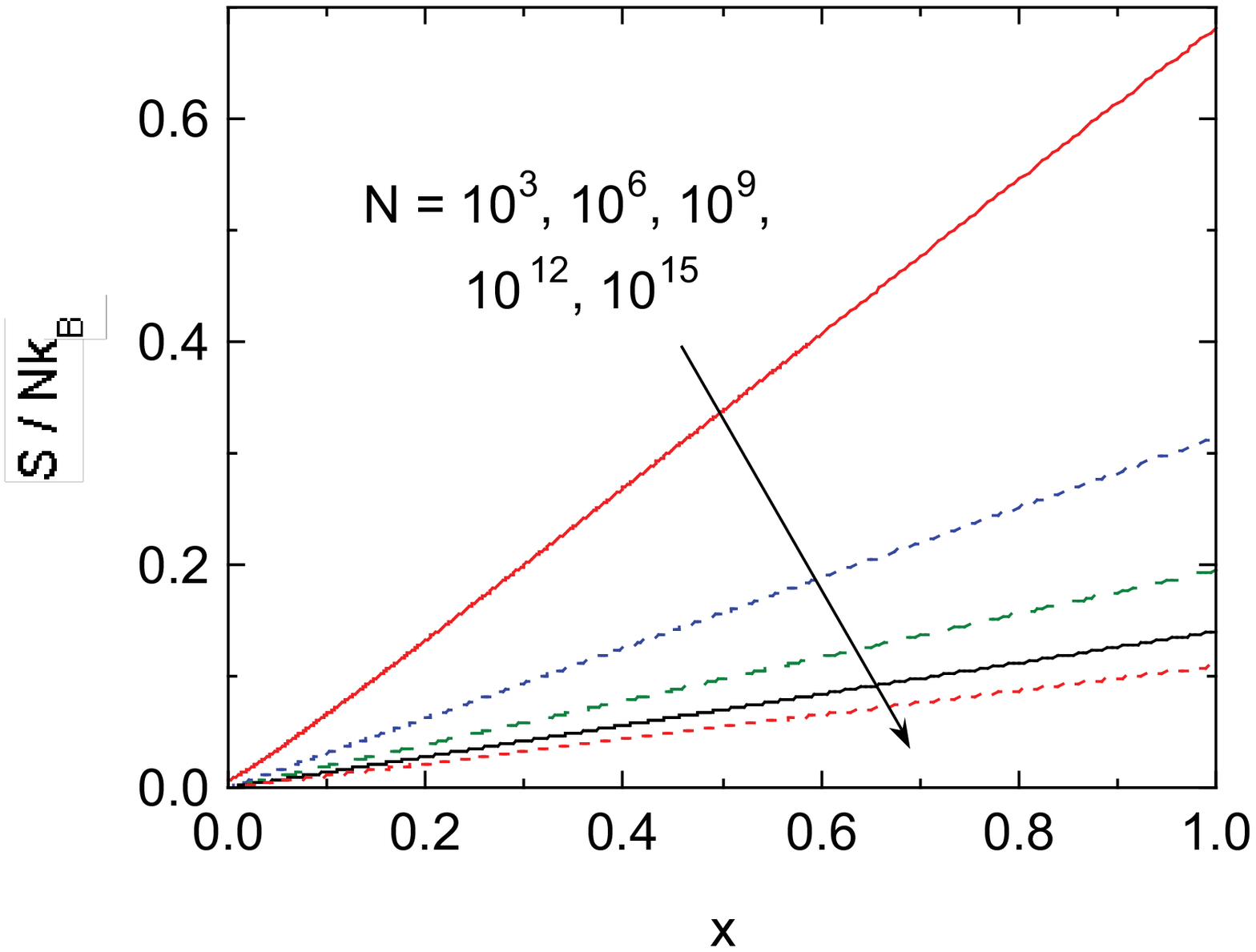}
\includegraphics[width=3.3in]{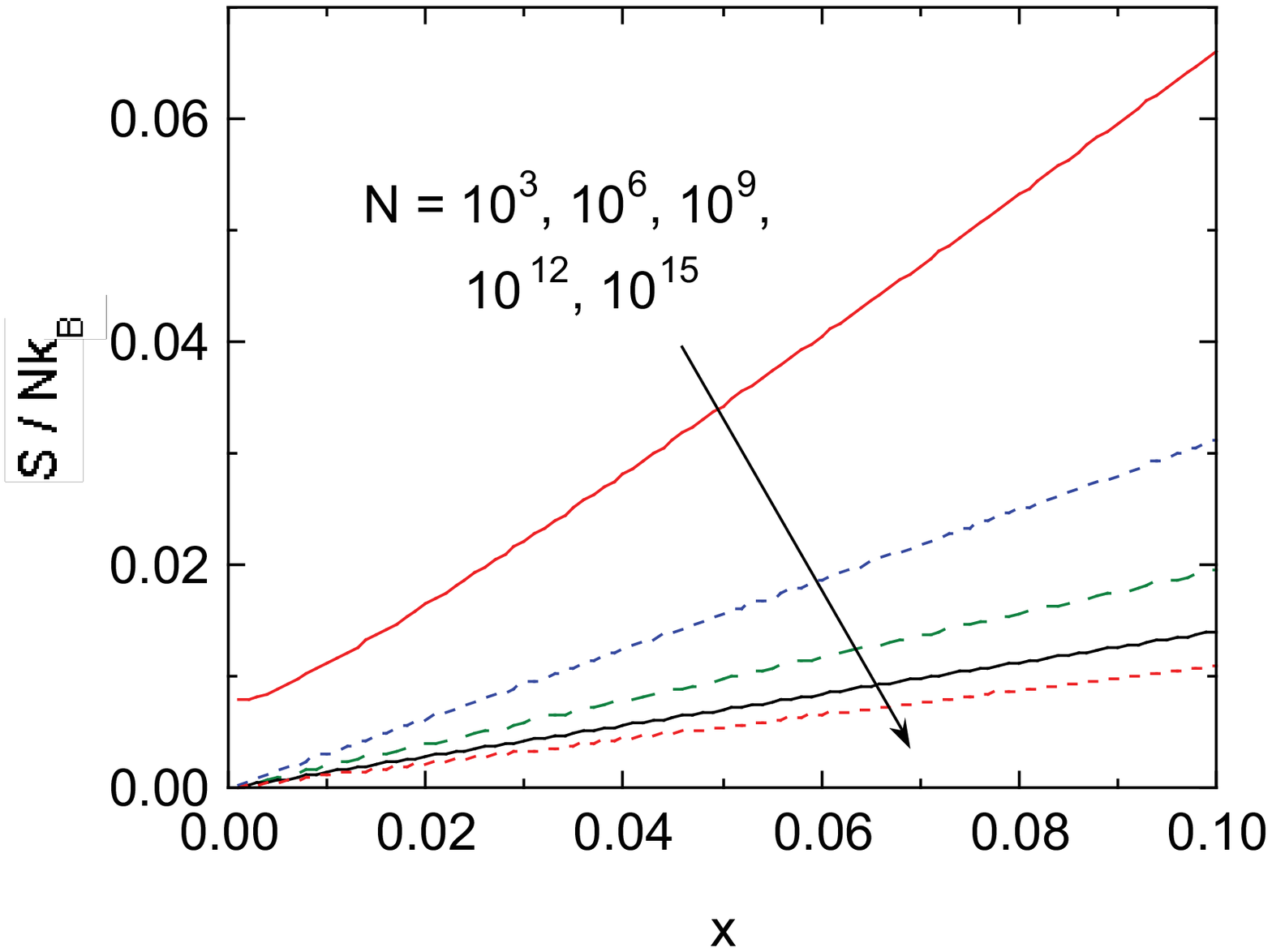}
\caption{(Color online) Normalized entropy per boson $S/N k_{\rm B}$ versus $x = \gamma t$ for (a) $N=10^3-10^{15}$ and $0.001\leq x\leq1$ and (b)~$N=10^6-10^{15}$ and $0.001\leq x\leq0.1$.}
\label{Fig:AllDataLoT_S}
\end{figure}

For each value of $N$ shown in Fig.~\ref{Fig:AllDataLoTHiT_CV}, $C_{\rm V}$ is approximately proportional to~$x$ for $x\lesssim 1$.  It is of interest to know whether or not the entropy $S(T = 0)=0$.  To determine that one must first calculate the Helmholtz free energy $F$, given by\cite{Huang1963,Reif1965}
\be
\frac{\bar{F}}{Nt}(x,N) = \frac{F}{Nk_{\rm B}T} = \ln z_{\rm unshifted}(x,N) - \frac{1}{N}\ln{\cal Z}(x,N),
\label{Eq:Fcalc}
\ee
where $\bar{F}=F/k_{\rm B}T_{\rm E}$ is the reduced free energy and the fugacity $z_{\rm unshifted}$ is for the actual unshifted energy levels as given in Eq.~(\ref{Eq:zunshifted}).  Then using the definition $F = U-TS$, one obtains
\be
\frac{S}{Nk_{\rm B}}(x,N) = \frac{\bar{U}}{Nt}(x,N) - \frac{\bar{F}}{Nt}(x,N),
\label{SGCE}
\ee
where $\frac{\bar{U}}{Nt}(x,N)$ was calculated above from Eq.~(\ref{Eq:Ucalc}) as a prerequisite for obtaining $C_{\rm V}$.

Following calculation of $\bar{F}/Nt$ from Eq.~(\ref{Eq:Fcalc}), $S(x,N)/Nk_{\rm B}$ was obtained from Eq.~(\ref{SGCE}) as shown for $N=10^3$ to $10^{15}$ in Fig.~\ref{Fig:AllDataLoT_S} in the $x$ ranges (a)~$0.001\leq x\leq 1$ and (b)~$0.001\leq x\leq 0.1$.  One sees that for $N = 10^6-10^{15}$, evidently $S(x\to0)\to0$, satisfying the third law of thermodynamics and also showing that the ground states for these $N$ values are nondegerate.  However,  in Fig.~\ref{Fig:AllDataLoT_S}(b) one also sees that $S(x\to0) =$~const~$>0$ for $N=10^3$, a surprising difference from the data for the larger $N$ values.  Therefore, the same result would presumably occur within the GCE formalism for $N=10^6-10^{15}$ at sufficiently small~$x$ with sufficiently high numerical resolution.  One anticipates that there is only one way to put all $N$ bosons into the nondegenerate ground state with $n_x=n_y=1$.  Hence the entropy at $T=0$ must be zero.  The nonzero entropy calculated for $N=10^3$ at $x\to0$ therefore demonstrates that the GCE formalism can give incorrect predictions for thermodynamic properties for finite~$N$ in the quantum regime with small~$x$, as found previously for the fluctuations in $N$ at low~$T$.\cite{Grossmann1997, Weiss1997, Wilkins1997, Holthaus2001, Mullin2003, Zannetti2015}  Indeed, in Sec.~\ref{Sec:CE} we show analytically using the CE formalism that the entropy for  $x\to0$ is identically zero for any finite~$N$\@.

\begin{table}
\caption{\label{Tab:S(xto0)} Reduced entropy $\frac{S}{Nk_{\rm B}}$ and compression factor $\tilde{\bar{p}} = \frac{pA}{Nk_{\rm B}T}$ for $x\to0$ versus $\log_{10}$ of the boson number $N$, obtained from Eqs.~(\ref{Eq:Sto0}) and~(\ref{Eq:pbarbarCalc}), respectively.}
\begin{ruledtabular}
\begin{tabular}{ccc}
$\log_{10}N$ 	& $\frac{S}{Nk_{\rm B}}(x\to0)$ & $\frac{pA}{Nk_{\rm B}T}(x\to0)$   \\
\hline
0  &  1.3863E+00  &  6.9315E$-$01  \\
1  &  3.3510E$-$01  &  2.3979E$-$01  \\
2  &  5.6102E$-$02  &  4.6151E$-$02  \\
3  &  7.9083E$-$03  &  6.9088E$-$03  \\
4  &  1.0210E$-$03  &  9.2104E$-$04  \\
5  &  1.2513E$-$04  &  1.1513E$-$04  \\
6  &  1.4816E$-$05  &  1.3816E$-$05  \\
7  &  1.7118E$-$06  &  1.6118E$-$06  \\
8  &  1.9421E$-$07  &  1.8421E$-$07  \\
9  &  2.1723E$-$08  &  2.0723E$-$08  \\
10  &  2.4026E$-$09  &  2.3026E$-$09  \\
11  &  2.6328E$-$10  &  2.5328E$-$10  \\
12  &  2.8631E$-$11  &  2.7631E$-$11  \\
13  &  3.0933E$-$12  &  2.9934E$-$12  \\
14  &  3.3235E$-$13  &  3.2236E$-$13  \\
15  &  3.5649E$-$14  &  3.4539E$-$14  \\
16  &  3.6841E$-$15  &  3.6841E$-$15  \\
17  &  3.9144E$-$16  &  3.9144E$-$16  \\
\end{tabular}
\end{ruledtabular}
\end{table}

In order to determine the source of the nonzero entropy at $x=0$ within the GCE formalism, we examine the contributions from each term in Eq.~(\ref{SGCE}) for the entropy at $x=0$.  From Eqs.~(\ref{Eq:zunshifted}) and~(\ref{Eq:zLimits}), one has
\be
\ln z_{\rm unshifted}(x\to0)= \frac{2a}{Nx}+\ln\left(\frac{N}{N+1}\right).
\label{Eq:zunshxto0}
\ee
For $x\to0$ only the ground state with $n_x=n_y=1$ is populated, and Eq.~(\ref{Eq:lnZGCE}) then gives
\be
\frac{1}{N}\ln{\cal Z}(x\to0) = \frac{1}{N}\ln(N+1).
\ee
From Eqs.~(\ref{Eq:zLimits}) and~(\ref{Eq:Ucalc0}) one obtains
\be
\frac{\bar{U}}{Nt}(x\to0)= \frac{2a}{N x}.
\ee
This term turns out to cancel the identical term in Eq.~(\ref{Eq:zunshxto0}).  Using these results and Eq.~(\ref{Eq:Fcalc}), Eq.~(\ref{SGCE}) gives
\be
\frac{S(x\to0)}{Nk_{\rm B}} = \ln\left(\frac{N+1}{N}\right) + \frac{1}{N}\ln(N+1),
\label{Eq:Sto0}
\ee
where the first term originated from $\ln z_{\rm unshifted}$ and the second came from $\ln{\cal Z}(x\to0)$, i.e., both terms originated from $F$\@.  Shown in Table~\ref{Tab:S(xto0)} is a list of values of $\frac{S(x\to0)}{Nk_{\rm B}}$ versus $N$ obtained using Eq.~(\ref{Eq:Sto0}).  The value for $N=10^3$ agrees with the value in Fig.~\ref{Fig:AllDataLoT_S}(b).  The value for $N=10^6$ is just below our resolution limit in Fig.~\ref{Fig:AllDataLoT_S}(b).  The results demonstrate that within the GCE formalism, $S(x\to0)$ is nonzero for all finite~$N$\@.  However, this result is not correct.  We prove analytically using the CE formalism in Sec.~\ref{Sec:FSCE} that $S(x\to0)$ is identically zero for any finite value of $N$\@.

\subsection{\label{Sec:pressure} Pressure}

\begin{figure}
\includegraphics[width=3.3in]{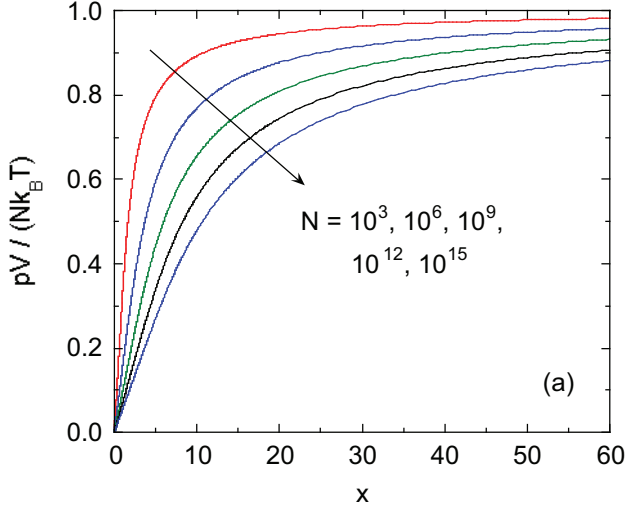}
\includegraphics[width=3.3in]{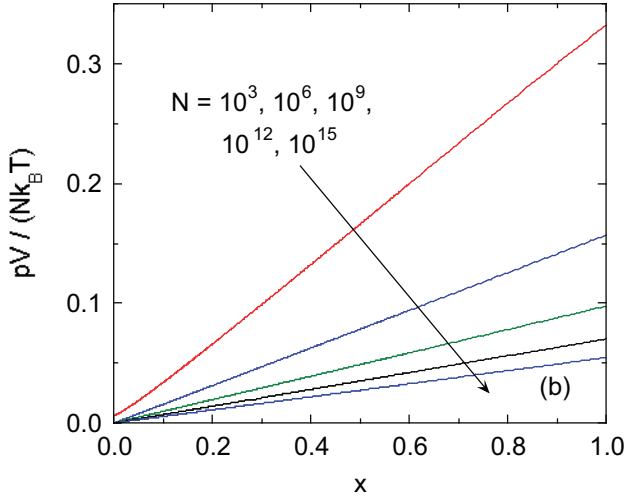}
\includegraphics[width=3.3in]{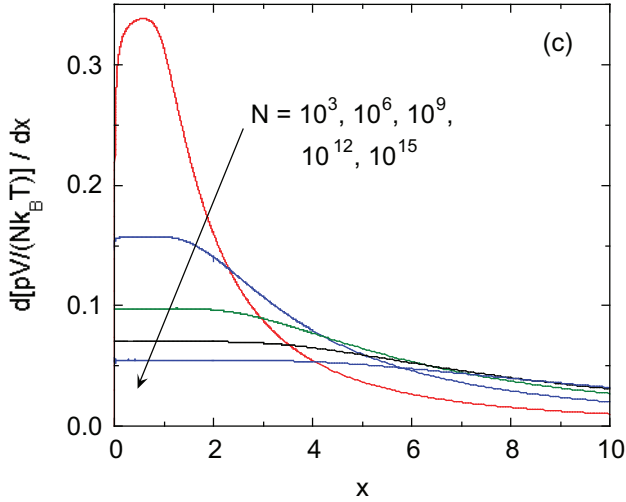}
\caption{(Color online) (a) Compression factor $\tilde{\bar{p}} = pV/N k_{\rm B}T$ versus the parameter $x = \gamma t$ for several $N$ values and $0.001\leq x\leq60$.  (b)~Expanded plot of the data in (a) for $0.001\leq x\leq1$.  (c)~The derivative $d\tilde{\bar{p}}/dx$ versus~$x$ for the same values of $N$ as in~(a) and~(b).}
\label{Fig:Pbarbar_vs_x}
\end{figure}

Within the GCE, the pressure~$p$ is given by\cite{Huang1963}
\bse
\label{Eqs:p}
\be
\frac{pA}{k_{\rm B}T} = \ln{\cal Z}.
\label{Eq:pVkT}
\ee
We define the reduced pressure $\bar{p}$ as
\be
\bar{p} = \frac{pv_{\rm E}}{k_{\rm B}T_{\rm E}} = \frac{t}{\gamma N}\ln[{\cal Z}(N,t,\gamma)].
\ee 
Another reduced pressure is
\be
\tilde{\bar{p}}(N,x) = \frac{\bar{p} \gamma}{t} = \frac{pV}{Nk_{\rm B}T} = \frac{1}{N}\ln[{\cal Z}(N,x)].
\label{Eq:tildeparp}
\ee
\ese
This quantity for a gas is sometimes called the ``compression factor'' in the literature (see, e.g., Ref.~\onlinecite{Johnston2014}).

Shown in Fig.~\ref{Fig:Pbarbar_vs_x}(a) are plots of $\tilde{\bar{p}}$ versus $x\equiv\gamma t$ for several values of $N$\@.  One sees that with increasing $x$, which corresponds to increasing area and/or temperature of the gas at fixed~$N$,  $\tilde{\bar{p}}$ approaches unity, as required since in these limits one must obtain the ideal gas law for which $\tilde{\bar{p}} = 1$.  An expanded plot of the data for $x\leq1$ is shown in Fig.~\ref{Fig:Pbarbar_vs_x}(b), where one sees that $\tilde{\bar{p}}(x\to0)=$~const for $N=10^3$.  We can obtain an exact value for $\tilde{\bar{p}}(x\to0)$ as follows.  For $T\to0$, the ground state is populated by all $N$ bosons.  Equation~(\ref{Eq:zLimits}) gives the fugacity as $z = N/(N+1)$.  Then Eq.~(\ref{Eq:lnZGCE}) gives $\ln{\cal Z}(t\to0)=\ln(N+1)$.  Using these results and Eq.~(\ref{Eq:tildeparp}) one obtains
\be
\tilde{\bar{p}}(N,x\to0) =\frac{\ln(N+1)}{N}.
\label{Eq:pbarbarCalc}
\ee
A list of $\tilde{\bar{p}}(N,x\to0)$ values versus~$N$ is given in Table~\ref{Tab:S(xto0)}.  For $N=1000$, one obtains $\tilde{\bar{p}}(N,x\to0)=6.9088\times10^{-3}$, in agreement with Fig.~\ref{Fig:Pbarbar_vs_x}(b).  For the larger $N$ values, $\tilde{\bar{p}}(N,x\to0)$ is too small to resolve on the scale of the figure.  The derivative $d\tilde{\bar{p}}/dx$ is plotted versus~$x$ in Fig.~\ref{Fig:Pbarbar_vs_x}(c).  For $N = 10^6$ to $10^{15}$, the data show regions of $x$ over which $d\tilde{\bar{p}}/dx=$~const and hence $\tilde{\bar{p}}$ is linear in~$x$ as also seen over the respective $x$ ranges with less precision in Fig.~\ref{Fig:Pbarbar_vs_x}(b).

\begin{figure}
\includegraphics[width=3.3in]{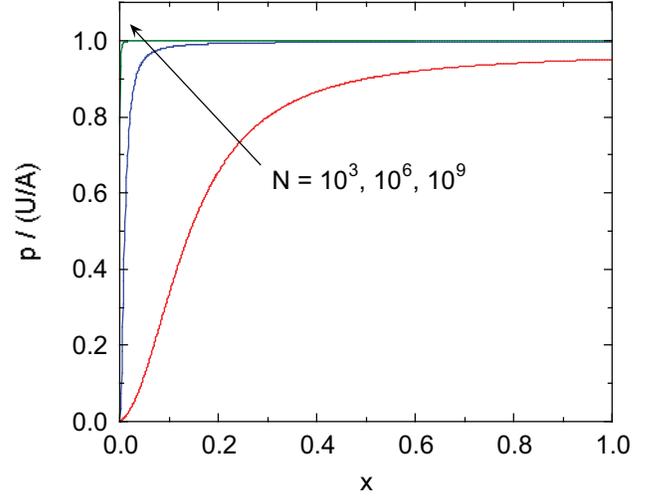}
\caption{(Color online) Ratio of the pressure $p$ to the energy density $U/A$ versus $x$ for boson numbers $N=10^3,\ 10^6$ and~$10^9$. The exact value obtained from the CE formalism is unity for all values of $x$ and for any finite~$N$. }
\label{Fig:p_on_U_ratio}
\end{figure}

The reduced pressure $\bar{p}$ is calculated from the above values of $\tilde{\bar{p}}(x)$ at fixed~$N$ obtained from Eq.~(\ref{Eq:tildeparp}) according to
\be
\bar{p}(\gamma,t) = \frac{t}{\gamma}\, \tilde{\bar{p}}(x).
\label{Eq:barp}
\ee
From dimensional considerations, one expects \mbox{$p\propto U/A$}.  As discussed later in Sec.~\ref{Sec:pCE}, the exact analytic result for the noninteracting 2D Bose gas obtained from the CE formalism is $p = U/A$, or $p/(U/A)=1$, for all $N$ and~$x$.  Within the GCE formalism, this ratio is equal to $\tilde{\bar{p}}/(U/Nt)$, where $\tilde{\bar{p}}$ is given by Eq.~(\ref{Eq:tildeparp}) and $U/Nt$ by Eq.~(\ref{Eq:UonNt}).  The ratio $p/(U/A)$ is plotted versus $x$ in Fig.~\ref{Fig:p_on_U_ratio} for $N=10^3,\ 10^6$ and $10^9$.  One sees that the GCE formalism gives incorrect $p/(U/A)$ ratios for $N=10^3$ and $10^6$, with the deviation from unity increasing with decreasing $N$ and~$x$.  Similar deviations must also occur for larger but finite~$N$ at lower~$x$ values than plotted.  These deviations from unity again illustrate the failure of the GCE formalism to accurately predict thermodynamic properties for finite $N$ at small~$x$ where significant BEC occurs.

Of particular interest for the thermodynamics are $\bar{p}$ versus $\gamma$ isotherms, $\bar{p}$ versus $t$ isochores and $\gamma$ versus $t$ isobars.  These relationships are generated parametrically from $\tilde{\bar{p}}(x)$ using Eq.~(\ref{Eq:barp}) and the definition $x=\gamma t$.  For a $\bar{p}$ versus~$t$ isochore, one chooses a particular fixed value of the reduced area $\gamma$ and $t$ is then obtained from $x$ according to $t(x) = x/\gamma$.  Similarly, for an isotherm, one chooses a particular value of $t$ and $\gamma$ is obtained as $\gamma(x) = x/t$.  In order to obtain a $\gamma$ versus $t$ isobar, one chooses a particular value of $\bar{p}$.  Then using $\gamma=x/t$, Eq.~(\ref{Eq:barp}) gives
\be
t(x) =\sqrt{\frac{x\bar{p}}{\tilde{\bar{p}}(x)}}
\ee
Once $t$ is determined for a given value of $x$, one uses $\gamma(x) = x/t(x)$ to obtain~$\gamma$ for that value of $t$.

\begin{figure}
\includegraphics[width=3.3in]{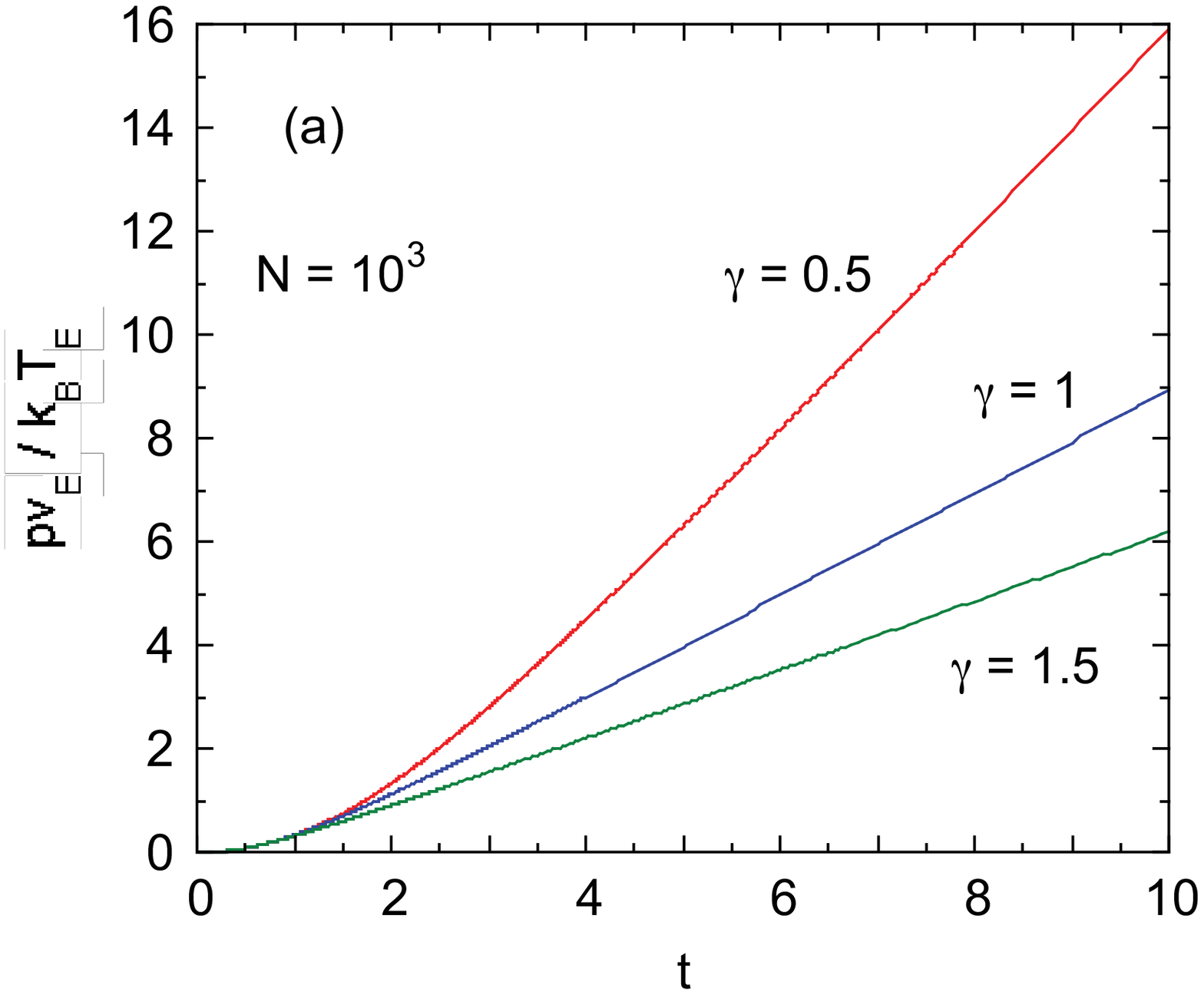}\vspace{-0.05in}
\includegraphics[width=3.3in]{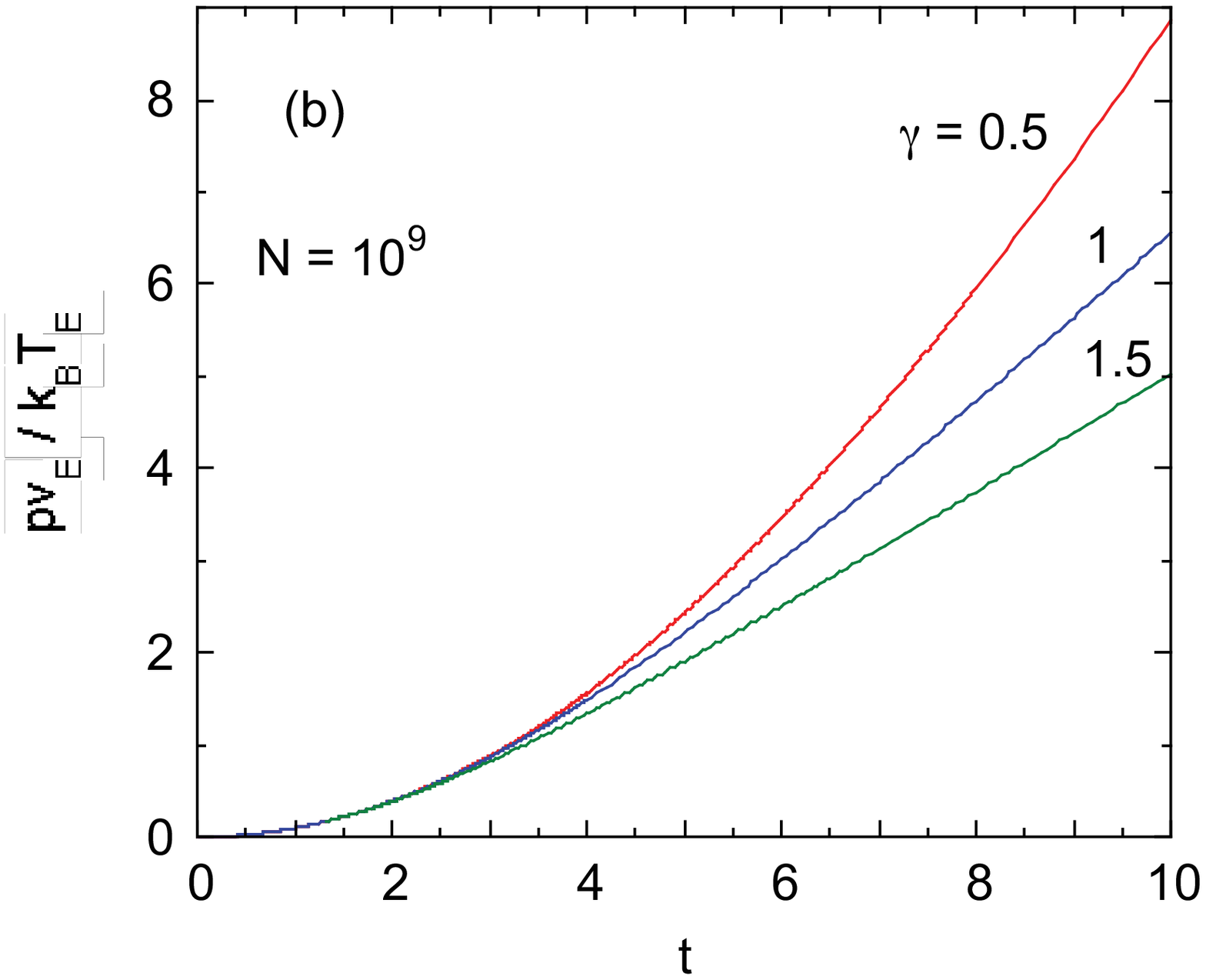}\vspace{-0.05in}
\includegraphics[width=3.3in]{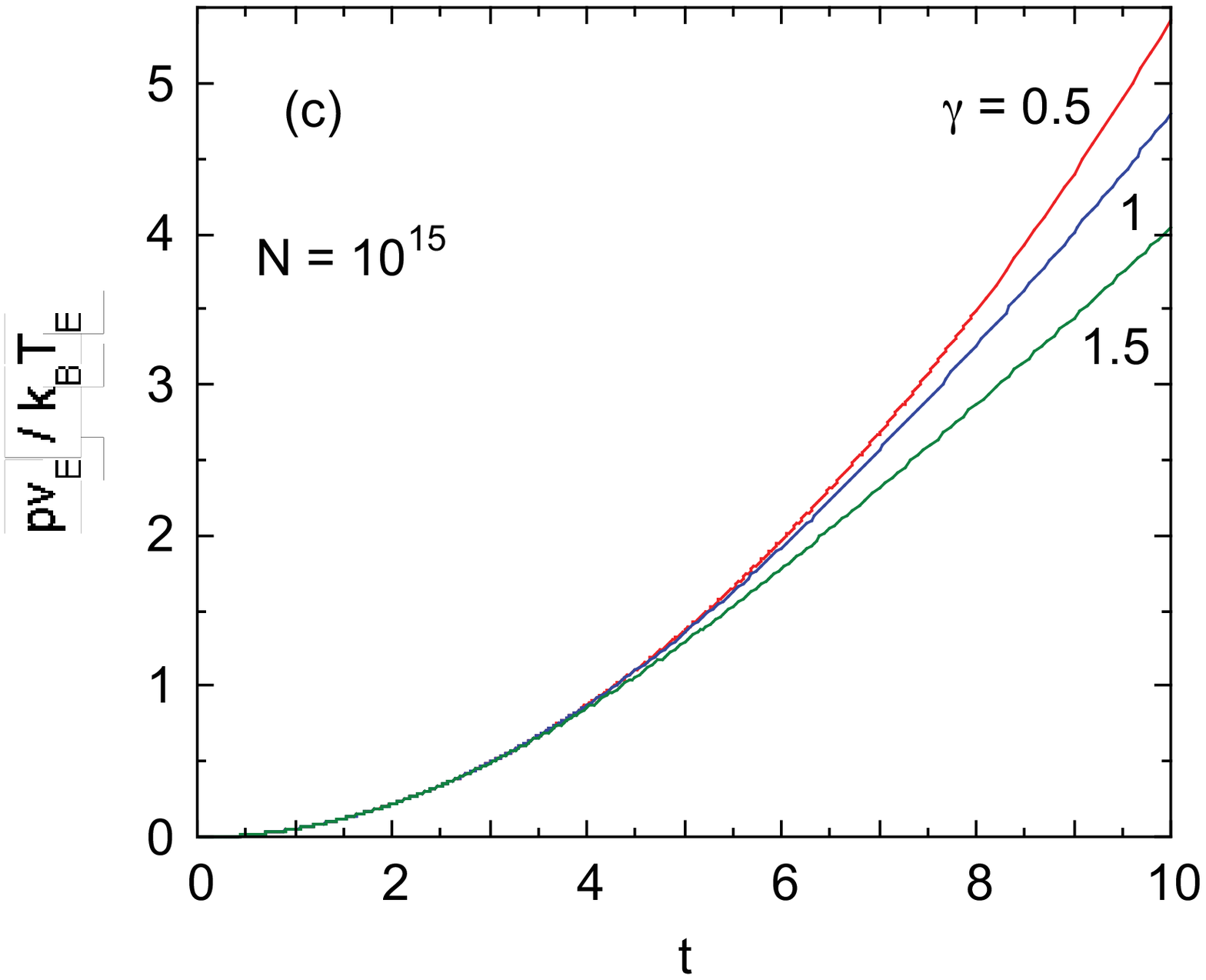}
\caption{(Color online) Reduced pressure $\bar{p} = p\,v_{\rm E}/(k_{\rm B}T_{\rm E})$ versus reduced temperature $t=T/T_{\rm E}$ for reduced area $\gamma = v/v_{\rm E} = 0.5,$ 1 and~1.5 for boson numbers (a)~$N = 10^3$, (b)~$N = 10^9$ and (c)~$N = 10^{15}$.  Note the different scales for the ordinates in (a), (b) and (c).}
\label{Fig:Pbar_vs_t}
\end{figure}

\begin{figure}
\includegraphics[width=3.3in]{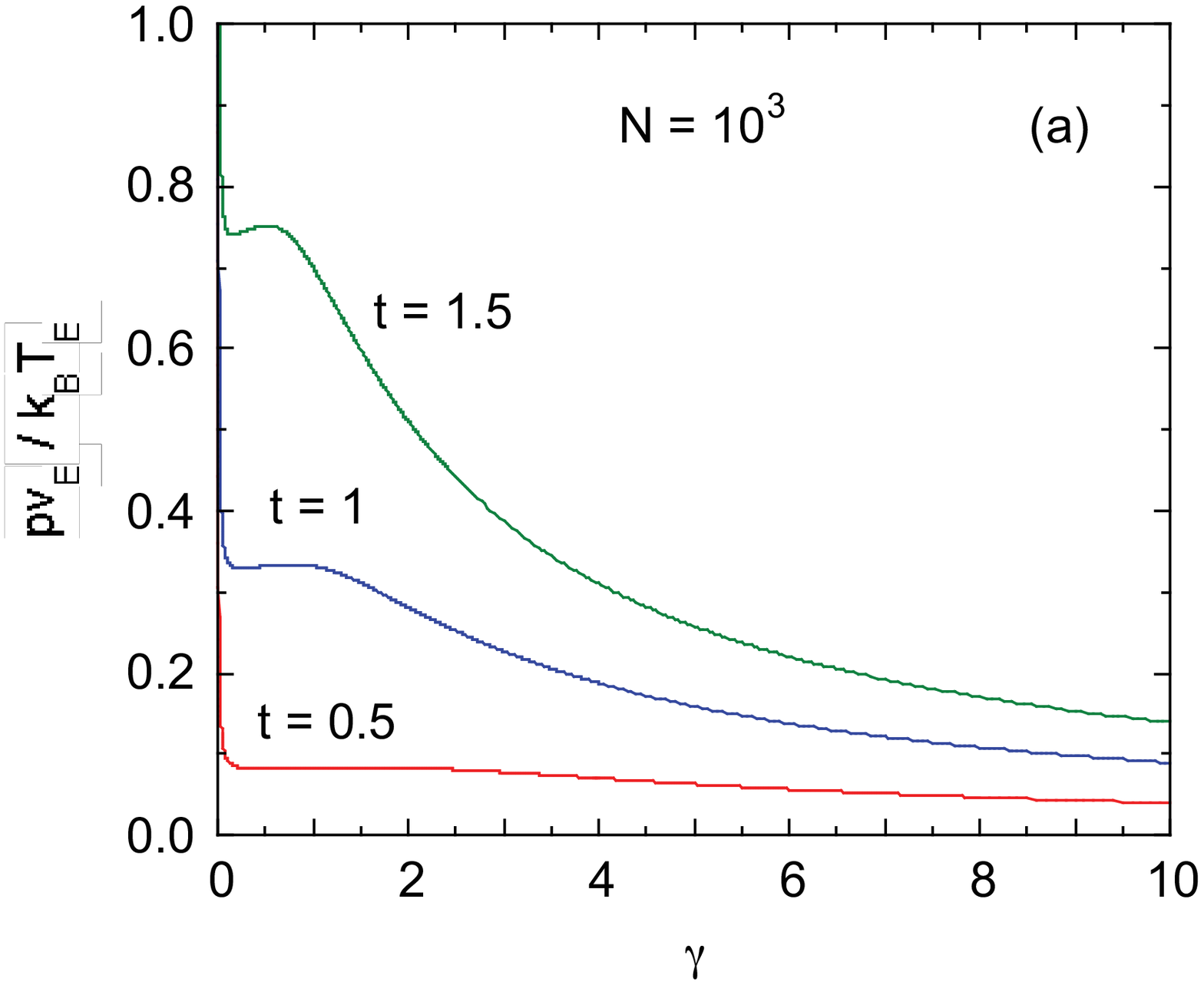}
\includegraphics[width=3.3in]{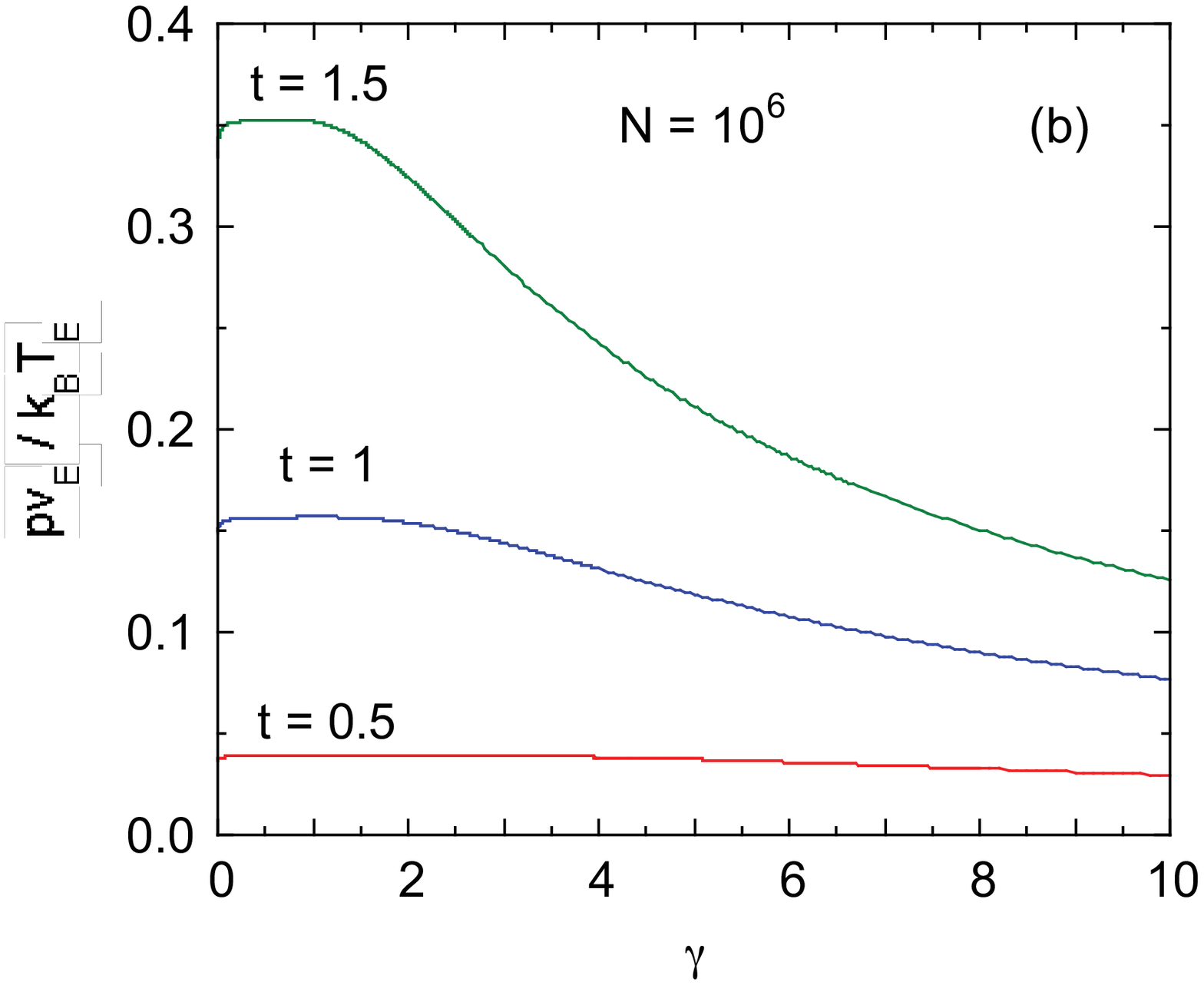}
\includegraphics[width=3.3in]{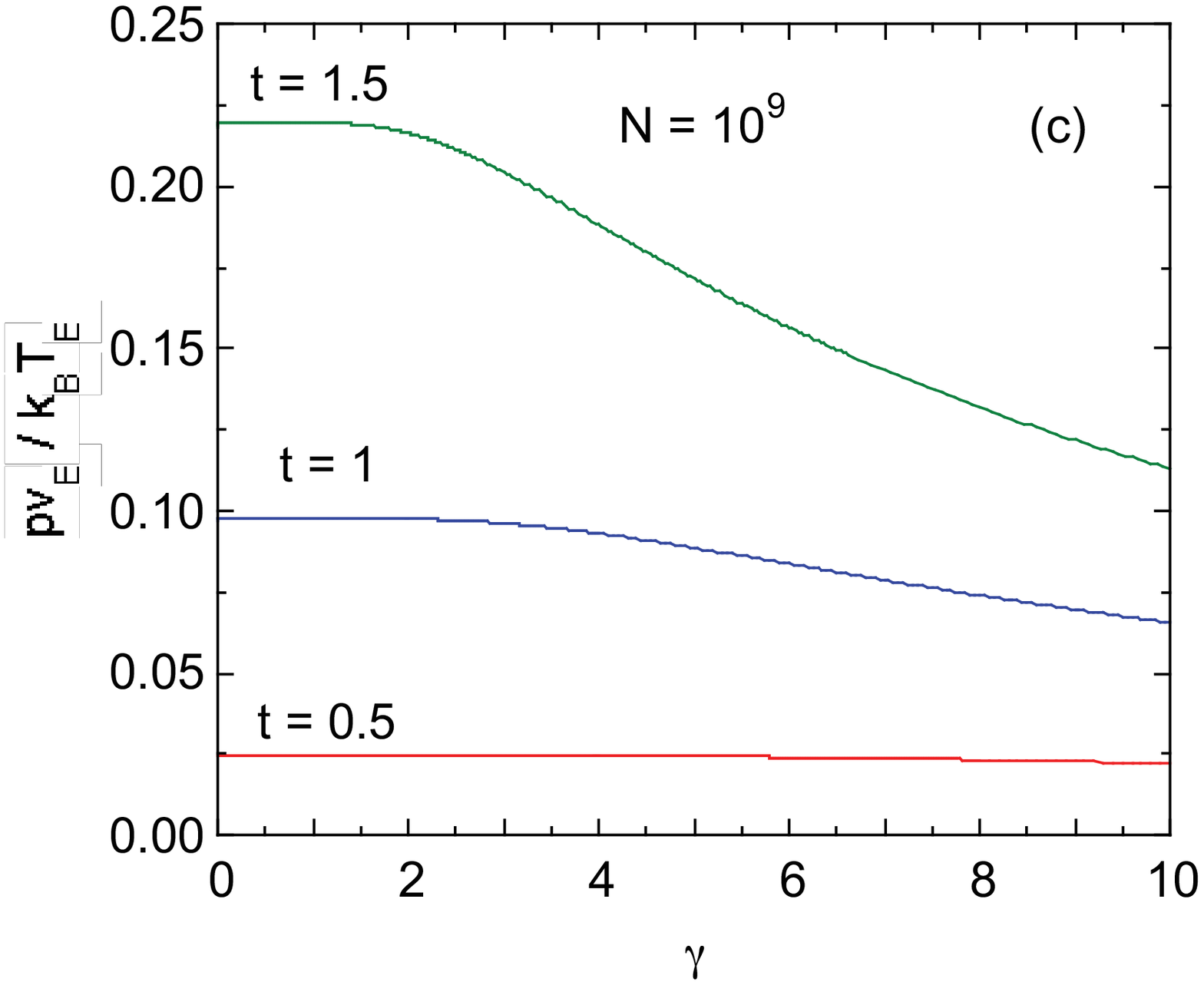}
\caption{(Color online) Reduced pressure $\bar{p} = p\,v_{\rm E}/k_{\rm B}T_{\rm E}$ versus reduced area $\gamma = v/v_{\rm E}$ for reduced temperatures $t=T/T_{\rm E} = 0.5$, 1 and~2 and boson numbers (a)~$N = 10^3$, (b)~$N = 10^6$ and (c)~$N = 10^9$.}
\label{Fig:Pbar_vs_gam}
\end{figure}

\begin{figure}
\includegraphics[width=3.3in]{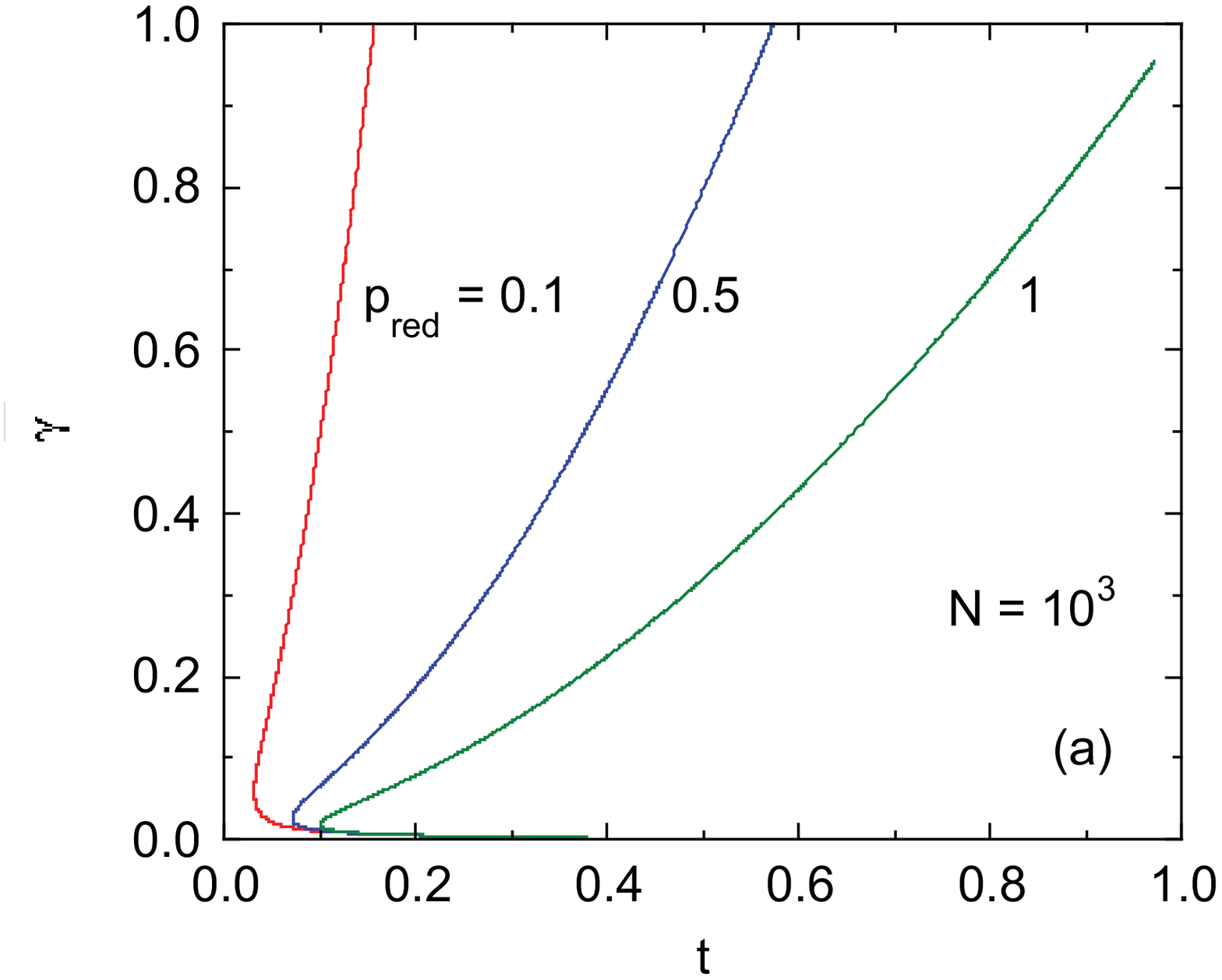}
\includegraphics[width=3.3in]{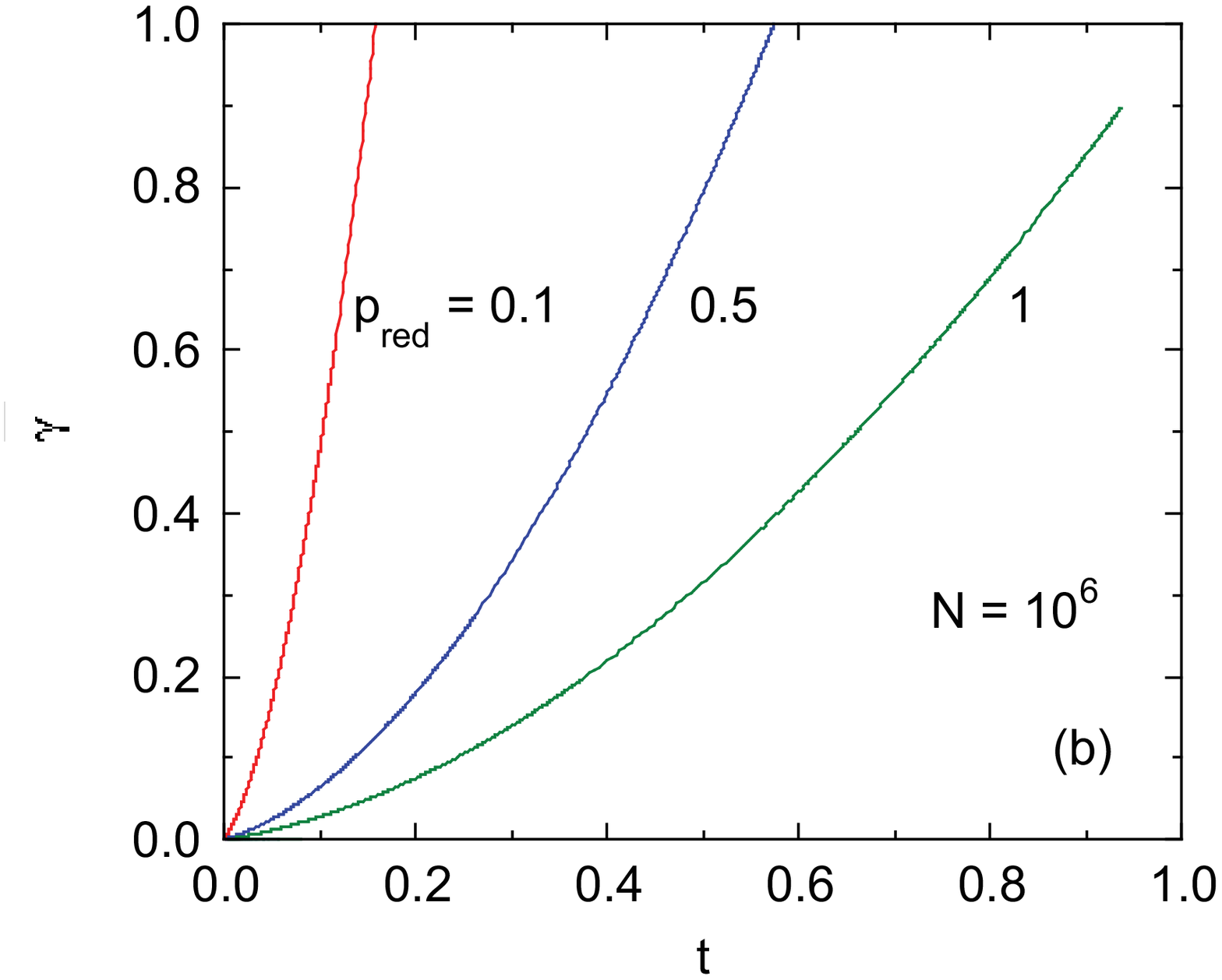}
\caption{(Color online) Reduced area $\gamma$ versus reduced temperature $t$ for reduced pressures $p_{\rm red} \equiv \bar{p} = 0.1$, 0.5 and~1 and boson numbers (a)~$N = 10^3$ and (b)~$N = 10^6$.  Isobars for larger $N$ are similar to those in~(b).}
\label{Fig:gam_vs_t}
\end{figure}

Isochores of $\bar{p}$ versus~$t$ with $\gamma=0.5$, 1 and~1.5 are plotted in Figs.~\ref{Fig:Pbar_vs_t}(a), \ref{Fig:Pbar_vs_t}(b) and \ref{Fig:Pbar_vs_t}(c) for $N=10^3,\ 10^9$ and $10^{15}$, respectively.  One sees that below an $N$-dependent temperature, the isochores for these three reduced areas for a given~$N$ are nearly the same.  This means that in the respective $t$ range, the pressure is nearly independent of area as will be seen explicitly in pressure versus area isotherms.  At higher temperatures, the pressure decreases with increasing reduced area.

Isotherms of $\bar{p}$ versus $\gamma$ at fixed $t=0.5,\ 1$ and~1.5 are shown in Fig.~\ref{Fig:Pbar_vs_gam} for $N=10^3$, $N=10^6$ and $N=10^9$.  The plots for $N=10^3$ and $N=10^6$ show unphysical regions at low temperatures with positive slope, corresponding to a negative isothermal compressibility $\kappa_{\rm T}$ according to its definition for a 2D system given by
\be
\frac{1}{\kappa_{\rm T}} = -A\frac{\partial p(T,A,N)}{\partial A}.
\label{Eq:KTDef}
\ee
Furthermore, the regions of $\gamma$ for which $d\bar{p}(\gamma)/d\gamma = 0$ for all three values of $N$ correspond to regions of infinite compressibility, which is unphysical for a finite noninteracting Bose gas. 

Reduced area~$\gamma$ versus reduced temperature~$t$ isobars with $\bar{p} = 0.1,$ 0.5 and~1 are shown in Figs.~\ref{Fig:gam_vs_t}(a) and~\ref{Fig:gam_vs_t}(b) for $N=10^3$ and $10^6$, respectively.  The thermal expansion coefficient $\alpha_{\rm p}$ is defined as
\bse
\be
\alpha_{\rm p} = \frac{1}{A}\left(\frac{\partial A}{\partial T}\right)_{\rm p}.
\ee
In dimensionless reduced form this becomes
\be
\bar{\alpha}_{\rm p} \equiv \alpha_{\rm p}T_{\rm E} = \frac{1}{\gamma}\left(\frac{\partial \gamma}{\partial t}\right)_{\rm \bar{p}}.
\label{Eq:alphabar}
\ee
\ese
The isobars for $N=10^3$ in Fig.~\ref{Fig:gam_vs_t}(a) exhibit unphysical regions of negative thermal expansion for $\bar{p}<1$ and small $\gamma$ that are not apparent in the isobars for $N=10^6$ and larger~$N$\@.

The above unphysical predictions of the GCE formalism for the thermodynamic properties at low values of $N$, $t$ and/or $\gamma$ at which significant BEC occurs are rectified in Sec.~\ref{Sec:CE} below when we consider the predictions of the CE formalism for the same properties.

\subsection{\label{Sec:kapalphCp} Isothermal Compressibility, Thermal Expansion Coefficient and Heat Capacity at Constant Pressure}

In dimensionless reduced units Eq.~(\ref{Eq:KTDef}) becomes
\bse
\be
\frac{\bar{\kappa}_{\rm T}}{\gamma^2} = \frac{1}{x^2\left[\frac{\tilde{\bar{p}}(x,N)}{x} - \frac{\partial\tilde{\bar{p}}(x,N)}{\partial x}\right]},
\label{Eq:kapbet2}
\ee
where the reduced isothermal compressibility $\bar{\kappa}_{\rm T} $ is
\be
\bar{\kappa}_{\rm T} = \left(\frac{k_{\rm B}T_{\rm E}}{v_{\rm E}}\right)\kappa_{\rm T}.
\ee
\ese
One also has
\be
\kappa_{\rm T}p = \bar{\kappa}_{\rm T}\bar{p} = \left(\frac{\bar{\kappa}_{\rm T}}{\gamma^2}\right)x\tilde{\bar{p}}.
\label{Eq:Kappap}
\ee
The ideal gas exhibits
\be
\kappa_{\rm T}p = 1\qquad {\rm (ideal~gas)},
\label{kapTlimit}
\ee
to which $\bar{\kappa}_{\rm T}\bar{p}$ for the Bose gas must asymptote for $x\to\infty$.

We now derive an expression for $\bar{\alpha}_{\rm p}(t)$ in terms of quantities already calculated.  Writing $\bar{p} = \bar{p}(t,\gamma,N)$, at constant pressure one has the differential
\be
d\bar{p}=0 = \frac{\partial\bar{p}(t,\gamma,N)}{\partial t}dt + \frac{\partial\bar{p}(t,\gamma,N)}{\partial\gamma}d\gamma,
\ee
yielding
\be
\left(\frac{\partial \gamma}{\partial t}\right)_{\rm \bar{p}} = -\frac{\frac{\partial\bar{p}(t,\gamma,N)}{\partial t}}{\frac{\partial\bar{p}(t,\gamma,N)}{\partial\gamma}}.
\label{Eq:dbdt}
\ee
Then using Eqs.~(\ref{Eq:tildeparp}), (\ref{Eq:alphabar}) and~(\ref{Eq:dbdt}) one obtains
\be
\frac{\bar{\alpha}_{\rm p}}{\gamma} = \left(\frac{\bar{\kappa}_{\rm T}}{\gamma^2}\right) \frac{\partial [x\,\tilde{\bar{p}}(x,N)]}{\partial x},
\label{Eq:baralpha}
\ee
where $\bar{\kappa}_{\rm T}/\gamma^2$ is given in Eq.~(\ref{Eq:kapbet2}).   Also, one has
\be
\alpha_{\rm p}T = x\frac{\bar{\alpha}_{\rm p}}{\gamma}.
\label{Eq:alphaT}
\ee
The ideal gas shows $\alpha_{\rm p}T = 1$, to which $x(\bar{\alpha}_{\rm p}/\gamma)$ for the Bose gas must approach for $x\to\infty$.

Finally, the difference $C_{\rm p}-C_{\rm V}$ between the heat capacities at constant pressure and constant volume satisfies\cite{Reif1965}
\be
C_{\rm p}-C_{\rm V} = \frac{TA\alpha_{\rm p}^2}{\kappa_{\rm T}}.
\ee
In dimensionless reduced parameters one obtains
\be
\frac{C_{\rm p}-C_{\rm V}}{Nk_{\rm B}} = x\,\frac{(\bar{\alpha}_{\rm p}/\gamma)^2}{(\bar{\kappa}_{\rm T}/\gamma^2)}.
\label{Eq:Cp-CV}
\ee
For the ideal gas this quantity equals unity, which the Bose gas must approach for $x\to\infty$.

\begin{figure}
\includegraphics[width=3.in]{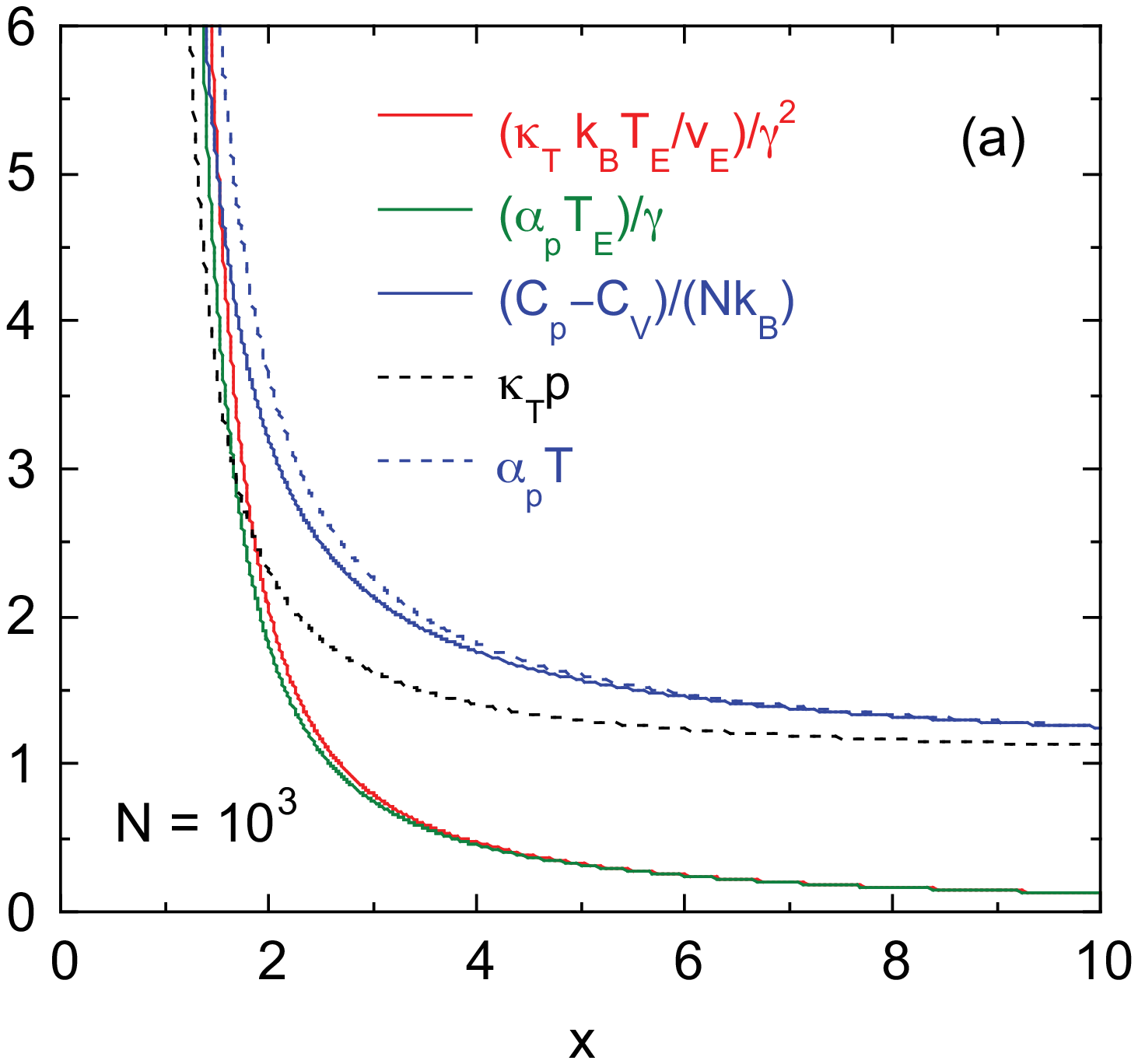}
\includegraphics[width=3.in]{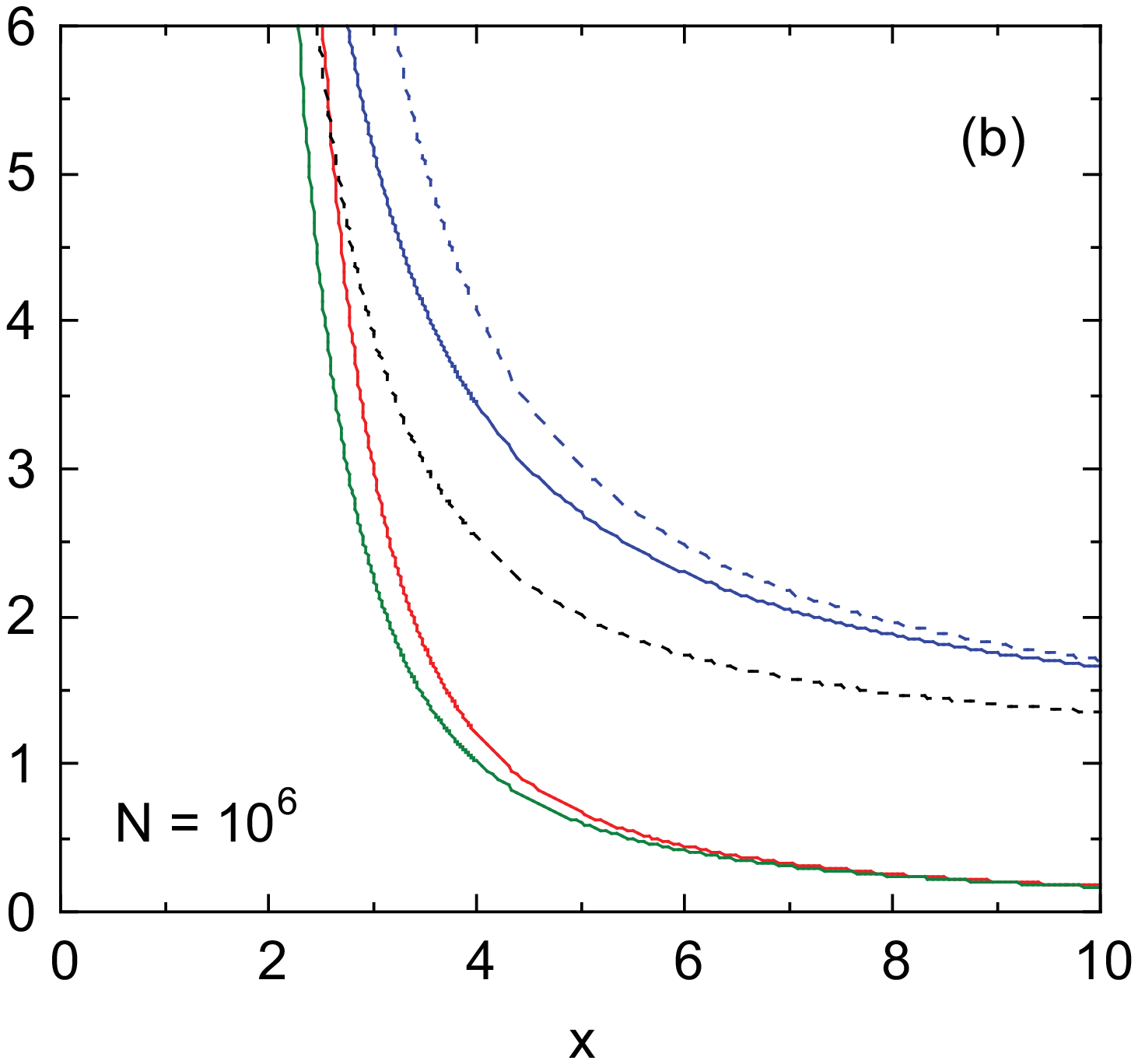}
\caption{(Color online) Plots of reduced isothermal compressibility $\bar{\kappa}_{\rm T}/\gamma^2$, thermal expansion coefficient $\bar{\alpha}_{\rm p}/\gamma$, and the difference $(C_{\rm p}-C_{\rm V})/Nk_{\rm B}$ between the heat capacity at constant pressure and at constant volume for $\gamma=1$ and boson number (a)~$N=10^3$ and (b)~$N=10^6$.  Also shown in each panel are the products $\alpha_{\rm p}T$ and $\kappa_{\rm T}p$ which for an ideal gas are both equal to unity, as verified in the respective high-$x$ limits in (a) and~(b).  The figure legend in~(a) also applies to~(b).}
\label{Fig:P_dPdx_alpha_kappa_Cp}
\end{figure}

Shown in Fig.~\ref{Fig:P_dPdx_alpha_kappa_Cp} are plots of $\bar{\kappa}_{\rm T}/\gamma^2$, $\bar{\alpha}_{\rm p}/\gamma$ and $(C_{\rm p}-C_{\rm V})/Nk_{\rm B}$ versus~$x$ for (a) $N=10^3$ and~(b) $N=10^6$.  All three quantities show unphysical divergences and then negative values as $x$ decreases into the BEC regime (not shown).  Also shown are the associated plots of $\kappa_{\rm T}p$ and $\alpha_{\rm p}T$ from Eqs.~(\ref{Eq:Kappap}) and~(\ref{Eq:alphaT}), respectively.  One sees that $(C_{\rm p}-C_{\rm V})/(Nk_{\rm B})$, $\kappa_{\rm T}p$ and $\alpha_{\rm p}T$ approach the same respective ideal gas value of unity at large~$x$, as required.

\section{\label{Sec:CE} Results: Canonical Ensemble}

In previous sections we pointed out a number of unphysical or unexpected predictions of the GCE formalism when $N$ and $x$ are both small (in the BEC regime) in addition to the known unphysically large fluctuations in $N$ at small~$x$ even in the thermodynamic limit.  In this section we resolve these problems by calculating the thermodynamics using the CE formalism which can give exact results for a finite system with fixed~$N$ in thermal contact with a temperature reservoir.  The partition function~$Q(N)$ and average number of bosons $\bar{n}_i$ in a given quantum state with energy $E_i$ are calculated recursively as described in Sec.~\ref{Sec:CEMethods}, and our calculations are carried out with a maximum boson number $N=1000$.  Some of the thermodynamic properties for $N=1000$ will be compared with the above unphysical and/or incorrect results predicted by the GCE formalism.

\subsection{\label{Sec:popstats} Population Statistics}

\begin{figure}
\includegraphics[width=3.3in]{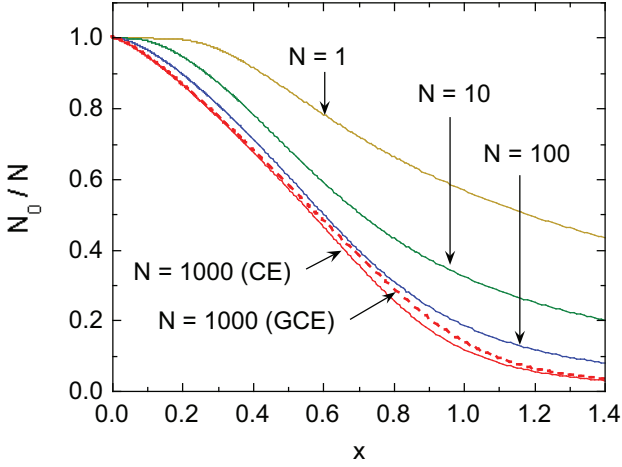}
\caption{(Color online) Fractional occupations $N_0/N$ of the ground state versus $x = \gamma t$ for four $N$ values as determined from the CE formalism using Eq.~(\ref{Eq:barni}) (solid curves).  The data for $N=1000$ from Fig.~\ref{Fig:AllDataLoT_N0N} calculated using the GCE formalism are shown for comparison (dashed curve).}
\label{Fig:AllDataLoCET_N0N}
\end{figure}

\begin{figure*}
\includegraphics[width=3.in]{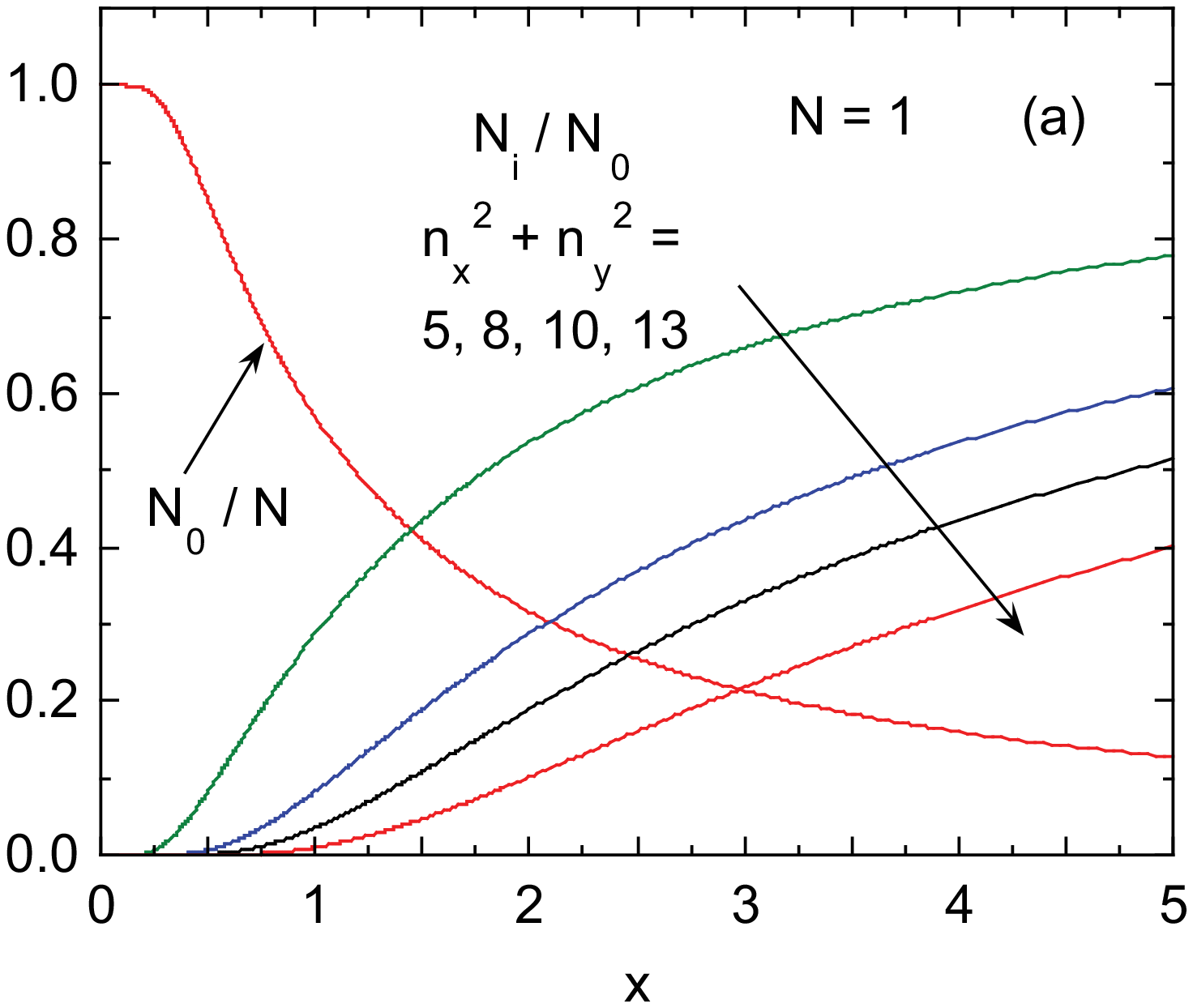}\hspace{0.2in}\includegraphics[width=3.in]{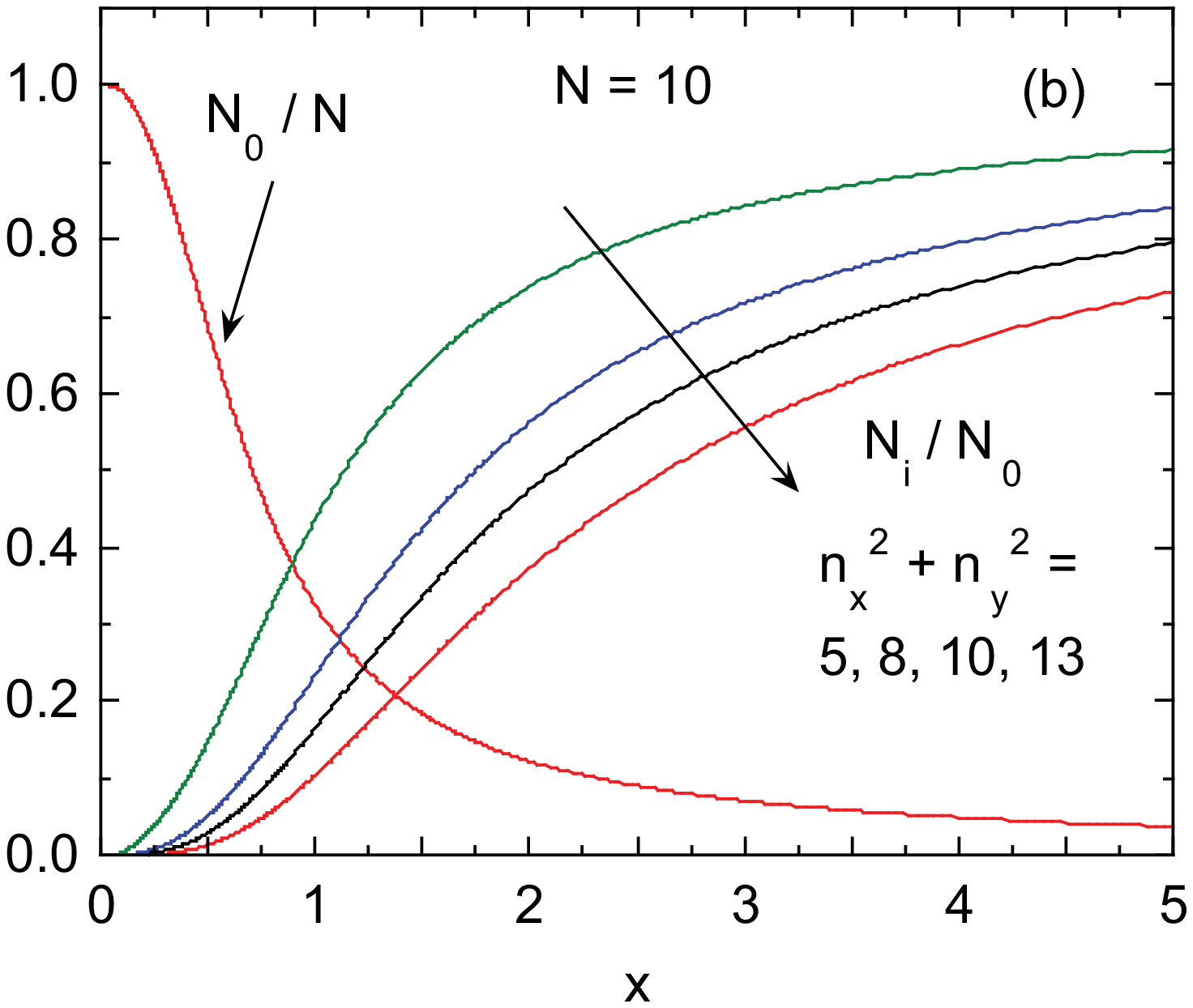}
\includegraphics[width=3.in]{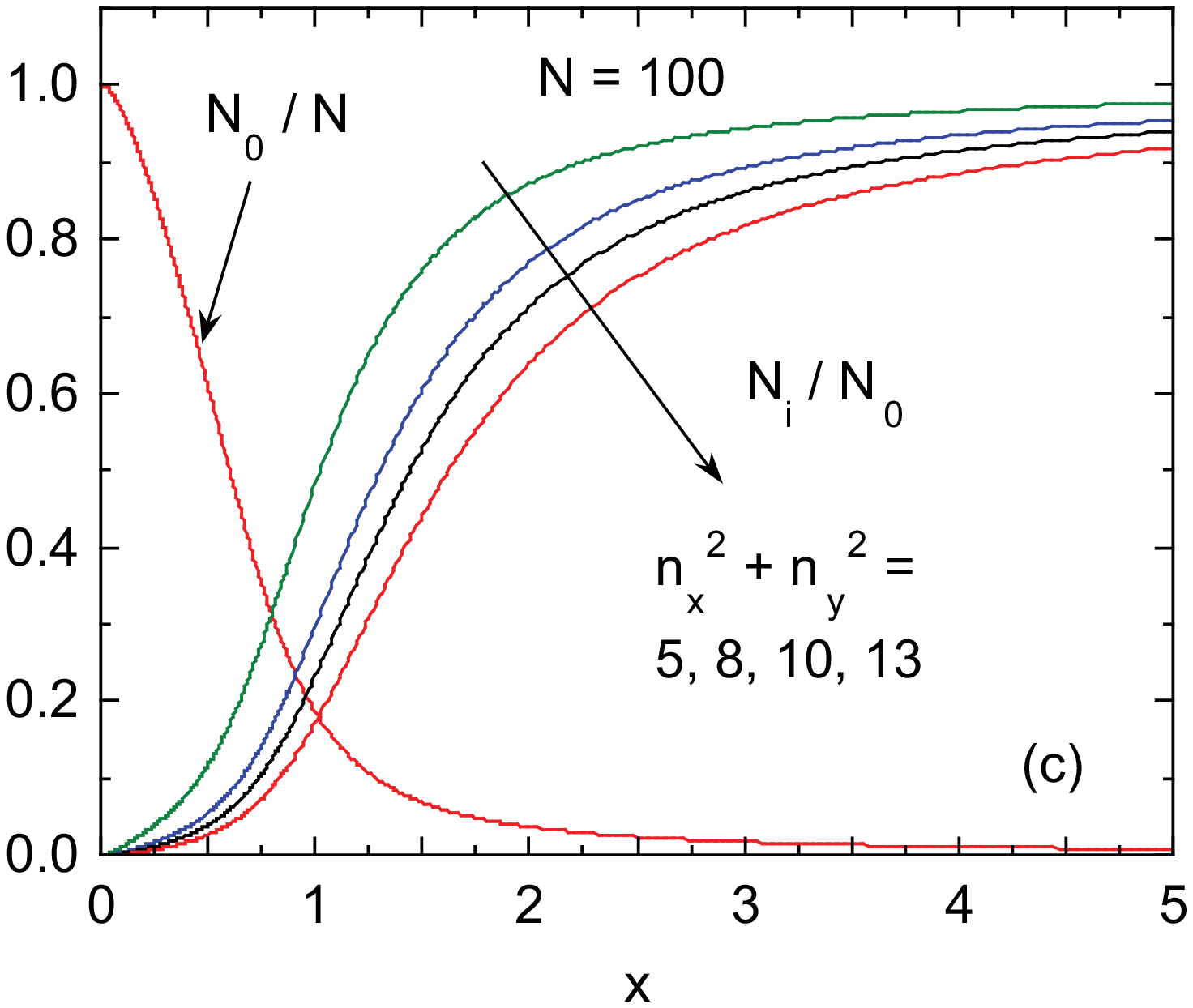}\hspace{0.2in}\includegraphics[width=3.in]{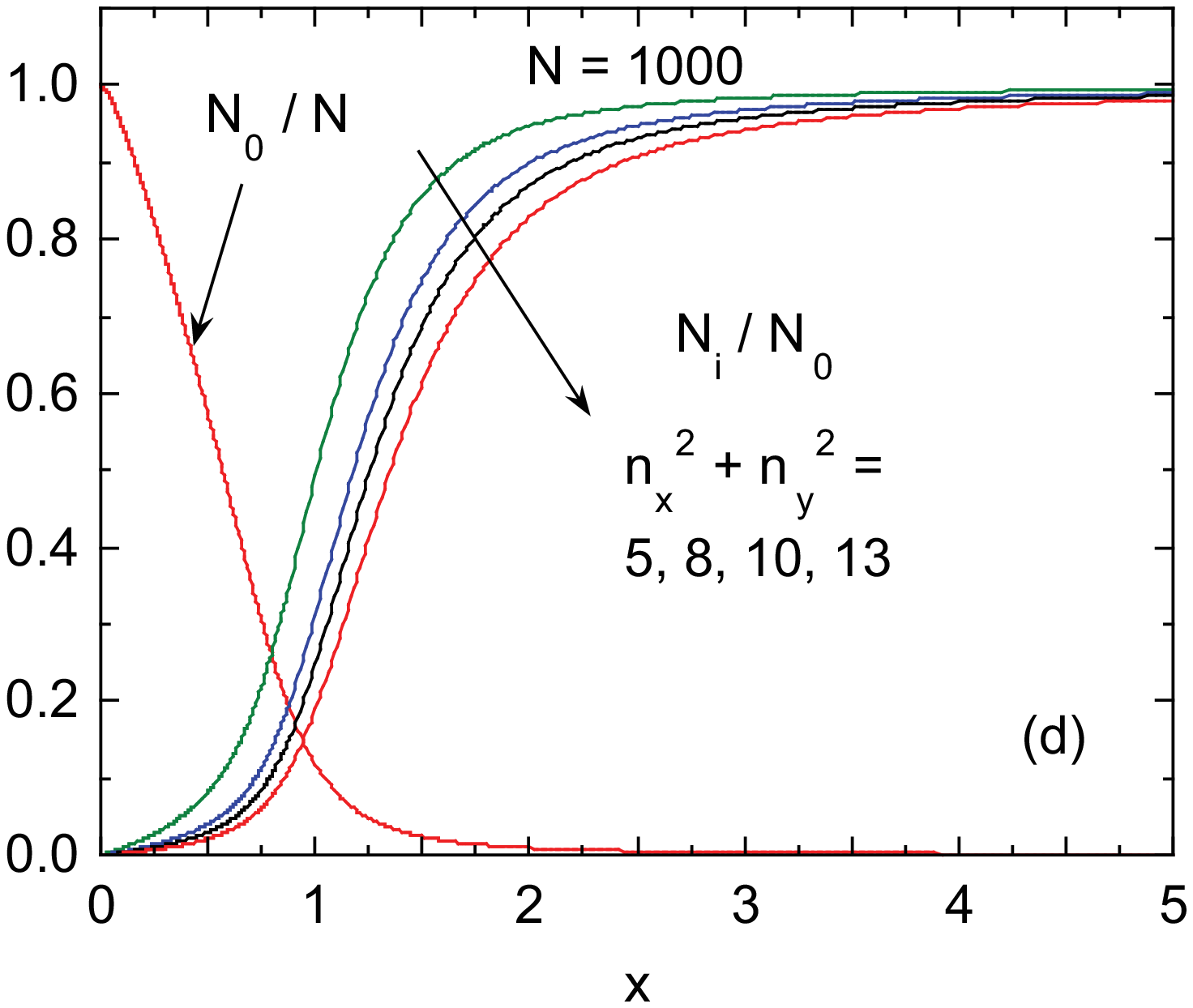}
\caption{(Color online) The ratio $N_0/N$ versus $x$ from Fig.~\ref{Fig:AllDataLoCET_N0N} and the ratios $N_i/N_0$ of the number of bosons $N_i$ occupying a state in each of the first four excited energy levels with quantum numbers $n_x^2+n_y^2 = 5$, 8, 10 and~13 in Eq.~(\ref{Eq:EonT}) to the ground-state occupation number~$N_0$ for (a) $N = 1$, (b) $N=10$, (c) $N = 100$ and (d) $N=1000$.  For the ground state $n_x^2+n_y^2 = 2$. }
\label{Fig:AllDataLoTCE_NiN0}
\end{figure*}

The fractional occupancies $N_0/N$ of the ground state for $N = 1$, 10, 100 and 1000 obtained using Eq.~(\ref{Eq:barni}) are plotted versus $x = \gamma t$ in Fig.~\ref{Fig:AllDataLoCET_N0N} (solid curves).  On comparing the data with those in Fig.~\ref{Fig:AllDataLoT_N0N} for larger $N$ values, one sees that the crossover between weak and strong increases of $N_0/N$ versus $x$ at $x=1$ becomes much less well defined for small~$N$\@.  The data for $N=1000$ from Fig.~\ref{Fig:AllDataLoT_N0N} obtained from the GCE formalism are shown as the dashed curve in Fig.~\ref{Fig:AllDataLoCET_N0N} for comparison.  One sees that the CE and GCE formalisms are in reasonably good agreement for this value of $N$\@.

The ratios $N_i/N_0$ of the populations of a quantum state in each of the first four excited energy levels $i=1-4$ to that in the ground state $N_0$ are plotted versus~$x$ in Fig.~\ref{Fig:AllDataLoTCE_NiN0}.  Compared with the larger-$N$ data in Fig.~\ref{Fig:AllDataLoT_NiN0}, the excited state populations for small $N$ approach the ground state population at much larger $x$ values than for larger~$N$\@.

\subsection{\label{Sec:FSCE} Helmholtz Free Energy,  Entropy and Internal Energy}
 
\begin{figure}
\includegraphics[width=3.3in]{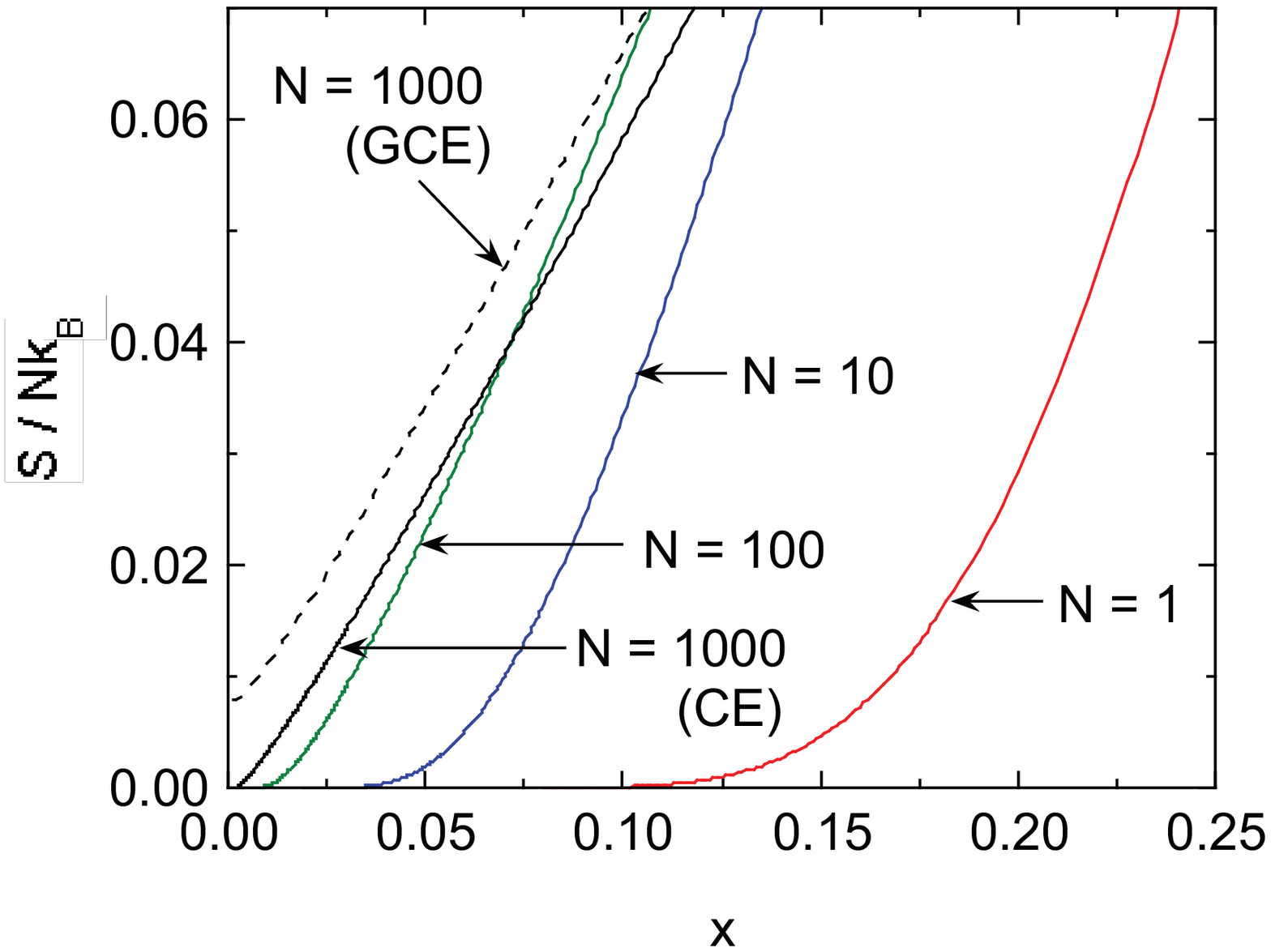}
\caption{(Color online) Normalized entropy per boson $S/N k_{\rm B}$ versus $x = \gamma t$ in the small-$x$ regime for  $N=1$ to 1000 obtained within the canonical ensemble (CE) formalism using Eq.~(\ref{Eq:SCE}) (solid curves).  Also shown is the incorrect prediction for $N=1000$ from Fig.~\ref{Fig:Pbar_vs_t}(a) obtained within the grand canonical ensemble (GCE) formalism (dashed curve).}
\label{Fig:AllDataLoTCE_S}
\end{figure}

Within the CE formalism, we use the same definitions of $t$ and $x$ as given above in Eqs.~(\ref{Eq:tDef}) and~(\ref{Eq:xDef}) and of the ratio $E/k_{\rm B}T$ in Eq.~(\ref{Eq:EonT2}), respectively.  To simplify notation we also define
\be
\overline{\ln Q}(x) = \frac{\ln Q(x)}{N},
\ee
where $Q(x)$ is calculated as described previously in Sec.~\ref{Sec:CEMethods}.  The reduced Helmholtz free energy $\overline{F}$ is given by
\be
\overline{F} = \frac{F}{Nk_{\rm B}T_{\rm E}} = -t\,\overline{\ln Q}(x)
\label{Eq:FRed}
\ee
and the reduced entropy $\overline{S}$ by
\be
\overline{S}(x) = \frac{S}{Nk_{\rm B}} = -\frac{\partial \bar{F}}{\partial t} = \overline{\ln Q}(x) + x\frac{d\overline{\ln Q}(x)}{dx}.
\label{Eq:SCE}
\ee

The exact entropy at $t\to0$ with fixed~$\gamma$ is easily obtained for any finite~$N$\@.  For $x\to0$, only the ground state term with $n_x=n_y=1$ in Eq.~(\ref{Eq:Q1Def}) is significant.  Furthermore, when calculating $Q(N)$ we must hold both $T$ and $A$ constant for each $N$\@.  Then the factor $Q_1(k)$ is
\bse
\be
Q_1(k,x\to0) = \exp\left(-\frac{2k}{g}\right),
\ee
where the expression for $g$ is given in Eqs.~(\ref{Eqs:EonkBTg}).  Inserting this into Eq.~(\ref{Eq:QNcalc}) and carrying out the sum over~$k$ yields
\bea
Q(N,x\to0) &=& \exp\left[-\frac{2N}{g}\right] = \exp\left[-\frac{2a(N)}{x}\right],\nonumber\\*
\overline{\ln Q}(N,x\to0) &=& \frac{\ln Q}{N} = -\frac{2a(N)}{N x}. \label{Eq:lnQNT0}
\eea
\ese
Then the reduced free energy is obtained from  Eq.~(\ref{Eq:FRed}) as
\be
\bar{F}(x=0) = \frac{2a(N)}{N\gamma},
\ee
where we used the definition $x = \gamma t$.  From the relation $\bar{S} = -\partial \bar{F}(t,\gamma,N)/\partial t$ one obtains the zero-temperature entropy
\be
\bar{S}(t=0) = 0,
\label{Eq:St0}
\ee
which is valid for arbitrary finite $N$\@.  This result makes physical sense, because there is only one way to put $N$ indistinguishable bosons into an orbitally nondegenerate ground state.

The reduced entropy within the CE formalism obtained from Eq.~(\ref{Eq:SCE}) is plotted versus $x$ at small~$x\leq 1$ for $N= 1$, 10, 100 and~1000 in Fig.~\ref{Fig:AllDataLoTCE_S}.  Also shown as the dashed curve are the data for $N=1000$ obtained from the GCE formalism in Fig.~\ref{Fig:AllDataLoT_S}(a).  One sees that the (incorrect) finite value of $\overline{S}$ for $t\to0$ and $N=1000$ obtained with the GCE formalism is corrected using the CE formalism.

The reduced internal energy $\overline{U}$ is 
\be
\overline{U} = \frac{U}{Nk_{\rm B}T_{\rm E}} = \bar{F} + t\bar{S} = t^2\gamma\frac{d\overline{\ln Q}}{dx},
\ee
or 
\be
\frac{U}{Nk_{\rm B}T} = x\frac{d\overline{\ln Q}}{dx},
\label{Eq:UonNkT}
\ee
where we used the relation $x=\gamma t$.  We find $C_{\rm V}(t\to0)=0$ and do not present plots of $C_{\rm V}(t)$ because they are similar to those in Fig.~\ref{Fig:AllDataLoTHiT_CV} obtained using the GCE formalism.

\subsection{\label{Sec:pCE} Pressure}

\begin{figure*}
\includegraphics[width=3.3in]{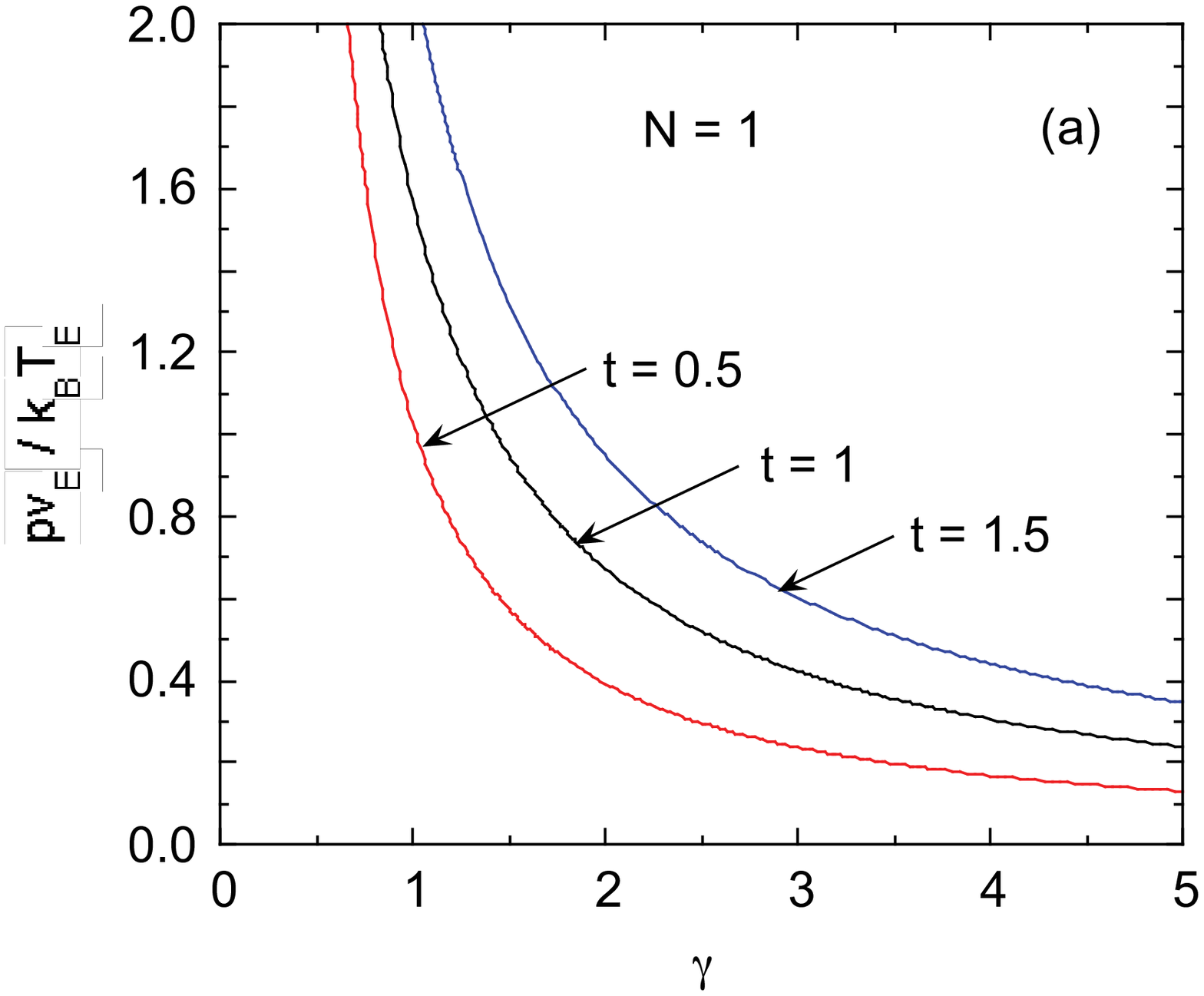}\hspace{0.2in}\includegraphics[width=3.3in]{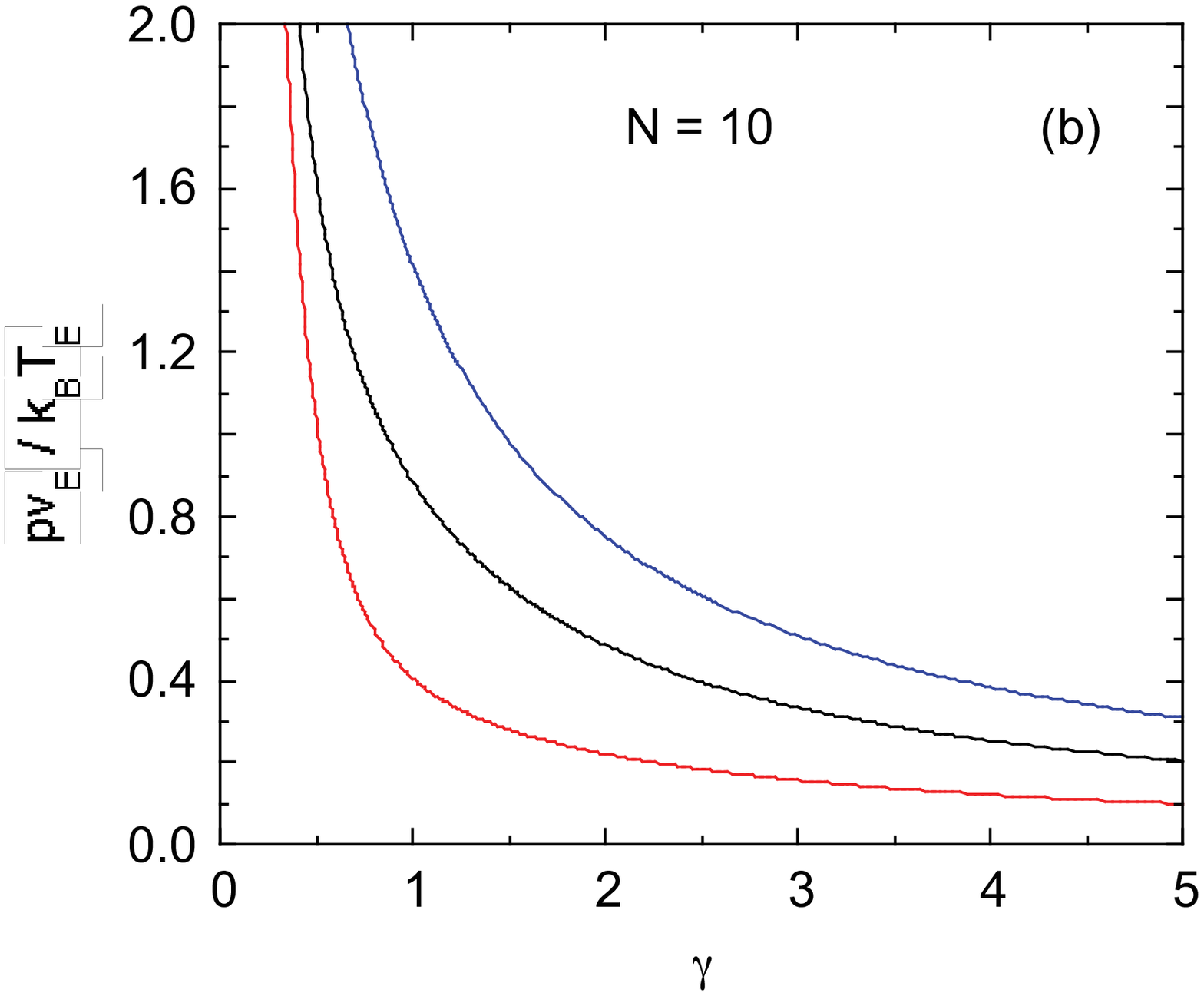}
\includegraphics[width=3.3in]{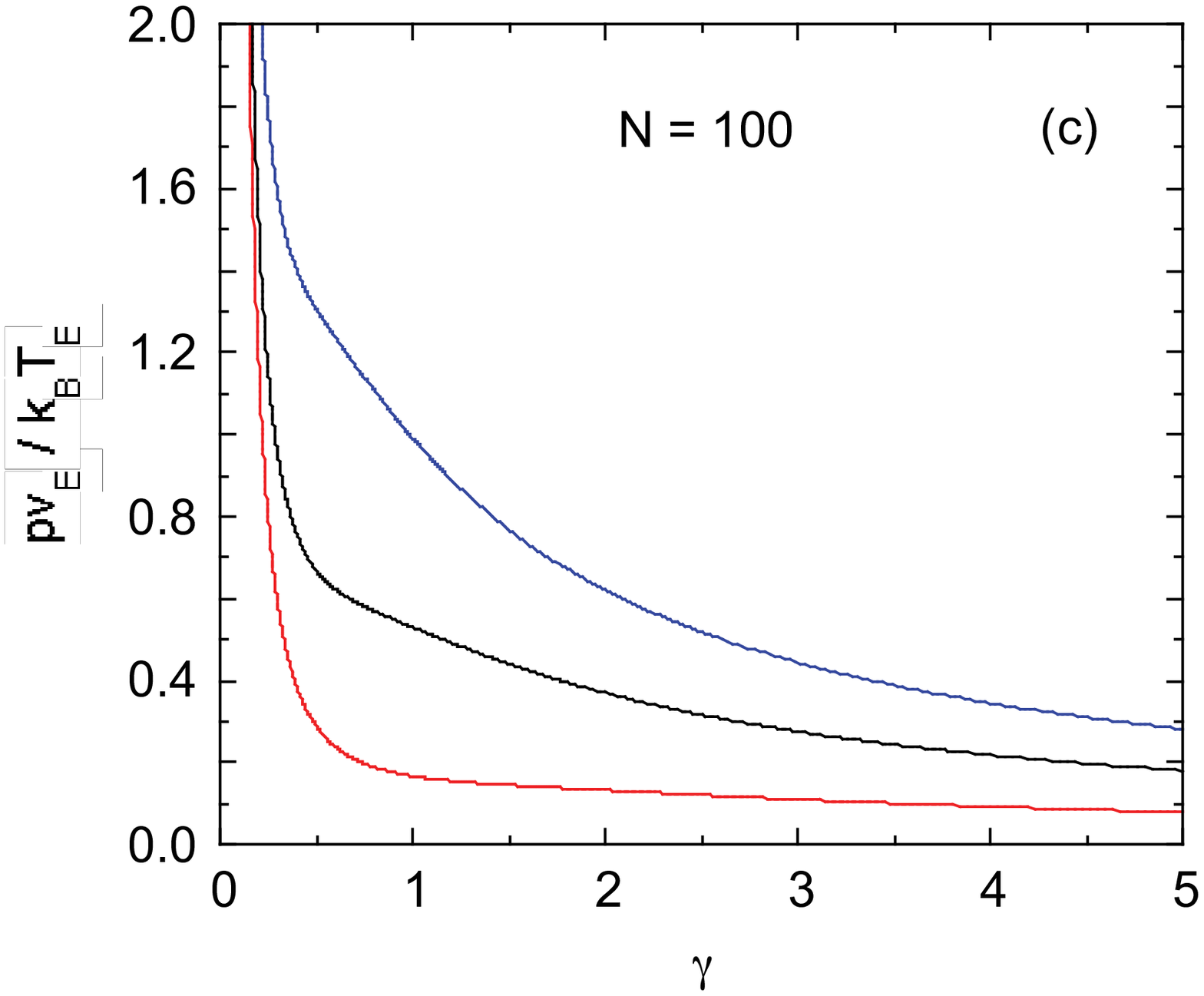}\hspace{0.2in}\includegraphics[width=3.3in]{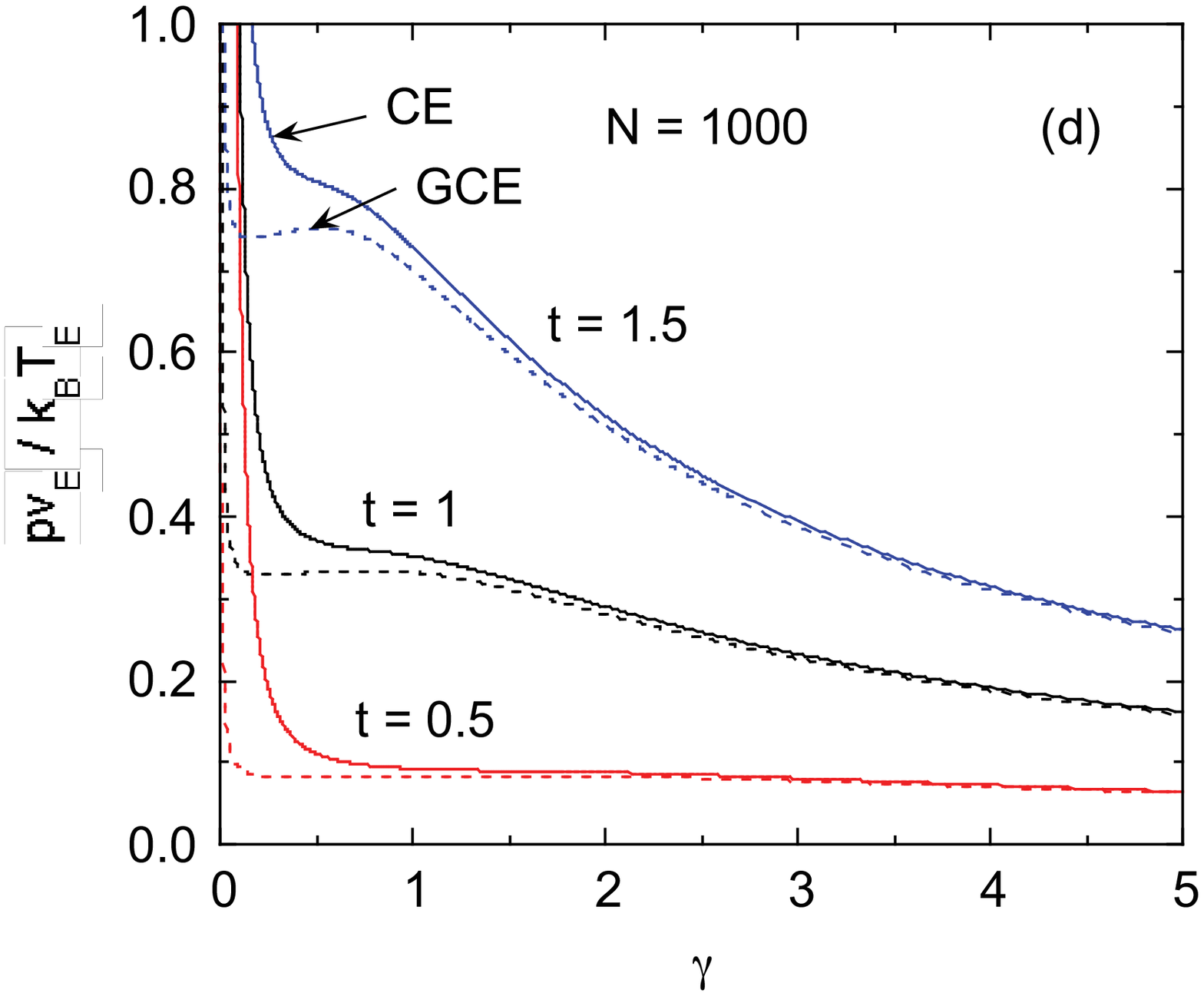}
\caption{(Color online) Reduced pressure $\bar{p} = p\,v_{\rm E}/k_{\rm B}T_{\rm E}$ versus reduced area $\gamma = v/v_{\rm E}$ isotherms at reduced temperatures $t=T/T_{\rm E} = 0.5$, 1 and~1.5 and boson numbers (a)~$N = 1$, (b)~$N = 10$, (c)~$N = 100$ and (d)~$N = 1000$ (solid curves) obtained using the canonical ensemble (CE) formalism.  Also shown in~(d) are the corresponding incorrect data obtained using the grand canonical ensemble (GCE) formalism for $N=1000$ in Fig.~\ref{Fig:Pbar_vs_gam}(a) (dashed curves).  Note that the scale of the ordinate in~(d) is different than in (a)--(c).}
\label{Fig:Pbar_vs_gamCE}
\end{figure*}  

The reduced pressure $\bar{p}$ within the CE formalism is given by
\be
\bar{p} = \frac{pV_{\rm E}}{Nk_{\rm B}T_{\rm E}} = -\left(\frac{\partial \bar{F}}{\partial\gamma}\right)_t = t^2\frac{d\overline{\ln Q}(x)}{dx}.
\label{Eq:barpQ}
\ee
The compression factor is 
\be
\tilde{\bar{p}} =\frac{pA}{Nk_{\rm B}T}= \bar{p}\,\frac{\gamma}{t} = x\frac{d\overline{\ln Q}(x)}{dx}.
\label{Eq:pbarbarCE}
\ee
Comparing Eqs.~(\ref{Eq:pbarbarCE}) and~(\ref{Eq:UonNkT}) demonstrates that 
\be
p = \frac{U}{A},
\ee
which says that the pressure is equal to the average energy density.  This type of relationship is expected from dimensional considerations.  For 3D Bose and Fermi gases in the thermodynamic limit, one obtains the similar expression $p = (2/3)U/V$, where $V$ is the volume of the gas.\cite{Huang1963}  Plots of $\tilde{\bar{p}} = \frac{pA}{Nk_{\rm B}T}$ versus~$x$ within the CE formalism are similar to those of the GCE formalism in Fig.~\ref{Fig:Pbarbar_vs_x}(a) and~\ref{Fig:Pbarbar_vs_x}(b) and are therefore not presented here.  

We solve Eq.~(\ref{Eq:barpQ}) parametrically using $x$ as an implicit parameter.  We first calculate $\overline{\ln Q}(x)$. Then at constant $\gamma$, one has $t=x/\gamma$ for a pressure versus temperature isochore, whereas at constant $t$ one has $\gamma=x/t$ for a pressure versus area isotherm.  Shown in Fig.~\ref{Fig:Pbar_vs_gamCE} are $\bar{p}$ versus $\gamma$ isotherms at $t=0.5$, 1 and~1.5 for $N=1$, 10, 100 and~1000 in panels (a), (b), (c) and (d), respectively.  As $N$ increases, a hump appears at the crossover area $\gamma\sim1$ for $N=100$ that is clearly defined by $N=1000$.  An important feature of these plots is that the slope is always negative.  This means that $\kappa_{\rm T}$ is always finite and positive.  This behavior is in contrast to the data for $N=1000$ in Fig.~\ref{Fig:Pbar_vs_gamCE}(d) obtained using the GCE formalism, where one sees maxima in $\bar{p}$ versus $\gamma$ at $\gamma\sim1$, which causes $\kappa_{\rm T}$ to exhibit an unphysical divergence on reducing $\gamma$ towards $\gamma\sim 1$ and then unphysical negative values at lower $\gamma$ values.

\begingroup
\squeezetable
\begin{table}[h]
\caption{\label{Tab:p0} Reduced pressure $\bar{p}$ for reduced areas $\gamma = 0.5$, 1 and~1.5 and reduced isothermal compressibility $\kappa_{\rm T}/\gamma^2$, all at zero temperature, versus~$N$ as predicted by the cononical ensemble formalism via Eqs.~(\ref{Eq:p0}) and~(\ref{Eq:kappaTCET0}), respectively.}
\begin{ruledtabular}
\begin{tabular}{ccccc}
	$\log_{10}N$	&	 $\bar{p}(t=0)$	& $\bar{p}(t=0)$	&  $\bar{p}(t=0)$ & $\kappa_{\rm T}(t=0)/\gamma^2$\\
 	& $\gamma = 0.5$  &  $\gamma = 1$  & $\gamma = 1.5$ \\
\hline
0  &  3.3231E+00  &  8.3077E$-$01  &  3.6923E$-$01  &  6.0185E$-$01  \\
1  &  8.8900E$-$01  &  2.2225E$-$01  &  9.8778E$-$02  &  2.2497E+00  \\
2  &  1.7417E$-$01  &  4.3542E$-$02  &  1.9352E$-$02  &  1.1483E+01  \\
3  &  2.8031E$-$02  &  7.0078E$-$03  &  3.1146E$-$03  &  7.1349E+01  \\
4  &  3.9907E$-$03  &  9.9768E$-$04  &  4.4341E$-$04  &  5.0116E+02  \\
5  &  5.2505E$-$04  &  1.3126E$-$04  &  5.8339E$-$05  &  3.8092E+03  \\
6  &  6.5526E$-$05  &  1.6382E$-$05  &  7.2807E$-$06  &  3.0522E+04  \\
7  &  7.8814E$-$06  &  1.9704E$-$06  &  8.7571E$-$07  &  2.5376E+05  \\
8  &  9.2284E$-$07  &  2.3071E$-$07  &  1.0254E$-$07  &  2.1672E+06  \\
9  &  1.0589E$-$07  &  2.6471E$-$08  &  1.1765E$-$08  &  1.8888E+07  \\
10  &  1.1959E$-$08  &  2.9897E$-$09  &  1.3287E$-$09  &  1.6724E+08  \\
11  &  1.3353E$-$09  &  3.3383E$-$10  &  1.4837E$-$10  &  1.4978E+09  \\
12  &  1.4722E$-$10  &  3.6804E$-$11  &  1.6357E$-$11  &  1.3585E+10  \\
13  &  1.6115E$-$11  &  4.0289E$-$12  &  1.7906E$-$12  &  1.2410E+11  \\
14  &  1.7499E$-$12  &  4.3747E$-$13  &  1.9443E$-$13  &  1.1429E+12  \\
15  &  1.8905E$-$13  &  4.7262E$-$14  &  2.1006E$-$14  &  1.0579E+13  \\
16  &  2.0311E$-$14  &  5.0777E$-$15  &  2.2568E$-$15  &  9.8470E+13  \\
17  &  2.1666E$-$15  &  5.4165E$-$16  &  2.4073E$-$16  &  9.2311E+14  \\
\end{tabular}
\end{ruledtabular}
\end{table}
\endgroup

As noted in the introduction, a nonzero pressure must occur at $t=0$ in a noninteracting Bose gas in a 2D box with Dirichlet boundary conditions because the ground state energy depends on the area.\cite{Grossman1995}  At $t=0$, all $N$ bosons are in the ground state with $n_x=n_y=1$.  From Eq.~(\ref{Eq:barEDef}), the energy of the ground state ($n_x=n_y=1$) containing $N$ bosons at $t=0$  is
\be
\frac{NE_0(t=0)}{k_{\rm B}T_{\rm E}} = \frac{2a(N)}{\gamma} = \frac{2a(N)A_{\rm E}}{A}.
\ee
Then the pressure~$p$ at $t=0$ is
\be
\frac{p(t=0)}{k_{\rm B}T_{\rm E}} = -\frac{\partial(NE_0/k_{\rm B}T_{\rm E})}{\partial A} = \frac{2a(N)A_{\rm E}}{A^2}
\ee
Using the definitions $\gamma=(A/N)/(A/N)_{\rm E} = A/A_{\rm E}$ and of $\bar{p}$ in Eq.~(\ref{Eq:barpQ}),  one obtains the reduced pressure
\be
\bar{p}(t=0) = \frac{2a(N)}{N\gamma^2}.
\label{Eq:p0}
\ee

\begin{figure*}
\includegraphics[width=3.3in]{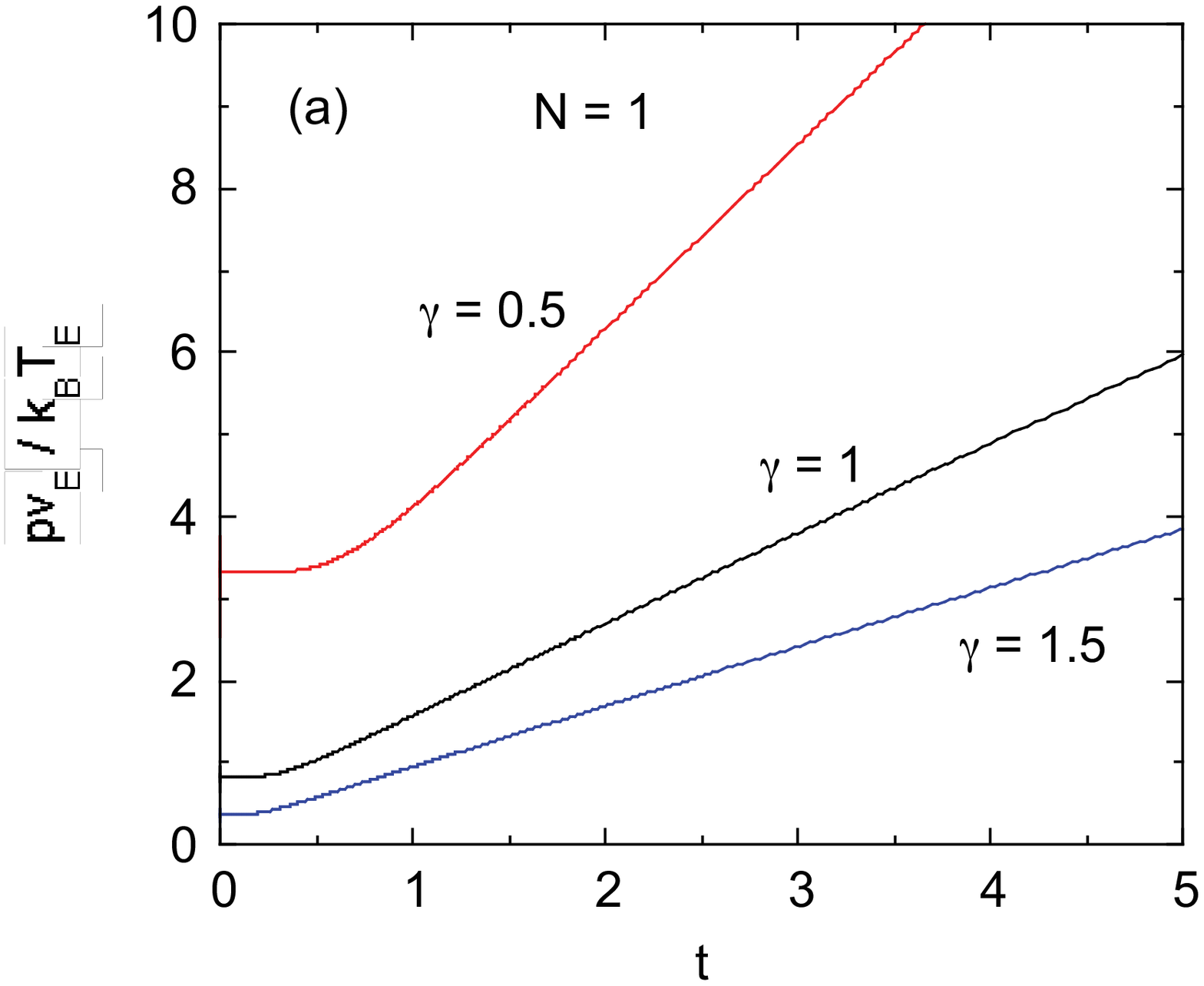}\hspace{0.2in}\includegraphics[width=3.3in]{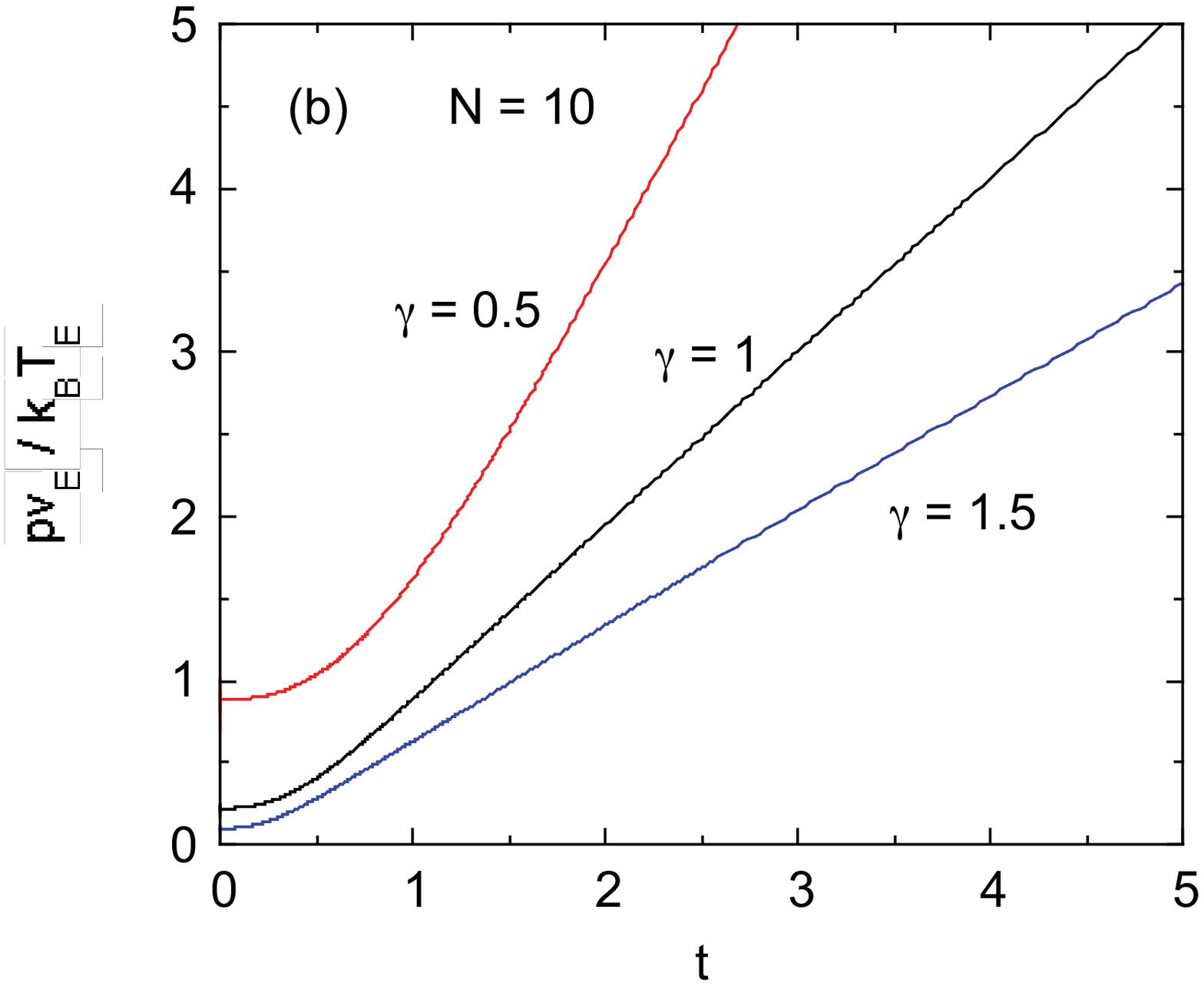}
\includegraphics[width=3.3in]{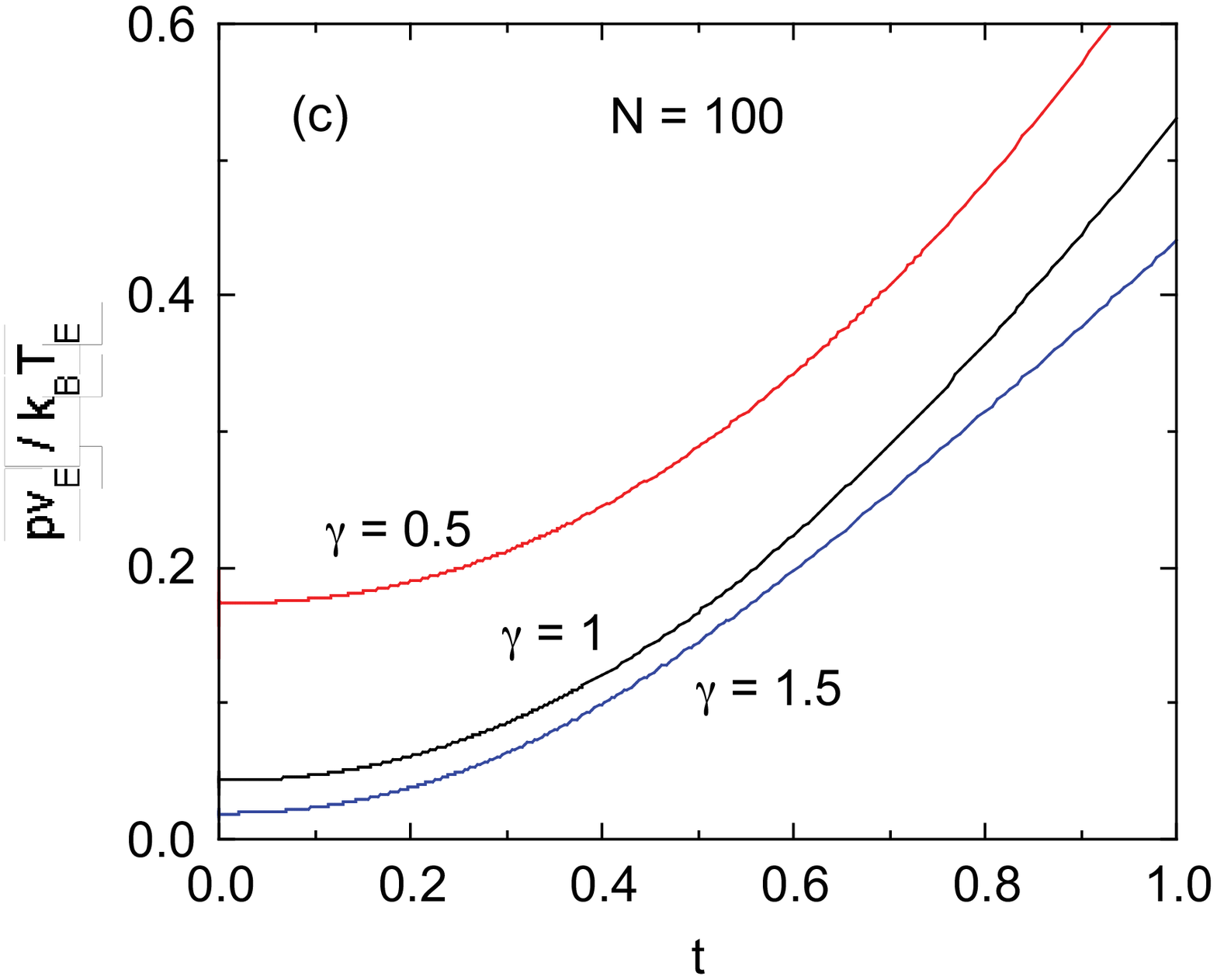}\hspace{0.2in}\includegraphics[width=3.3in]{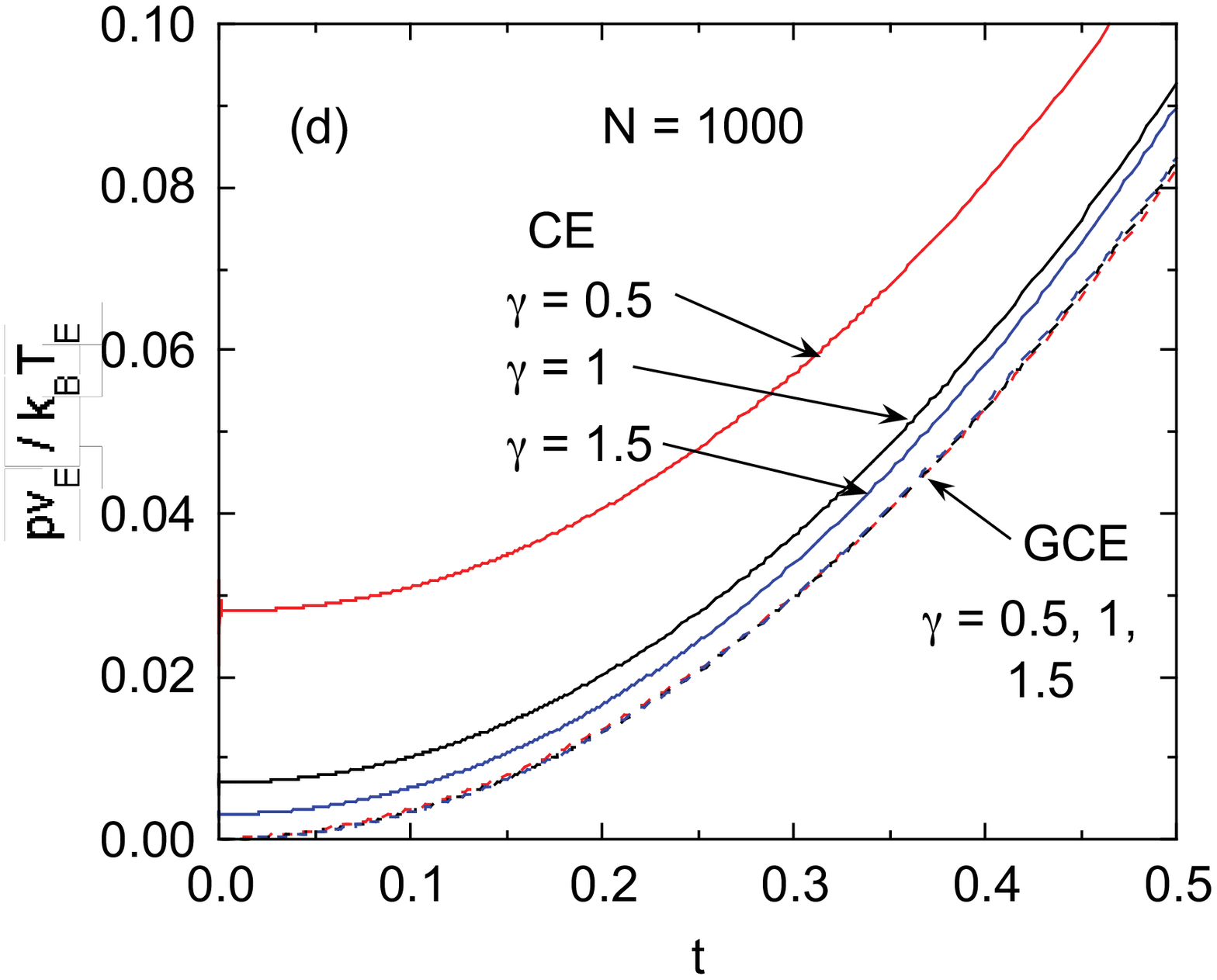}
\caption{(Color online) Reduced pressure $\bar{p} = p\,v_{\rm E}/(k_{\rm B}T_{\rm E})$ versus reduced temperature~$t$ isochores at low~$t$ at fixed reduced areas $\gamma = 0.5$, 1 and~1.5 and boson numbers (a)~$N = 1$, (b)~$N = 10$, (c)~$N = 100$ and (d)~$N = 1000$ (solid curves) obtained using the canonical ensemble (CE) formalism.  Also shown in~(d) are the corresponding data obtained using the grand canonical ensemble (GCE) formalism (dashed curves) from Fig.~\ref{Fig:Pbar_vs_t}(a).  The axis scales are different in each plot in order to emphasize the nonzero values of $\bar{p}(t\to0)$ obtained with the CE formalism.}
\label{Fig:Pbar_vs_tCE}
\end{figure*}

Values of $\bar{p}(t=0)$ for $N=10^0$ to~$10^{17}$ obtained from Eq.~(\ref{Eq:p0}) using the values of $a(N)$ in Table~\ref{Tab:aVSn} are listed in Table~\ref{Tab:p0} for $\gamma=0.5$, 1 and~1.5.  One sees that $\bar{p}(t=0)$ is quite large for small~$N$, but decreases rapidly as $N$ increases.  Isochores of $\bar{p}$ versus~$t$ for $\gamma = 0.5$, 1 and~1.5 obtained using Eq.~(\ref{Eq:barpQ}) are shown at low~$t$ for $N=1$, 10, 100 and~1000 in panels (a), (b), (c) and (d) of Fig.~\ref{Fig:Pbar_vs_tCE}, respectively.  One indeed sees that $\bar{p}(t\to0)$ decreases rapidly with increasing~$N$, with the $t=0$ values in agreement with those listed in Table~\ref{Tab:p0}.  Also shown in Fig.~\ref{Fig:Pbar_vs_tCE}(d) are the corresponding isochores in Fig.~\ref{Fig:Pbar_vs_t}(a) obtained from the GCE formalism (dashed curves), for which the incorrect limit $\bar{p}(t\to0)=0$ is obtained for each of the three $\gamma$ values.

\subsection{\label{Sec:kTapCpCE} Isothermal Compressibility, Thermal Expansion Coefficient, Heat Capacity at Constant Pressure}

\begin{figure*}
\includegraphics[width=3.in]{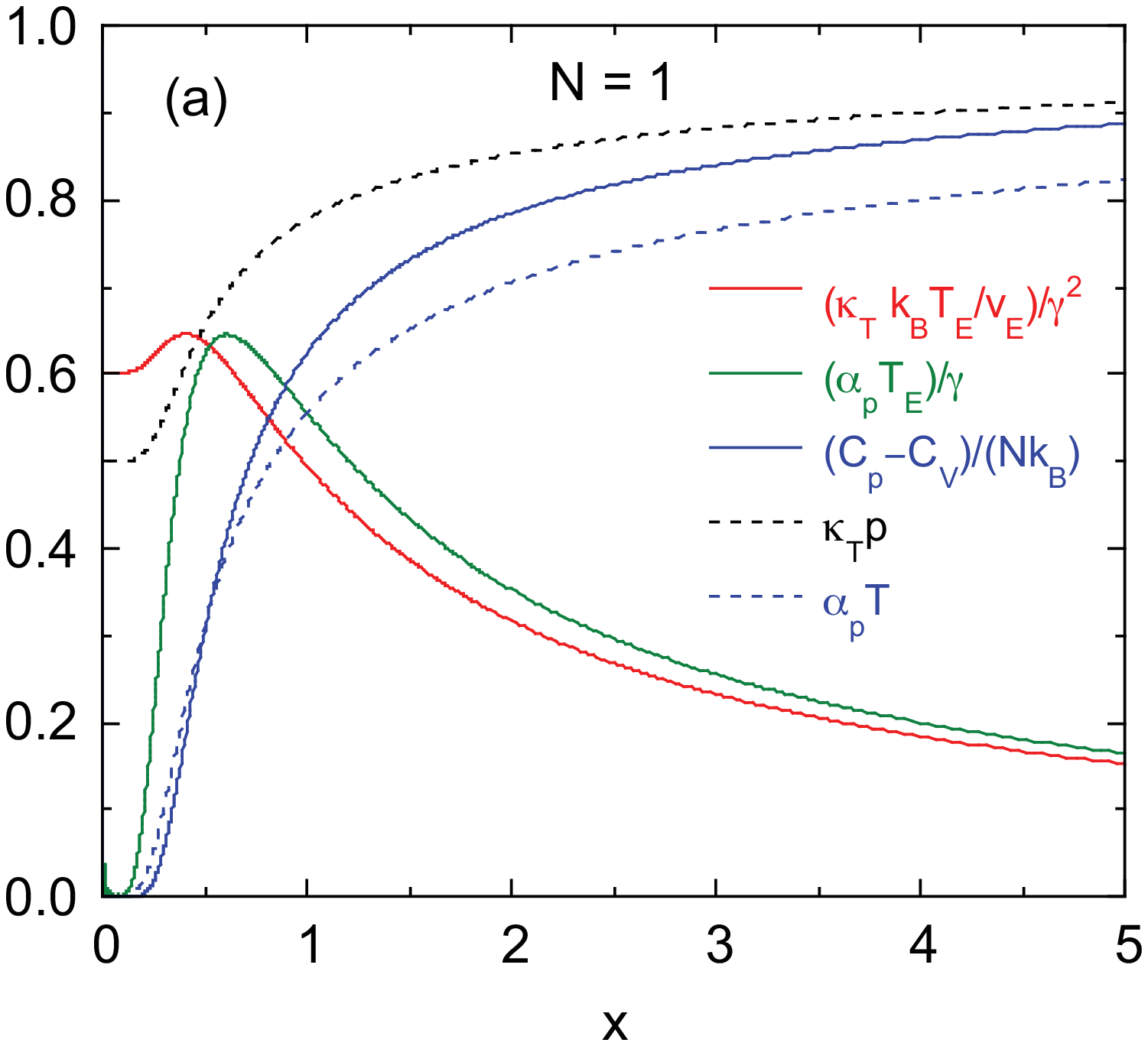}\hspace{0.2in}\includegraphics[width=3.in]{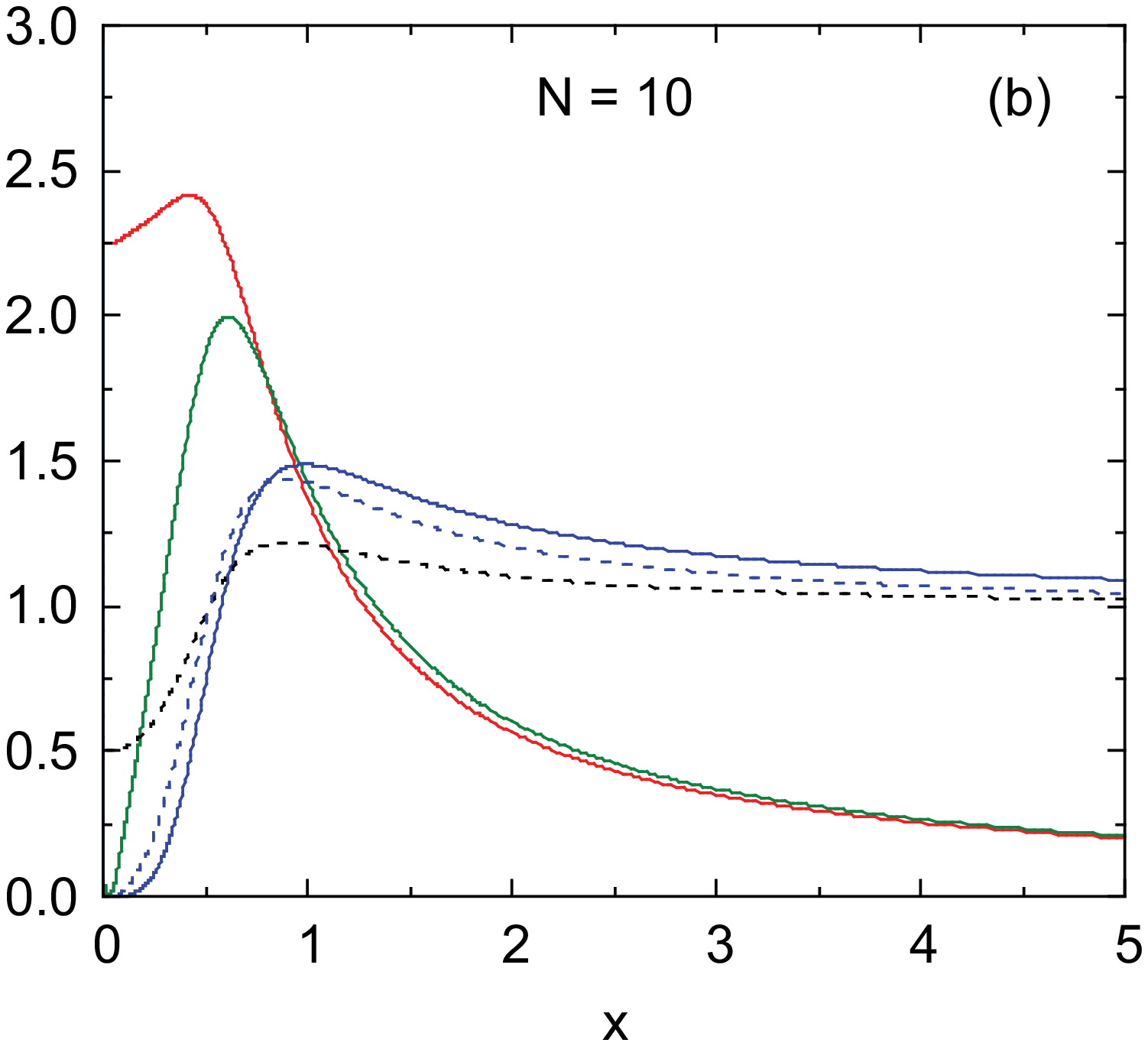}
\includegraphics[width=3.in]{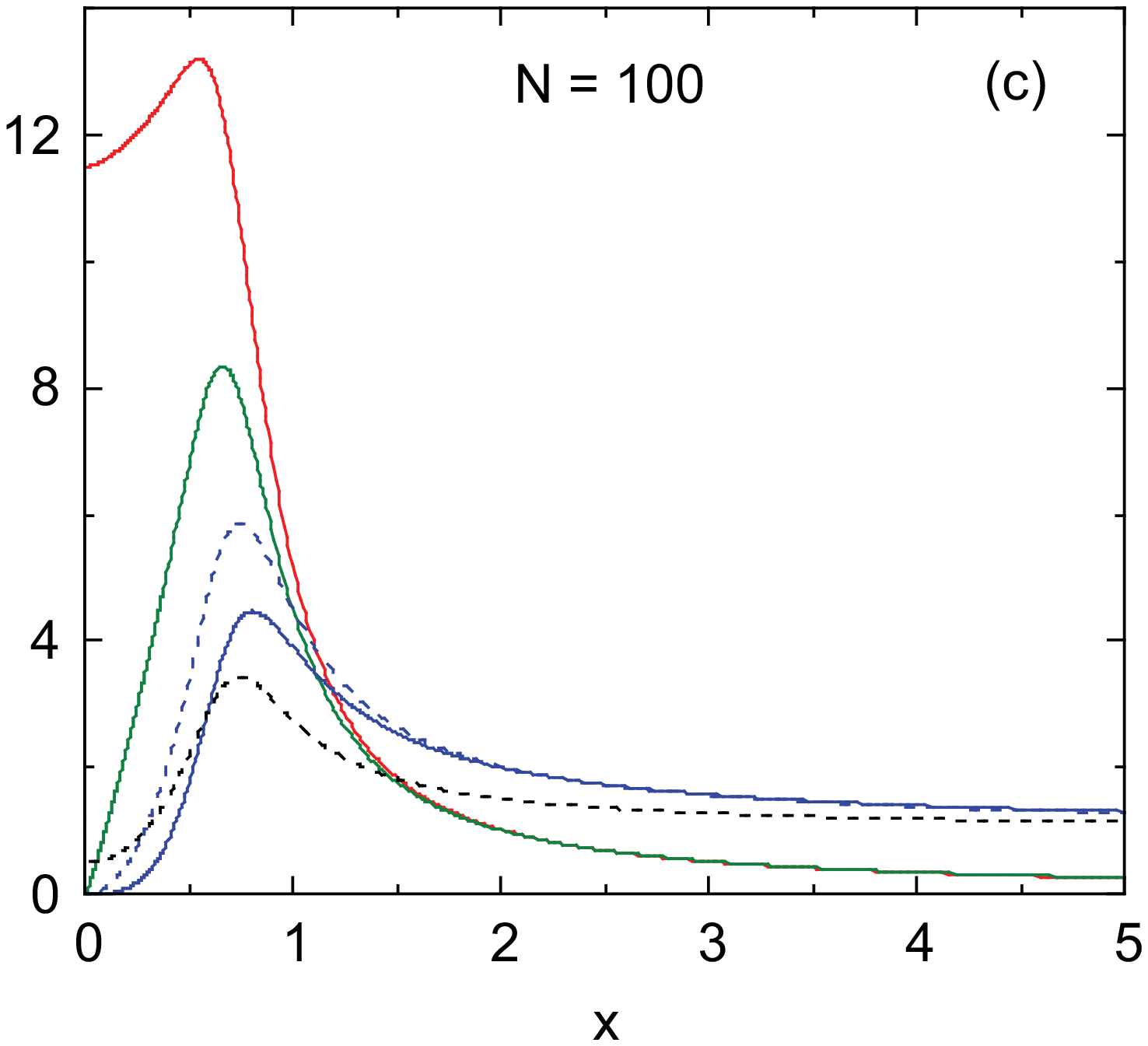}\hspace{0.2in}\includegraphics[width=3.in]{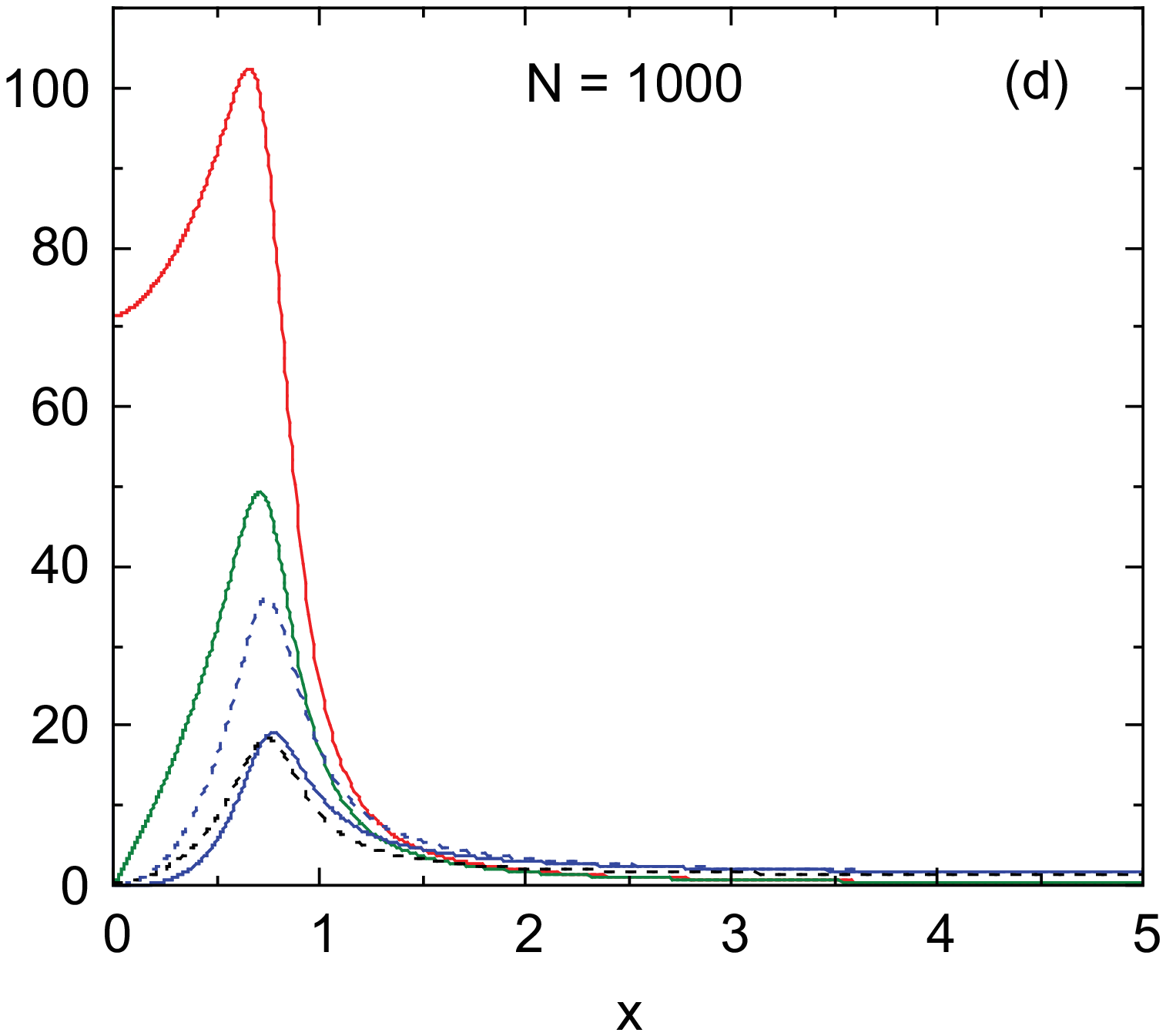}
\caption{(Color online) Reduced isothermal compressibility $\bar{\kappa}_{\rm T}/\gamma^2$, thermal expansion coefficient $\bar{\alpha}_{\rm p}/\gamma$, and normalized difference $(C_{\rm p}-C_{\rm V})/Nk_{\rm B}$ between the heat capacity at constant pressure and at constant volume versus $x=\gamma t$ for boson numbers (a)~$N=1$, (b) $N=10$, (c)~$N=100$ and (d)~$N=1000$.  Also shown in each panel are the products $\alpha_{\rm p}T$ and $\kappa_{\rm T}p$ which for an ideal gas are both equal to unity, as expected and found in the respective large-$x$ limits in (a)--(d).  The figure legend in (a) applies to all panels.  Note the different ordinate scales for each panel.}
\label{Fig:kappa_alpha_Cp_CE}
\end{figure*}

The reduced isothermal compressibility $\bar{\kappa}_{\rm T}/\gamma^2$ is
\be
\frac{\bar{\kappa}_{\rm T}}{\gamma^2}(x) = \kappa_{\rm T}\frac{Nk_{\rm B}T_{\rm E}}{\gamma^2V_{\rm E}}= -\frac{1}{\gamma^3\left(\frac{\partial \bar{p}}{\partial\gamma}\right)_t} = -\frac{1}{x^3\frac{d^2\overline{\ln Q}(x)}{dx^2}},
\label{Eq:kappaCE}
\ee
where the large-$x$ ideal gas limit $\kappa_{\rm T}p =\bar{\kappa}_{\rm T}\bar{p} (x\to\infty)=1$ is expected.

The reduced thermal expansion coefficient $\bar{\alpha}_{\rm p}/\gamma$ is
\be
\frac{\bar{\alpha}_{\rm p}}{\gamma}(x) = \frac{\alpha_{\rm p}T_{\rm E}}{\gamma} = -\frac{1}{\gamma^2}\frac{\left(\frac{\partial\bar{p}}{\partial t}\right)_\gamma}{\left(\frac{\partial\bar{p}}{\partial\gamma}\right)_t} = -\frac{2\frac{d\overline{\ln Q}}{dx}}{x^2\frac{d^2\overline{\ln Q}}{dx^2}} - \frac{1}{x},
\label{Ea:alphaCE}
\ee
where the large-$x$ limit is expected to be the ideal gas value $\alpha_{\rm p}T = \bar{\alpha}_{\rm p}x(x\to\infty)=1$.  The normalized difference between the heat capacities at constant pressure $C_{\rm p}$ and constant volume (constant area) $C_{\rm V}$ is given by
\be
\frac{C_{\rm p}-C_{\rm V}}{Nk_{\rm B}}(x) = x\,\frac{(\bar{\alpha}_{\rm p}/\gamma)^2}{\bar{\kappa}_{\rm T}/\gamma^2}.
\label{Eq:CpCVCE}
\ee
All of these quantities are plotted versus $x$ in Fig.~\ref{Fig:kappa_alpha_Cp_CE} for $N = 1$, 10, 100 and~1000 in panels (a)--(d), respectively.  One sees that with the CE formalism, one does not encounter the unphysical divergences and other inaccuracies discussed above that occur with the GCE formalism at small $x$ and~$N$ values where BEC comes significant.

Using Eq.~(\ref{Eq:lnQNT0}) for $\overline{\ln Q}(x\to0)$ together with the general definition for $\bar{\kappa}_{\rm T}/\gamma^2$ in Eq.~(\ref{Eq:kappaCE}), one obtains the zero-temperature limit at fixed~$\gamma$ given by
\be
\frac{\bar{\kappa}_{\rm T}}{\gamma^2}(t\to0) = \frac{N}{4a(N)}.
\label{Eq:kappaTCET0}
\ee
A list of values of $\frac{\bar{\kappa}_{\rm T}}{\gamma^2}(t\to0)$ versus $N$ obtained using Eq.~(\ref{Eq:kappaTCET0}) is given in the fifth column of Table~\ref{Tab:p0}, where to calculate these we used the $a(N)$ values in Table~\ref{Tab:aVSn}.  Similarly, we find that $(\bar{\alpha}_{\rm p}/\gamma) (t=0)=0$ and hence $(C_{\rm p}-C_{\rm V})(t=0)=0$ using Eq.~(\ref{Eq:CpCVCE}).  These zero-temperature results are in agreement with the $x=0$ limits of the respective plots in Fig.~\ref{Fig:kappa_alpha_Cp_CE}.

\section{\label{Summary} Summary}

We confirmed the literature result that BEC does not occur in the thermodynamic limit at finite temperature~$T$ in a noninteracting Bose gas confined to a 2D box.  However, as also previously reported for finite $N$, BEC does occur in 2D, where the ground state boson occupation is $N_0/N\to1$ at fixed area~$A$ for $T\to0$, but without any phase transition occurring.\cite{Ingold1998}  The lack of a phase transition is confirmed from the analytic behavior of the calculated $C_{\rm V}(T)$ upon traversing the characteristic temperature $T_{\rm E}$. Thus the parameter $T_{\rm E}$ that we define corresponds to a crossover temperature between weak and strong increases in $N_0/N$ and in the low-lying excited states with decreasing~$T$ at fixed~$A$ and not to a phase transition temperature.  We find that $T_{\rm E}$ decreases with increasing $N$ according to $T_{\rm E}\sim 1/\log(N)$ at fixed area per boson $(A/N)_{\rm E}$ yielding $T_{\rm E}(N\to\infty)=0$. Hence BEC is precluded at finite~$T$ in the thermodynamic limit in 2D whereas it does occur at low~$T$ with finite $N$ in the absence of a BEC phase transition, a perhaps counterintuitive result.

The main contribution of this paper is a comprehensive and detailed study of the thermodynamic properties of noninteracting bosons in a 2D box with Dirichlet boundary conditions.  Such a study  has not been carried out before to our knowledge and is therefore a benchmark for future studies on similar systems.  We used both the GCE and CE formalisms for the calculations.  The GCE formalism generally gives accurate results for the thermodynamic properties at large $N$ and large values of the product $TA$, but fails to give correct results for small $N$ at small $TA$ values where significant BEC occurs.  Such failures of the GCE formalism in the latter ranges of parameters include incorrect predictions of nonzero entropy and zero pressure, strong deviations of the ratio of the pressure to the energy density $p/(U/A)$ from the exact CE value of unity, and divergent and/or negative values of $\kappa_{\rm T}$, $\alpha_{\rm p}$ and $C_{\rm p}$.  These incorrect behaviors predicted by the GCE formalism are revealed using the CE formalism which permits numerically and analytically exact results to be obtained, albeit at comparatively small~$N$\@.  Thus apart from the specific study reported here, we hope that the present results will be more generally useful because they illustrate several generic shortcomings of the GCE formalism in predicting the thermodynamic properties of finite quantum boson systems. 

\acknowledgments

DCJ is grateful to Professor Fanlong Ning and the Department of Physics of Zhejiang University for the gracious hospitality during the visit at which this work was initiated.

\end{document}